\documentclass[11pt,a4paper]{article}
\usepackage[utf8]{inputenc}
\usepackage[english]{babel}
\usepackage{jheppub}
\usepackage{doi}

\usepackage{graphicx}
\usepackage{subcaption}

\usepackage{tikz}
\usetikzlibrary{decorations.markings}
\usetikzlibrary{decorations.pathmorphing}
\usetikzlibrary{positioning}
\usetikzlibrary{cd}

\usepackage{amsmath,amssymb,amsthm,amscd,graphicx,mathabx}
\usepackage{physics}
\usepackage{dsfont}
\usepackage{seqsplit}
\usepackage{graphicx}
\usepackage{subcaption}
\usepackage{supertabular}

\newcommand{\be}{\begin{equation}}
	\newcommand{\ee}{\end{equation}}
\newcommand{\ba}{\begin{align}}
	\newcommand{\ea}{\end{align}}
\newcommand{\ben}{\begin{eqnarray}\displaystyle}
	\newcommand{\een}{\end{eqnarray}}

\providecommand{\ceil}[1]{\left \lceil #1 \right \rceil }
\providecommand{\floor}[1]{\left \lfloor #1 \right \rfloor }

\def\mylist#1 {\ifx!#1\else\makebox[4em][r]{#1} \expandafter\mylist\fi}

\newcommand{\refeq}[1]{(\protect\ref{#1})}
 
\title{TBA Equations and Quantization Conditions}
\author[a]{Yoan Emery}
\affiliation[a]{Université de Genève,\\
	30, quai Ernest-Ansermet
	CH-1211 Genève 4, Switzerland}
\emailAdd{yoan.emery@unige.ch}
\abstract{
	It has been recently realized that, in the case of polynomial potentials, the exact WKB method can be reformulated in terms of a system of TBA equations. In this paper we study this method in various examples. We develop a graphical procedure due to Toledo, which provides a fast and simple way to study the wall-crossing behavior of the TBA equations. When complemented with exact quantization conditions, the TBA equations can be used to solve spectral problems exactly in Quantum Mechanics. We compute the quantum corrections to the all-order WKB periods in many examples, as well as the exact spectrum for many potentials. In particular, we show how this method can be used to determine resonances in unbounded potentials.
}
 
\begin{document}
 
\maketitle
\flushbottom


\section{Introduction}
\label{chap:Intro}
The one-dimensional time independent Schrödinger equation, describing a non-relativistic particle with energy $E$ in a potential $V(q)$,
\begin{equation}
\label{eq:Schro}
	-\hbar^2 \psi''(q)+(V(q)-E)\psi(q) = 0
\end{equation}
has always been an inexhaustible source of useful toy models in physics. However, solving it exactly is a tricky and subtle matter, depending on the form of the potential. Solving \eqref{eq:Schro} for generic potentials is still an active research subject, even though almost a hundred years have passed since his first inception \cite{PhysRev.28.1049}. Applying perturbation theory on \refeq{eq:Schro} yields a profusion of different perturbative series --~corresponding to different classical configurations~-- which are divergent most of the time. Resumming and organizing these divergent series into a non-perturbative and coherent picture of the quantum system of interest was achieved mainly during the late 70s and 80s through \emph{Exact asymptotics}: the seemingly paradoxical idea that drastically divergent series can be used to perform fully exact computations. Providing a comprehensive and impartial bibliography or a complete review of this rich subject is far outside the scope of the present work. We will instead briefly present some of the key developments we deem directly relevant for the understanding of this paper. In order to fix the notations, we will also provide a lightening review of the exact WKB method in section \ref{chap:WKB&ResurgentQM}.

One of the pioneering work using WKB techniques in order to investigate divergences in the anharmonic quartic oscillator is from Bender and Wu \cite{PhysRev.184.1231}. Yet, a better starting point for the ``exact'' story is Balian-Bloch's idea that we can reconstruct Quantum Mechanics from the complex classical trajectories \cite{BALIAN1974514}. Leveraging this idea with resumming techniques (Borel resummation), one can develop an exact version of the WKB method. Using Écalle's Resurgence theory \cite{ecalle1981fonctions}, the exact WKB method can be further refined and anchored to rigorous grounds. This resurgent program was first applied on one-dimensional Quantum Mechanics mainly by Voros \cite{10.1007/3-540-09532-2_85,voros1981spectre,Voros1983,Voros:1986}, then was further refined by Delabaere, Dillinger and Pham \cite{AIF_1993__43_1_163_0,doi:10.1063/1.532206,AIHPA_1999__71_1_1_0} and led to powerful exact results, like exact connection formulae and exact quantization conditions (EQC). These EQC are relating the resummed WKB periods to the energy of the quantum system, thus allowing to solve exactly the spectral problem that is \refeq{eq:Schro} provided that we know the exact periods. The road map is pretty straightforward in principle: given a quantum system, compute the all order WKB periods, resum them, then extract the spectrum from the EQC. \\

We were stuck with this three step program until more recent progress. Dorey and Tateo realized in \cite{Dorey:1998pt} that certain quantities in Quantum Mechanics 
were satisfying functional equations arising in the seemingly unrelated topic of integrable models. Indeed, these functional equations are strongly reminiscent of solutions of the non-linear integral equations appearing in the context of the thermodynamic Bethe ansatz (TBA). One can find a fairly modern review in e.g. \cite{vanTongeren:2016hhc}. This correspondence between ordinary differential equation (in the form of the Schrödinger equation for this instance) and integrable  models is called the ODE/IM correspondence. By solving these TBA equations, one can (among other quantities) obtain the resummed WKB periods, thus taking care of the step one and two of our recipe above in one go. However, the method presented in \cite{Dorey:1998pt} is quite limited: one can only apply it on pure potentials of the form $x^{2M}$ (and is then generalized to potentials of the form $\abs{x}^\alpha$). 
See \cite{Dorey:2007zx} for a more recent review, from Dorey, Dunning and Tateo, extending the ODE/IM correspondence to some other interesting examples beyond pure potentials (like $V(x) = x^{2M}+\frac{l(l+1)}{x^2}$ ).

In QM, when one want to reconstruct the exact WKB periods using only classical periods, asymptotic behaviors and discontinuities of the functions obtained by Borel-resummation of the quantum-periods, a Riemann-Hilbert problem arises. A generalization of the ODE/IM correspondence presented above can be realized as the fact that the solutions of the aforementioned Riemann-Hilbert problem are precisely given in term of a TBA system.  Actually, this description of the WKB periods in the language of monodromy and resurgence theory is known since Voros' work \cite{Voros1983} and is named ``analytical bootstrap''. However, it did not led to alternative computational methods in QM before these ideas were promoted to the method described in this paper, first developed by Ito, Mariño and Shu in \cite{Ito:2018eon}, reformulating \cite{Gaiotto:2014bza} in a pure quantum mechanical and resurgent framework.

In order to understand what motivated \cite{Ito:2018eon}, a good starting point is the seminal works of Gaiotto, Moore and Neitzke \cite{Gaiotto:2008cd,Gaiotto:2009hg} in the context of four-dimensional $\mathcal{N}=2$ gauge theories. They are deriving integral equations solving a Riemann-Hilbert problem in term of the $\mathcal{X}$ map defined in \cite{Gaiotto:2008cd}, the discontinuities of which are given by Kontsevich-Soibelman symplectomorphisms when crossing a BPS ray. See the Kontsevich-Soibelman wall-crossing formula, originally proposed in \cite{Kontsevich:2008fj}. These integral equations can in fact be restated as a version of the TBA in \cite{Zamolodchikov:1989cf}, as pointed out by Zamolodchikov (see appendix E in \cite{Gaiotto:2008cd}). To connect these results with standard Quantum Mechanics, the final step is to realize, as Gaiotto in \cite{Gaiotto:2014bza}, that the conformal limit of the TBA equations for the Hitchin system obtained in \cite{Gaiotto:2008cd,Gaiotto:2009hg} leads to TBA equations that are solving Schrödinger-type spectral problems. Moreover, they have applications outside Dorey and Tateo's ODE/IM correspondence, limited to pure potentials. See \cite{Dumas:2020zoz} for recent numerical experiments testing the predictions of \cite{Gaiotto:2014bza}.

Before finishing this introduction, we want to establish yet another fascinating connection: the Hitchin problem studied in \cite{Gaiotto:2008cd,Gaiotto:2009hg} is related to the geometrical problem of computing minimal surfaces in $AdS$, as explained by Alday and Maldacena in \cite{Alday:2009yn} and as precised below. See also \cite{Alday:2010vh}. The study of minimal surfaces in $AdS$ is motivated by the study of the holographic dual QFT through gauge/gravity duality, since we can then leverage integrability and compute interesting quantities (see \cite{Kazakov:2004qf}). To be more accurate, computing the minimal surfaces in $AdS$ delimited by a particular polygonal contour on the boundary allow us to compute the scattering amplitude or Wilson loop expectation value. Provided that these models are integrable, we can write a Hitchin system leading to a system of TBA equations. A precise example would be the quantum sigma model describing strings in $AdS_5 \times S^5$, the classical limit of which consists of strings moving in $AdS_5$\footnote{We can forget about the worldsheet fermions and five sphere in this limit.} and is integrable. This model is corresponding (through $AdS/CFT$) to four-dimensional $\mathcal{N}=4$ super Yang Mills, integrable in the planar limit. Another interesting example, more closely related to the TBA system of interest in this paper, is the study of minimal surfaces in $AdS_3$ delimited by a polygonal closed contour on the boundary. This problem simplifies and reduces to the $Z_2$ projection of a $SU(2)$ Hitchin problem arising in \cite{Gaiotto:2008cd,Gaiotto:2009hg} in the context of four-dimensional $\mathcal{N}=2$ gauge theories, as described in the previous paragraph. They are numerous interesting papers on the subject, including (but not limited to) \cite{Alday:2009yn,Alday:2010vh,Hatsuda:2010cc,Toledo:2014koa,ToledoJonathan2016}.

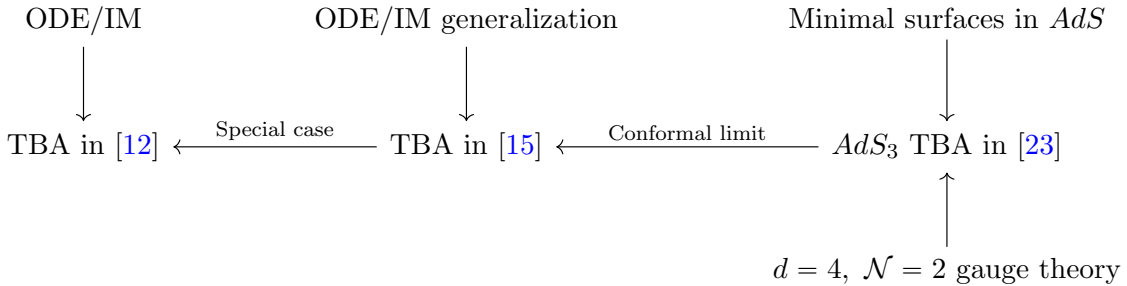
\begin{figure}
	\centering
	\begin{tikzcd}[row sep=1 cm, column sep = 1.7cm]
		\text{ODE/IM}\arrow[d]&\text{ODE/IM generalization}\arrow[d] &\text{Minimal surfaces in}~AdS\arrow[d] \\
		\text{TBA in \cite{Dorey:1998pt}}& \text{TBA in \cite{Ito:2018eon}}\arrow[l,"\text{Special case}",swap]&AdS_3 \text{ TBA in \cite{Alday:2010vh}}\arrow[l,"\text{Conformal limit}",swap]\\
		& & d=4,~\mathcal{N}=2\text{ gauge theory} \arrow[u]
	\end{tikzcd}
	\caption{Implication diagram for the TBA systems mentioned in the introduction.}
	\label{fig:TBAdiagram}
\end{figure}

As we already explained, the conformal limit of the TBA equations in \cite{Gaiotto:2008cd,Gaiotto:2009hg} brings us back to the TBA equations in \cite{Ito:2018eon}. Therefore, this property is transposed to the TBA equations arising in the context of minimal surfaces in $AdS_3$ (explicitly written in their integral form in \cite{Alday:2010vh,ToledoJonathan2016} for example). Additionally, the TBA equations in \cite{Ito:2018eon} are obtained using the generalized ODE/IM correspondence and work in principle for any polynomial potential. Consequently, when restricting to pure potentials, one should be able to recover the TBA equation of the standard ODE/IM correspondence found in \cite{Dorey:1998pt,Dorey:2007zx} as a special case. The relations between all the aforementioned TBA equations are contained in the diagram of figure \ref{fig:TBAdiagram}.\\


In section \ref{chap:WKB&TBA}, we review the general theory of \cite{Ito:2018eon}. The exact WKB method will be presented, and a rigorous link between the TBA system and the Schrödinger problem \refeq{eq:Schro} will be established in two different way. First, we will derive the TBA equations as solutions of a Riemann-Hilbert problem. Then, we will be directly deriving functional equations from \refeq{eq:Schro}, leading to the appropriate TBA system. This section is not containing any original work per sei, but is containing the necessary information to make this paper self contained as well as to fix the notations. In section \ref{chap:wallCrossing}, we explain how to analytically continue the TBA equations obtained in section \ref{chap:WKB&TBA} outside the regime where all the turning points are real i.e. outside the so called ``minimal chamber''; this procedure is called ``wall-crossing''. We  provide (inspired by Toledo's work in \cite{ToledoJonathan2016}) a fast and simple diagrammatic procedure that allows us to carry out wall-crossing from any chamber to any other. In addition, we will also use this procedure in order to write the quantum mechanical TBA system originating from the ODE/IM generalization for arbitrary degree polynomial potentials in the maximal chamber. We then justify and clarify how it is implying the TBA system originating from the standard ODE/IM correspondance as a special case. In section \ref{chap:res}, we present some of the results obtained solving the TBA equations numerically and we compare them to known quantities computed using more standard Quantum Mechanical techniques. In appendix \ref{app:num}, we discuss the methods used in order to solve the TBA integral equations numerically.  In appendix \ref{app:EQC}, we describe how to extract exact quantization conditions from connection formulae. Finally, in the appendices \ref{app:qperiods} and \ref{app:CDil}, we provide the standard quantum mechanical techniques we used to verify our TBA results. In appendix \ref{app:qperiods}, we explain how one can compute the quantum corrections to the WKB periods using differential operators. In appendix \ref{app:CDil}, we describe how to extract the bounded or resonant spectrum from a Hamiltonian with polynomial potential by expressing it in the harmonic oscillator basis.

\section{WKB periods and TBA equations}
\label{chap:WKB&TBA}

In this section, we will prove constructively -- reviewing the ideas of \cite{Ito:2018eon} -- that the spectral problem \refeq{eq:Schro} can be related to a TBA system. A summary of the exact WKB method will be presented in section \ref{chap:WKB&ResurgentQM}. For a comprehensive review of the WKB method, see also \cite{MarcosAQM}. We will then construct the TBA system in \ref{chap:TBAasRiemannHilbert} and \ref{chap:TBAasODEIM}, by looking at Voros' Riemann-Hilbert problem arising from the resurgent analysis of \refeq{eq:Schro} for the former, and writing functional equations arising by massaging \refeq{eq:Schro} directly for the later.

\subsection{The WKB method and resurgent Quantum mechanics}
\label{chap:WKB&ResurgentQM}

The one-dimensional time independent Schrödinger equation \refeq{eq:Schro} can be written as
\begin{equation}
\label{eq:Schro2}
\hbar^2 \psi''(q)+p^2(q)\psi(q) = 0
\end{equation}
where $p(q) = \sqrt{E-V(q)}$ is the classical momentum. The $\hbar \to 0$ limit cannot be taken directly in \refeq{eq:Schro2} since it clearly become algebraic in this limit. Nonetheless, we can write the following ansatz for the wavefunction
\begin{equation}
\label{eq:WKBansatz}
\psi(q) = \exp\left(\frac{i}{\hbar} \int^{q} Q(\bar{q}) d\bar{q}\right)
\end{equation}
By plugging \refeq{eq:WKBansatz} into \refeq{eq:Schro2}, we transform it into the Riccati equation for $Q(q)$:
\begin{equation}
\label{eq:Riccati}
Q^2(q)- i \hbar Q'(q) = p^2(q)
\end{equation}
We can solve \refeq{eq:Riccati} by writing $Q(q)$ as a (formal\footnote{i.e. we will not adress the convergence issues: \refeq{eq:formalSeriesQ} is never meant to be a convergent series in $\hbar$!}) power series in $\hbar$,
\begin{equation}
\label{eq:formalSeriesQ}
Q(q) = \sum_{k \in \mathbb{N}} Q_k(q) \hbar^k
\end{equation}
Solving for the $Q_k(q)$ recursively, one finds
\begin{equation}
Q_{n+1}(q) = \frac{1}{2 Q_0(q)}\left(i Q_n'(q)-\sum_{k=1}^{n} Q_k(q)Q_{n+1-k}(q) \hbar^k\right),~~ \text{with}~~ Q_0(q)=p(q)
\end{equation}
By splitting the formal series \refeq{eq:formalSeriesQ} in odd and even powers of $\hbar$, such that $Q(q)=P(q)+Q_{\text{odd}}(q)$, i.e.
\begin{equation}
\label{eq:formalSeriesP}
P(q) = \sum_{k \in \mathbb{N}} p_k(q) \hbar^{2k},~ \text{with}~~ p_0(q)=p(q)~~\text{and}~~Q_{\text{odd}}(q) = \sum_{k \in \mathbb{N}} Q_{2k+1}(q) \hbar^{2k+1}
\end{equation}
one realizes that \refeq{eq:Riccati} splits into two equations. The odd equation is allowing us to solve $Q_{\text{odd}}(q)$ in terms of $P(q)$ alone. In fact, one finds that $Q_{\text{odd}}(q)$ is a total derivative:
\begin{equation}
Q_{\text{odd}}(q) = \frac{i\hbar}{2}\frac{d}{dq}\log P(q)
\end{equation}
and one can reexpress the problem without this redundancy. The WKB ansatz \refeq{eq:WKBansatz} becomes then
\begin{equation}
\label{eq:WKBansatz2}
\psi(q) = \frac{1}{\sqrt{P(q)}}\exp\left(\frac{i}{\hbar} \int^{q} P(\bar{q}) d\bar{q}\right)
\end{equation}
Geometrically, we can interpret $P(q)dq$ as a meromorphic differential on the so called ``WKB curve'':
\begin{equation}
\label{eq:WKBcurve}
y^2 = 2(E-V(q))
\end{equation}
In this paper, we will treat the case where $V(q)$ is a polynomial of degree $d$. In this case, \refeq{eq:WKBcurve} is a hyperelliptic curve defining a Riemann surface $\Sigma_\text{WKB}$ of genus $\floor{\frac{d-1}{2}}$. This curve is characterized by a set of moduli (the energy $E$ and the parameters of the polynomial $V(q)$). The basic objects appearing in the WKB method are the \emph{quantum periods} or \emph{WKB periods}. They consist of the periods of the meromorphic differential $P(q)dq$ integrated along the one-cycles $\gamma \in H_1(\Sigma_\text{WKB})$:
\begin{equation}
\label{eq:WKBperiods}
\Pi_\gamma(\hbar)=\oint_\gamma P(q)dq
\end{equation}
As $P(q)$, the quantum periods can be expressed as (formal) power series in $\hbar^2$:
\begin{equation}
\label{eq:WKBperiodsSeries}
\Pi_\gamma(\hbar)=\sum_{n \in \mathbb{N}} \Pi_\gamma^{(n)} \hbar^{2n}~,~~\text{where}~~~~ \Pi_\gamma^{(n)} = \oint_\gamma p_n(q)dq
\end{equation}
For an efficient algorithm computing the coefficients $ \Pi_\gamma^{(n)}$, see appendix \ref{app:qperiods}. As stated in the introduction, the formal series used as basic objects in Quantum Mechanics are almost always divergent. This is the case with the WKB periods which are typically diverging like double-factorials:
\begin{equation}
\label{eq:WKBperiodsDiv}
\Pi_\gamma^{(n)} \sim (2n)!
\end{equation}
Thus, we need to promote them into a meaningful function using resummation techniques. In this paper, we will use the Borel resummation \cite{ASENS_1899_3_16__9_0} procedure as defined below. First, we define the \emph{Borel transform} of the WKB periods as
\begin{equation}
\label{eq:BorelTransform}
\hat{\Pi}_\gamma(\xi) = \sum_{n \in \mathbb{N}} \frac{\Pi_\gamma^{(n)}}{(2n)!} ~ \xi^{2n}
\end{equation}
such that this new power series has finite radius of convergence\footnote{Usually, we define the borel transform of $f(z)=\sum f^{(n)} z^n$ as $\hat{f}(\xi) = \sum_{n \in \mathbb{N}} \frac{1}{n!}f^{(n)} ~ \xi^{n}$, which has finite radius of convergence if and only if $f$ is of Gevrey-1 type, i.e. $f^{(n)}<\abs{C^n n!}$. But we can recover our case by a simple change of variable $z \mapsto \hbar^2$ (at the cost of introducing new monodromies in the complex $z$-plane).}. The \emph{Borel resummation} of a series is the Laplace transform of its Borel transform. In the context of the quantum periods, one gets
\begin{equation}
\label{eq:BorelResum}
\mathcal{B}(\Pi_\gamma)(\hbar) = \frac{1}{\hbar} \int_{\mathbb{R}^+} e^{-\xi/\hbar}~\hat{\Pi}_\gamma(\xi)~ d\xi
\end{equation}
These definitions are coming from the well known integral definition of the Gamma function\footnote{Indeed, $\Gamma(n+1)=n!=\int_{\mathbb{R}^+} \bar{\xi}^n e^{-\bar{\xi}} d\bar{\xi}$ such that $1=\frac{1}{n!}\int_{\mathbb{R}^+} \bar{\xi}^n e^{-\bar{\xi}} d\bar{\xi}$.}. The integral \refeq{eq:BorelResum} is called \emph{Borel summable} if it converges for $\hbar$ sufficiently small. Let us denote that \refeq{eq:BorelTransform} can be analytically continued in the complex $\xi$-plane. It can display various type of singularities (poles and branch cuts typically). Now, let us assume that a singularity is present on the positive real axis. Then, the Borel resummation \refeq{eq:BorelResum} will hit it and will be undefined for any $\hbar$: it will not be Borel summable anymore! This happens everywhere in Quantum Mechanics. In order to circumvent this apparent problem and ``dodge'' the singularities, we will now introduce some of the key ingredients of resurgence.\\

If diverging series occurring in the context of differential equations can appear undefined hence useless, they are containing in fact a lot of useful information on the system of interest. In the same spirit, if singularities in the Borel complex $\xi$-plane are obstacles to Borel summability, their discontinuities are containing a crucial amount of information nonetheless: it's not a bug, it's a feature! These discontinuities are the result of the Borel transform of the all-order WKB periods after all. In fact, we will show in the following that they are involving only the WKB periods and are relating them together. First, let's generalize \refeq{eq:BorelResum} by allowing to integrate along \emph{any} direction in the complex plane. The \emph{directional Borel resummation along the direction} $\varphi$  is defined as the directional Laplace transform of the Borel transform:
\begin{equation}
\label{eq:BorelResumAlongPhi}
\mathcal{B}_\varphi(\Pi_\gamma)(\hbar) = \frac{1}{\hbar} \int_{e^{i \varphi}\mathbb{R}^+} e^{-\xi/\hbar}~\hat{\Pi}_\gamma(\xi)~ d\xi
\end{equation}
where $\int_{e^{i \varphi}\mathbb{R}^+}$ denotes an integration path between $0$ and $e^{i \varphi} \infty$ i.e. along the complex ray forming an angle $\varphi$ with the real axis. As before, \refeq{eq:BorelResumAlongPhi} is \emph{Borel summable} -- \emph{along the direction $\varphi$} this time -- if and only if the integral \refeq{eq:BorelResumAlongPhi} converges. Now, let's assume that the Borel transformation $\hat{\Pi}_\gamma$ has a singularity along the direction $\varphi$. Then, the value of the directional Borel transform will jump when we cross $\varphi$: there will be a discontinuity. It defines two possible directional Borel resummation: just above and just below the critical angle $\varphi$, or
\begin{equation}
\label{eq:LateralBorelResum}
\mathcal{B}_{\varphi\pm}(\Pi_\gamma)(\hbar)=\lim_{\delta \to 0} \mathcal{B}_{\varphi\pm \delta}(\Pi_\gamma)(\hbar)
\end{equation}
which are called the \emph{lateral Borel resummations}. From them, we can define the \emph{median Borel resummation}, which is the average of the lateral resummation above and below $\varphi$:
\begin{equation}
\label{eq:meidanBorelResum}
	\mathcal{B}_{\varphi\,\text{med}}(\Pi_\gamma)(\hbar) =\frac{1}{2}\left(\mathcal{B}_{\varphi+}(\Pi_\gamma)(\hbar)+\mathcal{B}_{\varphi-}(\Pi_\gamma)(\hbar)\right)
\end{equation}
One can also measure the previously mentioned discontinuity by subtracting them:
\begin{equation}
\label{eq:Dicontinuity}
\text{disc}_\varphi(\Pi_\gamma)(\hbar) = \mathcal{B}_{\varphi+}(\Pi_\gamma)(\hbar)-\mathcal{B}_{\varphi-}(\Pi_\gamma)(\hbar)
\end{equation}
Let's note that the $\mathcal{B}_{\varphi\pm}$ and $\text{disc}_\varphi$ can be regarded as operators acting on  formal power series $\Pi \in WKB$. From this point of view, one can define the \emph{Stokes automorphism} by the commutativity of the  diagram\footnote{In the literature, the inverse convention (the $\mathfrak{S}$ arrow is swapped, corresponding to our $\mathfrak{S}^{-1}$) is often used. This is the case in \cite{AIHPA_1999__71_1_1_0} for example, but we are following the conventions in \cite{Ito:2018eon}.}
\begin{equation}
\label{eq:StokesAutomorphismDiagram}
\begin{tikzcd}
	WKB\arrow[rr,"\mathfrak{S}_{\varphi}"]\arrow[rd,"\mathcal{B}_{\varphi+}"]&&WKB \arrow[dl,"\mathcal{B}_{\varphi-}",swap] \\
	&\mathcal{S}&
\end{tikzcd}
\end{equation}
where $\mathcal{S}$ denotes the space of ``proper functions'' leading to resurgent solutions of the Schrödinger equation through \refeq{eq:WKBansatz2}. In other words,
\begin{equation}
\label{eq:StokesAutomorphism}
\mathfrak{S}_{\varphi} = \mathcal{B}_{\varphi+} \circ \mathcal{B}_{\varphi-}^{-1} = 1+\text{disc}_\varphi \circ \mathcal{B}_{\varphi-}^{-1}= (1-\text{disc}_\varphi \circ \mathcal{B}_{\varphi+})^{-1}
\end{equation}
In the following, we purposefully gloss out most of the technical details. The reader can find rigorous statements and proofs in \cite{AIHPA_1999__71_1_1_0}. In essence, we can define connection paths relating the $WKB$ elements between different Stokes regions. The action of the connection cycle $\gamma$ is simply a multiplication by a \emph{Voros symbol}. A special subset
of Voros symbols are Voros multipliers $\cal{V_\gamma}$. At the end of the day, they simply end up being the exponent of our WKB periods:
\begin{equation}
\label{eq:VorosMultiplier}
\mathcal{V}_\gamma = \exp\left(\frac{1}{\hbar} \Pi_\gamma\right)
\end{equation}
These definitions lead to the \emph{Delabaere-Pham formula} (theorem 2.5.1 in \cite{AIHPA_1999__71_1_1_0}), that is expressing a relation between Voros multipliers, thus between quantum periods. Let's consider the case where all the $d$ turning points of the potential are on the real axis, and the $d-1$ cycles are encircling classically allowed and classically forbidden regions in succession (one can find such an example in figure \ref{fig:plotTripleWellMinC} and \ref{fig:cyclesSexticMinC}). We call this subspace of the moduli the \emph{minimal chamber} for reason we will develop later. In this regime, we will be interested in the $\phi=0$ direction and will be omitting the index $\phi$. The Delabaere-Pham formula then yields
\begin{equation}
\label{eq:DPformula}
\mathfrak{S} \mathcal{V}_{\gamma_a} =  \prod_{\{1,\ldots,d-1\}\setminus\{a\}} \left(1+\mathcal{V}_{\gamma_b}^{-1}\right)^{\langle\gamma_b,\gamma_a\rangle} \mathcal{V}_{\gamma_a}
\end{equation}
where $\langle\gamma_b,\gamma_a\rangle$ is the intersection of cycles\footnote{i.e. the algebraic number of times $\gamma_a$ is crossing the Stokes line associated with the Voros symbol $\mathcal{V}_{\gamma_b}$, with signature $+$ (resp. $-$) when it is crossing it from right to left (resp. left to right). See \cite{AIHPA_1999__71_1_1_0} for the detail.}. As an application of \refeq{eq:DPformula}, let's look at $a=1$, corresponding to the first cycle from the left (the first red cycle in figure \ref{fig:cyclesSexticMinC} for example). In this case,  $\langle\gamma_2,\gamma_1\rangle = +1$ is the only non trivial cycle intersection and \refeq{eq:DPformula} reduces to
\begin{equation}
\label{eq:DPformulaFirstCycle}
\mathfrak{S} \mathcal{V}_{\gamma_1} =   \left(1+\mathcal{V}_{\gamma_2}^{-1}\right) \mathcal{V}_{\gamma_1}
\end{equation}
Applying the operator identity \refeq{eq:DPformulaFirstCycle} on our WKB periods yields an alternative formulation of the Delabaere-Pham formula:
\begin{equation}
\label{eq:DPformulaDisc}
\text{disc}\left(\Pi_{\gamma_1}\right)(\hbar)= -i \hbar \log\left(1+\exp\left(-\frac{i}{\hbar}\Pi_{\gamma_2}(\hbar)\right)\right)
\end{equation}
As a result, $\Pi_{\gamma_1}$ is not Borel summable along the direction $\phi=0$. This is an inevitable consequence of \refeq{eq:DPformulaDisc}, that is measuring the precise discontinuity when we cross the real axis, hence revealing the presence of singularities along this ray. \refeq{eq:DPformulaDisc} also contains a very resurgent statement: this discontinuity in the Borel resummation of the quantum period $\Pi_{\gamma_1}$ is \emph{entirely} encoded in another quantum period, $\Pi_{\gamma_2}$. This is the same situation for the general case \refeq{eq:DPformula}, which is linking all the quantum periods together (as long as they don't have a null intersection of cycle), through their Voros symbol and the action of the Stokes automorphism.\\

One of the goals of the resurgent analysis of Quantum Mechanics is the complete determination of the discontinuity structure of the quantum periods we just outlined. However, in order to solve Schrödinger spectral problems exactly, one need to find a way to use the all-order resumed quantum periods in a meaningful way. This needed key ingredients typically takes the form of functional equations relating all the WKB periods together, i.e.
\begin{equation}
\label{eq:genericEQC}
\mathcal{Q}\left(\mathcal{V}_{\gamma_1},\ldots,\mathcal{V}_{\gamma_r}\right) = 0
\end{equation}
which are called \emph{exact quantization conditions}. A standard way to derive them is to consider that the wave function is decaying at $\pm \infty$ along a ray in the complex plane, applying the Voros-Silverstone connection formula at each turning points, as is done in \cite{MarcosAQM} for example. We will provide a brief presentation of this method in appendix \ref{app:EQC}. An alternative and equivalent way to derive EQC is to relate Stokes regions, crossing Stokes lines around turning points as is done in \cite{voros1981spectre,AIF_1993__43_1_163_0,doi:10.1063/1.532206,AIHPA_1999__71_1_1_0} (Knoll-Schaeffer connection method). For a recent paper exploring the connections between EQC arising in the exact WKB context and the other nonperturbative approach to Quantum Mechanics, i.e. saddle point analysis in the Euclidean path integral formulation, see \cite{Sueishi:2020rug}.
Typically, there are multiple EQC corresponding to different lateral Borel resummations. A very non-trivial check for this kind of EQC is to see if we can go from one EQC to the others by the use of the Delabaere-Pham formula \eqref{eq:DPformula}. We will show a concrete example of this in the context of the cubic oscillator in section \ref{chap:cubicOsc}.\\

The periods are depending on $\hbar$ but also on the moduli of the WKB curve, including the energy. This means that the constraint \refeq{eq:genericEQC} is selecting a codimension one -- we will assume discrete\footnote{In this paper, we will only consider bounded or resonant potentials, leading to discrete spectra.} -- submanifold in the moduli$\times \hbar$ space. If we fix every moduli parameters excepted the energy, the EQC \refeq{eq:genericEQC} is drawing an infinite discrete family of curves, parametrized by $(\hbar,E)$ and labeled by the quantum number $n$. Normally, one wants to compute the energy spectrum as a function of $\hbar$: $E_n(\hbar)$, slicing the $(\hbar,E)$ plane along the vertical axis, leaving an infinite tower of energies. But one can notice that nothing is preventing us from doing the contrary, reversing last relation, thus extracting the ``Planck spectrum'' $\hbar_n(E)$. In fact, as we shall soon see, the unknown functions intervening in the TBA equations -- main actors of the present paper -- are functions of the variable $\theta=-\log(\hbar)$. It is then more natural to adopt the later convention and compute what we will be calling the ``Voros spectrum'' $\theta_n(E)$, as in \cite{Ito:2018eon}. To connect a TBA result with a standard quantum mechanical control result, we can simply check that 
\begin{equation}
\label{eq:checkE}
E^\text{(c)}_n\left(\exp(-\theta_n(E))\right) =E^\text{(c)}_n\left(\hbar_n(E))\right) = E
\end{equation}
where $E$ is to be tough as a parameter of the WKB curve and $E^\text{(c)}(\hbar)$ is the control computation that take a $\hbar$ value as an input and output the standard energy spectra. For our purpose, we will use the methods decribed in appendix \ref{app:CDil}, basically diagonalizing the Hamiltonian in the harmonic oscillator basis, in order to compute the spectra $E^\text{(c)}_n$ numerically and control our results. In section \ref{chap:res}, we will typically present tables of the levels $E^\text{(c)}_n$, normalized to $1$ in the sense that we divide them by the WKB curve parameter $E$, using the Voros level $m$, i.e.
\begin{equation}
\label{eq:checkEnrom}
\frac{1}{E} E^\text{(c)}_n\left(\hbar_m(E))\right)
\end{equation}
such that we can check that its diagonal is indeed respecting \begin{equation}
\label{eq:checkEnromDiag}
\frac{1}{E} E^\text{(c)}_n\left(\hbar_n(E))\right) = 1
\end{equation}
within the numerical precision.

\subsection{The TBA equations as a solution of a Riemann-Hilbert problem}
\label{chap:TBAasRiemannHilbert}

\begin{figure}
	\centering
	~ 
	\begin{subfigure}[b]{0.48\textwidth}
		\includegraphics[width=\textwidth]{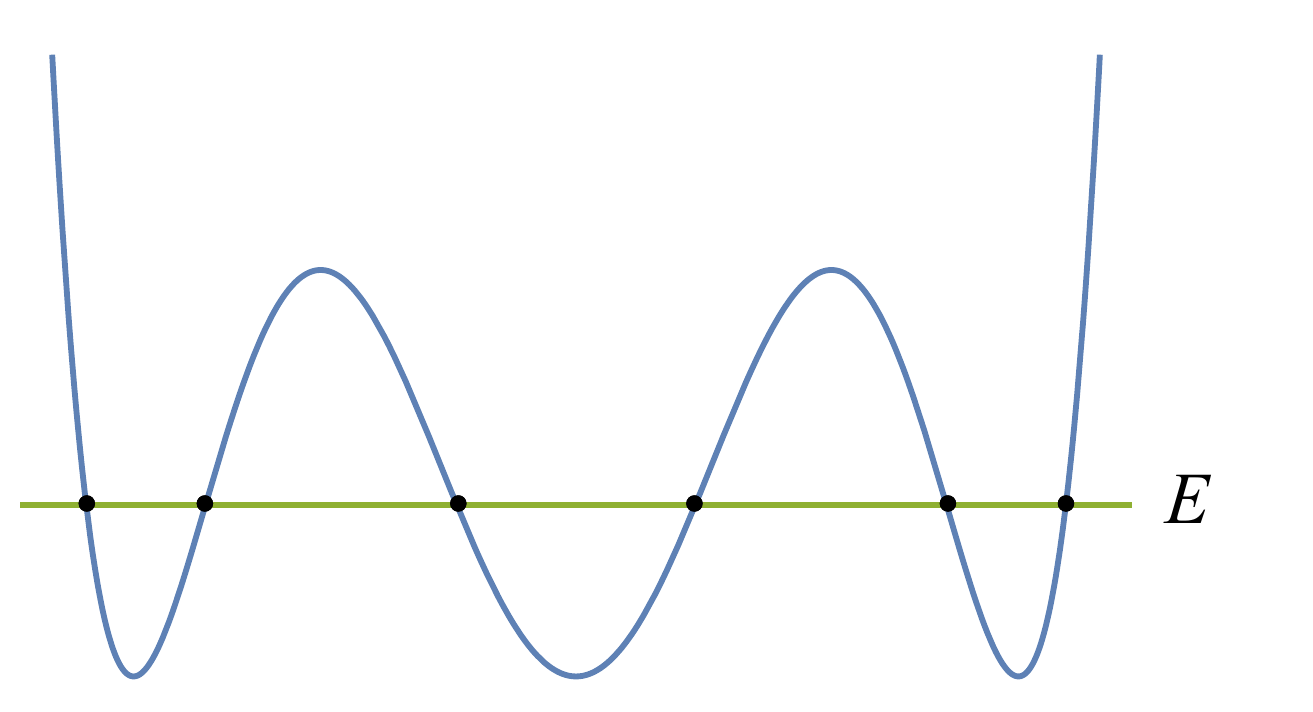}
		\caption{Energy bellow the wells}
		\label{fig:plotTripleWellMinC}
	\end{subfigure}
	~ 
	\begin{subfigure}[b]{0.48\textwidth}
		\includegraphics[width=\textwidth]{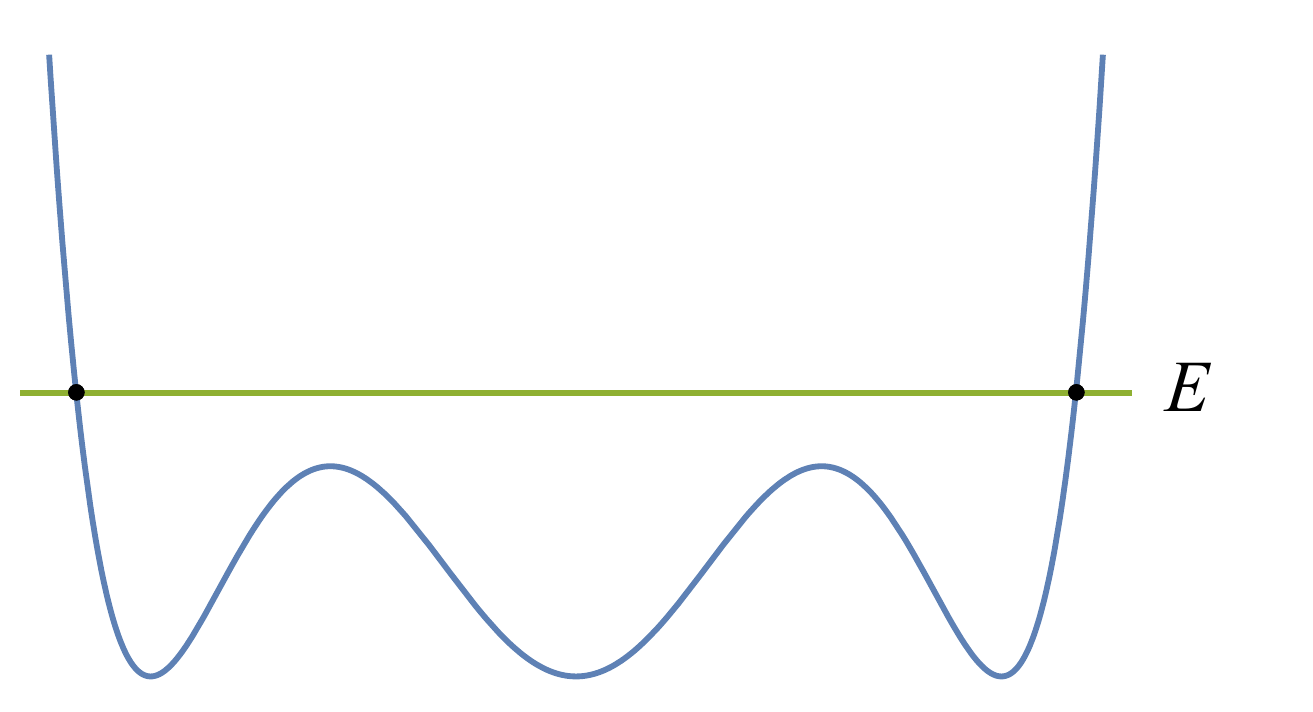}
		\caption{Energy above the wells}
		\label{fig:plotTripleWellMaxC}
	\end{subfigure}
	~ 
	\begin{subfigure}[b]{0.48\textwidth}
		\includegraphics[width=\textwidth]{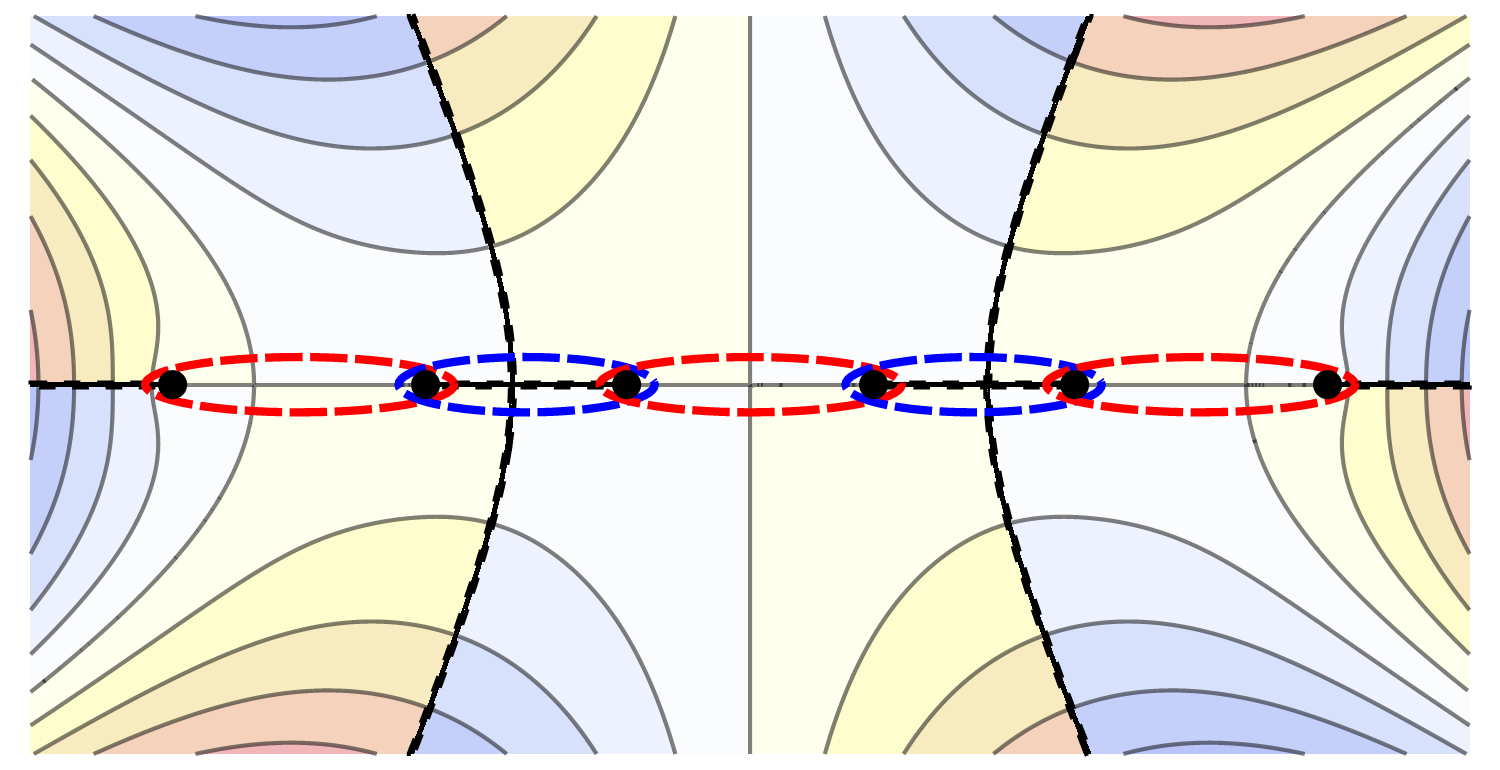}
		\caption{Minimal chamber (5 TBA equations)}
		\label{fig:cyclesSexticMinC}
	\end{subfigure}
	~ 
	\begin{subfigure}[b]{0.48\textwidth}
		\includegraphics[width=\textwidth]{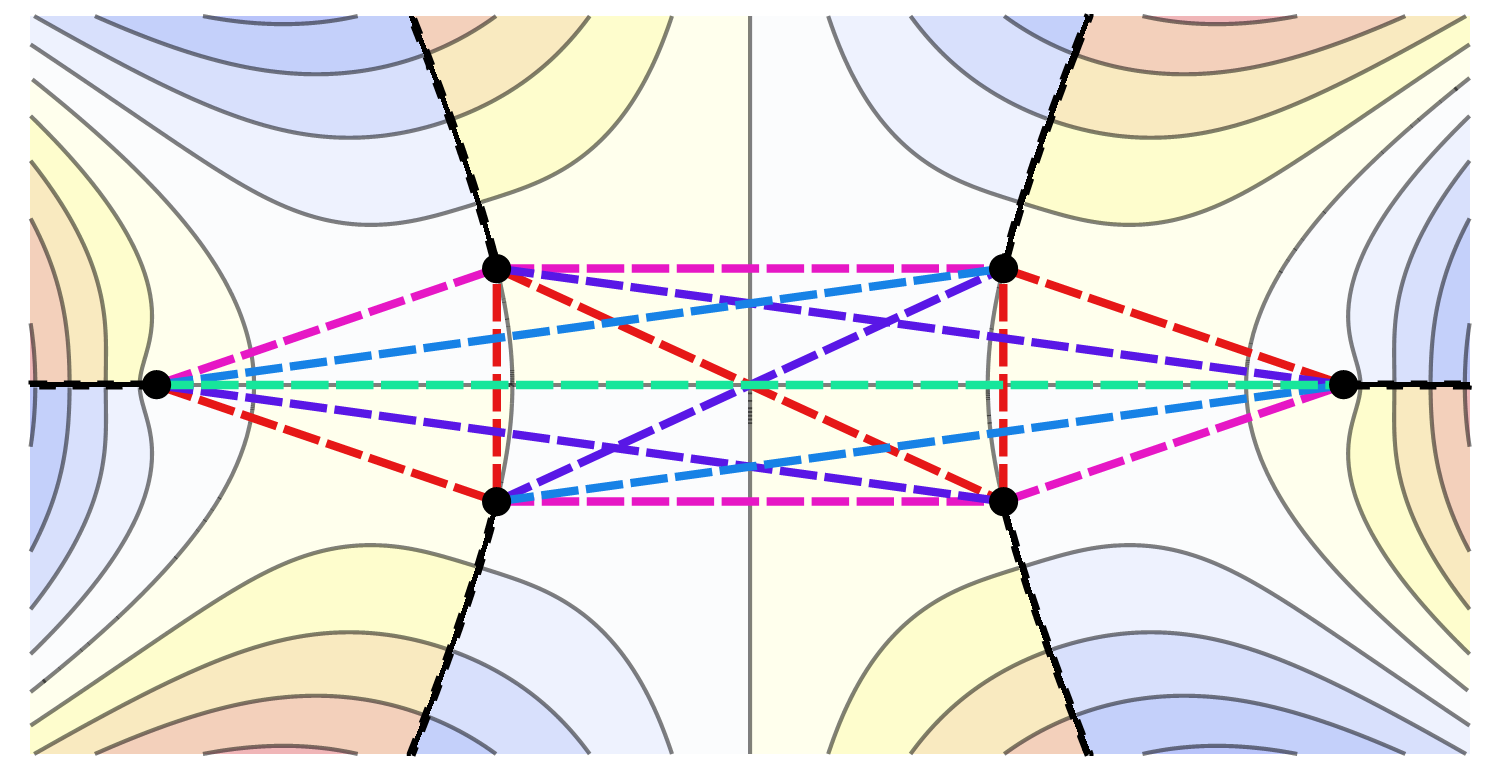}
		\caption{Maximal chamber (15 TBA equations)}
		\label{fig:cyclesSexticMaxC}
	\end{subfigure}
	\caption{In (a-b), we plot the triple well potential (blue curve) along with the energy (green line) and real turning points (black dots). In (a), the energy is below the wells and there are 6 real turning points since the potential is a sextic polynomial. If we increase the energy above the wells, we end up in the case (b) where the turning points are moved into the complex plane. In (c-d), we plot $\Im p(x)$ for the different energy configurations and the associated relevant cycles along wich we are integrating in order to get the periods. The branch cuts are drawn as black squiggly lines and the turning points as black dots. In (c) we draw the cycles of the periods corresponding to the classically allowed (resp. forbidden) region as dashed red (resp. blue) cycles. The cycles in (c) are corresponding to the red cycles in (d) after analytical continuation. The cycles in (d) are squished into lines between turning points for clarity.}\label{fig:Potential&Cycles}
\end{figure}

The goal of this paper is to provide an exact non-perturbative solution to the Schrödinger spectral problem \ref{eq:Schro} for arbitrary polynomial potential. As we already mentioned, if the degree of $V(q)$ is $d$, the WKB curve \refeq{eq:WKBcurve} for this problem  is a hyperelliptic curve defining a Riemann surface $\Sigma_\text{WKB}$ of genus $\floor{\frac{d-1}{2}}$. At first, let us assume we are in a very special region of the moduli space: the ``minimal chamber'', in which all the turning points of $V(q)$ are real and distinct. An example of such a configuration can be found in the figure \ref{fig:plotTripleWellMinC}, with the relevant cycles shown in figure \ref{fig:cyclesSexticMinC}. We shall extend the result found here later, in the section \ref{chap:wallCrossing}, by analytical continuation. In the minimal chamber, we can always organize our turning points $q_i$, $i \in \{1,\ldots,d\}$ such that
\begin{equation}
\label{eq:orderingTP}
q_1<\ldots<q_d
\end{equation} 
The only relevant $d-1$ periods in this regime are obtained integrating along the cycles $\gamma_{a,a+1}$,  $a \in \{1,\ldots,d-1\}$,  where $\gamma_{i,j}$ denotes a cycle encircling the tuning point $q_i$ and $q_j$. In the following, $\gamma_{i,j}$ will often be omitted in expressions of the form $Q_{\gamma_{i,j}}$, preferring $Q_{i,j}$ for shortness of notation.\\

In order to make the computation more tractable, we will use the ``\emph{masses}'' conventions, defined below, as long as we are in the minimal chamber (i.e. in the rest of this section), following the conventions in \cite{Ito:2018eon}. The name \emph{masses} is reminiscent of similar quantities in two-dimensional integrable theories and is borrowed for this reason. The masses $m_{a,a+1}$ are defined as following, depending on the parity of $a$:
\begin{align}
	\label{eq:masses}
\begin{split}
	m_{2k-1,2k} &= \oint_{\gamma_{2k-1,2k}}p(q)dq=\Pi^{(0)}_{2k-1,2k}\\
m_{2k,2k+1} &= i \oint_{\gamma_{2k,2k+1}}p(q)dq= i~\Pi^{(0)}_{2k,2k+1}
\end{split}
\end{align}
with the branch cut prescriptions and the cycle orientations chosen so that \refeq{eq:masses} are all real and positive.
Note that we are however switching for an equivalent ``\emph{periods}'' formulation (i.e. using the classical periods $\Pi^{(0)}_\gamma$), following the conventions in \cite{ToledoJonathan2016}, once we are analytically continuing the problem outside of the minimal chamber.\\

The starting point for deriving the TBA equation is the Delabaere-Pham formula \refeq{eq:DPformula}. The WKB periods in the classically allowed region (corresponding to the red cycles in figure \ref{fig:cyclesSexticMinC}) are not Borel summable, as stated by \refeq{eq:DPformula}, and the discontinuity is expressed only in terms of the other periods. In the minimal chamber, \refeq{eq:DPformula} simplifies considerably and involves only the (at most 2, 1 for the extremities) ``tangent periods'' corresponding to encircling the classically forbidden region (corresponding to the blue cycles in figure \ref{fig:cyclesSexticMinC}). In the language of Stokes automorphisms,
\begin{equation}
\label{eq:DPformulaMinC}
\mathfrak{S} \mathcal{V}_{2k-1,2k} =   \left(1+\mathcal{V}_{2k-2,2k-1}^{-1}\right)^{\langle 2k-2,2k-1\rangle}\left(1+\mathcal{V}_{2k,2k-1}^{-1}\right)^{\langle 2k,2k-1\rangle} \mathcal{V}_{2k-1,2k}
\end{equation}
which translates into
\begin{align}
\begin{split}
\label{eq:DPformulaDiscMinC}
\text{disc}\left(\Pi_{2k-1,2k}\right)(\hbar)=& -i \hbar \langle 2k-2,2k-1\rangle \log\left(1+\exp\left(-\frac{i}{\hbar}\Pi_{2k-2,2k-1}(\hbar)\right)\right)\\
&-i \hbar \langle 2k,2k-1\rangle \log\left(1+\exp\left(-\frac{i}{\hbar}\Pi_{2k,2k-1}(\hbar)\right)\right)
\end{split}
\end{align}
where the intersection $\langle a,b\rangle$ is either 0 or 1;  $\langle 0,1\rangle = 0$, $\langle d,d+1\rangle = 0$ and $\langle a,a+1\rangle = 1$. Note that this formula is intended to work only on a ray of the complex plane, the real axis in this case. Nonetheless, remember that we are considering a formal power series in $\hbar^2$, such that a similar formula also holds for the negative real axis. Finally, we can repeat this analysis almost verbatim along the imaginary axis  (positive and negative) : let's notice that a rotation $\hbar \to i \hbar$ is simply exchanging the classically allowed and classically forbidden region, hence exchanging the odd and the even cycles in \refeq{eq:DPformulaMinC} and \refeq{eq:DPformulaDiscMinC}. Thus, we can unify these Delabaere-Pham formulae in one unique formula along the direction $\phi=\pi/2$ by defining the ``\emph{pseudo-energies}'' or ``$\epsilon$-\emph{functions}'' as
\begin{align}
\label{eq:regEpsilon}
\begin{split}
\lim_{\delta \to 0}\epsilon_{2k-1,2k}\left(\theta+\frac{i \pi}{2} \pm i \delta\right)&=\frac{i}{\hbar} \mathcal{B}_{\pm}\left(\Pi_{2k-1,2k}\right)(\hbar) \\
\epsilon_{2k,2k+1}\left(\theta\right)&=\frac{i}{\hbar} \mathcal{B}\left(\Pi_{2k,2k+1}\right)(\hbar)
\end{split}
\end{align}
where $\theta=\exp(-\hbar)$. Notice that, because of \eqref{eq:regEpsilon}, if we find the $\epsilon$-functions, we also find the Borel resumed all order WKB periods. The resulting unified Delabaere-Pham formula is then
\begin{equation}
\label{eq:discepsilon}
\text{disc}_{\pi/2}~ \epsilon_{a,a+1} = L_{a-1,a}(\theta)+L_{a+1,a+2}(\theta)
\end{equation}
with $L_{a,b}(\theta)= \log\left(1+\exp\left(-\epsilon_{a,b}(\theta)\right)\right)$ and, because of the intersection cycles,  $L_{0,1}(\theta)=L_{d,d+1}(\theta)=0$. A similar formula can be obtained along the $-\pi/2$ direction. The equation \refeq{eq:discepsilon}, together with the asymptotic property 
\begin{equation}\label{eq:asymptoticPropEps}
 \epsilon_{a,a+1}(\theta) - m_{a,a+1}\exp(\theta) = \order{\exp(-\theta)}~~\text{as}~~\theta\to\infty
\end{equation}
produce a Riemann-Hilbert problem, the solution of which is given by 
\begin{equation}
\label{eq:TBAsysMinC}
\epsilon_{a,a+1}(\theta) = m_{a,a+1}\exp(\theta) - K\star L_{a-1,a}(\theta)-K\star L_{a+1,a+2}(\theta)
\end{equation}
where the convolution $\star$ with the kernel $K$ is defined as
\begin{equation}
\label{eq:CoshKernel}
K \star f (\theta) =  \int_{\mathbb{R}} K(\theta,\bar{\theta}) f(\bar{\theta}) d\bar{\theta} = \frac{1}{2 \pi}\int_{\mathbb{R}} \frac{ f(\bar{\theta})~ d\bar{\theta}}{\cosh(\theta-\bar{\theta})}
\end{equation}
One can very easily notice that the TBA system \refeq{eq:TBAsysMinC} is indeed the ``conformal limit'' (i.e. $2 \cosh(\theta) \mapsto \exp(\theta)$) of the $AdS_3$ TBA system (appearing in e.g. \cite{Alday:2010vh,ToledoJonathan2016}), as stated in the introduction. Since the ``wall-crossing'' procedure follow the same pattern in both cases, this fact will stay true outside the minimal chamber.\\

As an example, let's write the TBA system for the sextic potential in the minimal chamber, consisting of $5$ TBA equations. An example of such moduli and energy configuration can be observed in figure \ref{fig:plotTripleWellMinC}. According to \refeq{eq:TBAsysMinC}, the TBA system solving the associated Riemann-Hilbert problem is
\begin{align}
\begin{split}
\epsilon_{1,2}(\theta)=& m_{1,2}\exp(\theta) - K\star L_{2,3}(\theta)\\
\epsilon_{2,3}(\theta)=& m_{2,3}\exp(\theta) - K\star L_{1,2}(\theta)- K\star L_{3,4}(\theta)\\
\epsilon_{3,4}(\theta)=& m_{3,4}\exp(\theta) - K\star L_{2,3}(\theta)- K\star L_{4,5}(\theta)\\
\epsilon_{4,5}(\theta)=& m_{4,5}\exp(\theta) - K\star L_{3,4}(\theta)- K\star L_{5,6}(\theta)\\
\epsilon_{5,6}(\theta)=& m_{5,6}\exp(\theta) - K\star L_{4,5}(\theta)
\end{split}
\end{align}
This TBA system is corresponding to the configuration of cycles in figure \ref{fig:cyclesSexticMinC} in the sense that each relevant cycle can be associated in a one to one relation to an $\epsilon$-function since they are encoding the Borel resummed WKB periods, periods that are obtained by integration along the said cycles.

\subsection{The TBA equations as a generalization of the ODE/IM correspondence}
\label{chap:TBAasODEIM}


In this section, we repeat the arguments presented in \cite{Ito:2018eon} with a few changes of notation. We want to study the Schrödinger equation \eqref{eq:Schro} provided that the potential is a degree $d$ polynomial. Getting rid of the $q^{d-1}$ term by a shift and scaling $u_d\mapsto 1$, it is always possible to rewrite \eqref{eq:Schro} in the canonical form
\begin{equation} \label{eq:SchroODEIM}
\left(	-\hbar^2 \partial_q^2 + q^d + \sum_{a=0}^{d-2} u_k q^k\right)\psi(q,u,\hbar) = 0
\end{equation}
where $u$ is the vector of $u_a$, $a\in \{1,\ldots,d-1\}$. \eqref{eq:SchroODEIM} is analogous to the $\hbar=1$ Schrödinger equation
\begin{equation} \label{eq:SchroODEIMz}
\left(	- \partial_z^2 + z^d + \sum_{a=0}^{d-2} b_k z^k\right)\tilde{\psi}(z,b) = 0
\end{equation}
obtained by the scaling
\begin{equation} \label{eq:scalingODEIS}
q \mapsto \hbar^{\frac{2}{d+2}}z~,\quad u_a \mapsto \hbar^{\frac{2(d-a)}{d+2}} b_a
\end{equation}
First, let's study the equation \eqref{eq:SchroODEIMz}. It is connecting with the usual ODE/IM correspondence in the particular case where $b_0 \neq 0$ but $b_a=0$ for $a\neq 0$ (i.e.  the potential is a pure polynomial). The WKB expansion of \eqref{eq:SchroODEIMz} yields for the exponentially decaying solution at positive infinity
\begin{equation}  \label{eq:WKBsolS0}
\tilde{y}(z,b) \sim \frac{1}{\sqrt{2i}} z^{n_d(b)} \exp\left(-\frac{2}{d+2} z^{\frac{d+2}{2}}\right)
\end{equation}
where
\begin{equation} 
n_d(b) = \begin{cases}
-\frac{d}{4} &d \text{ odd}\\
-\frac{d}{4}-B_{\frac{d+2}{2}}(b) &d \text{ even}
\end{cases}
\end{equation}
and the coefficients $B_n(b)$ defined such that
\begin{equation}
	\sqrt{1+\sum_{a=0}^{d-2} b_a z^{a-d}} = 1+ \sum_{n>0} B_n(b) z^{-n}
\end{equation} 
 \eqref{eq:SchroODEIMz} is invariant under the Symanzik rotations
 \begin{equation}  \label{eq:SimanzikRot}
 z \mapsto \omega z~,\quad b_a \mapsto \omega^{d-a} b_a ~,\quad \omega = e^{\frac{2 \pi i}{d+2}}
 \end{equation}
which acts on the coefficients $B_n(b)$ as
\begin{equation}
	B_{\frac{d+2}{2}}(\omega^{-k(d-a)} b_a) = (-1)^k B_{\frac{d+2}{2}}( b_a)
\end{equation}
With the use of Symanzik rotation, we can extend the solution \refeq{eq:WKBsolS0}, valid in $\mathcal{S}_0$, to the $d+2$ sectors $\mathcal{S}_k$ defined as
\begin{equation}
\mathcal{S}_k = \left\{ z \in \mathbb{C} \mid \abs{\arg(z)-\frac{2 \pi k}{d+2}}<\frac{\pi}{d+2} \right\}
\end{equation}
Explicitly, the extended solutions $\tilde{y}_k$ are
\begin{equation}
	\tilde{y}_k(z,b) = \omega^{\frac{k}{2}} \tilde{y}(\omega^{-k}z,\omega^{-k(d-a)} b_a)
\end{equation}
A general property of Wronskians is that the Wronskian of two solutions is independent of $z$:
\begin{equation} \label{eq:WronskianTilde}
\tilde{W}_{k_1,k_2}(b) = W(\tilde{y}_{k_1}(z,b),\tilde{y}_{k_2}(z,b))
\end{equation}
where we remind that the Wronskian of two differentiable functions is defined as $W(f(z),g(z)) = f(z) g'(z) - f'(z) g(z)$. Furthermore, it is not too hard to evaluate \eqref{eq:WronskianTilde} when $k_2=k_1+1=k+1$:
\begin{equation}
\tilde{W}_{k,k+1}(b) =\begin{cases}
1 &d \text{ odd}\\
\omega^{(-1)^k}B_{\frac{d+2}{2}}(b) &d \text{ even}
\end{cases}
\end{equation}
Now, let us introduce $f^{[n]}(z,b)$ as a short hand notation for the function with rotated arguments
\begin{equation}
f^{[n]}(z,b) = f(\omega^{-\frac{n}{2}}z,\omega^{-\frac{n}{2}(d-a)})
\end{equation}
such that one can prove 
\begin{equation}
\tilde{W}_{k_1+1,k_2+1}(b)=\tilde{W}_{k_1,k_2}^{[2]}(b)
\end{equation}
and deduce the following Plücker type relation
.
\begin{equation}\label{eq:Plucker}
\tilde{W}_{k_1,k_2}^{[2]}\tilde{W}_{k_1,k_2}=\tilde{W}_{k_1+1,k_2+1}\tilde{W}_{k_1,k_2}=-\tilde{W}_{k_1+1,k_2}\tilde{W}_{k_2+1,k_1}-\tilde{W}_{k_1+1,k_1}\tilde{W}_{k_2,k_2+1}
\end{equation}
We will now introduce the $Y$-functions associated with \eqref{eq:SchroODEIMz} as
\begin{align}\label{eq:YfunctionTilde}
	\begin{split}
		\tilde{Y}_{2n}(b)&=\left(\frac{\tilde{W}_{-n,n}\tilde{W}_{-n-1,n+1}}{\tilde{W}_{n,n+1} \tilde{W}_{-n-1,-n}}\right)(b)\\
		\tilde{Y}_{2n+1}(b)&=\left(\frac{\tilde{W}_{-n-1,n}\tilde{W}_{-n-2,n+1}}{\tilde{W}_{n,n+1}\tilde{W}_{-n-2,-n-1}}\right)^{[1]}(b)
	\end{split}
\end{align}
with $n\in \mathbb{N} $. Using \eqref{eq:Plucker} repeatedly, one can find the functional equations
\begin{equation} \label{eq:YsysTilde}
	\tilde{Y}_a^{[+1]}(b)\tilde{Y}_a^{[-1]}(b) = \left(1+\tilde{Y}_{a+1}(b)\right)\left(1+\tilde{Y}_{a-1}(b)\right)
\end{equation}
By the definition \eqref{eq:YfunctionTilde}, $\tilde{Y}_a=0$. Because $\tilde{y}_{n+d+2}(z) ~\propto~ \tilde{y}_{n}(e^{-2\pi i}z)=\tilde{y}_{n}(z)$, we also have $\tilde{Y}_d=0$. As a result, we have $d-1$ $Y$-functions leading to a $Y$-system of the $A_{d-1}$-type, reproducing in the pure potential case the $Y$-system of the usual ODE/IM correspondence, as stated above. \\

However, because of the rotated arguments (introduced through the $f^{[n]}$ notation), the moduli in the r.h.s of \eqref{eq:YsysTilde} are not the same than in the l.h.s. and the $Y$-system is not closed. In order to circumvent this problem, let's go back to the equation \eqref{eq:SchroODEIM}, including the extra parameter $\hbar$. First, we can relate the solutions $\tilde{y}(z,b)$ and $y(q,u,\hbar)$ using the scaling \eqref{eq:scalingODEIS}:
\begin{equation}
	y(q,u,\hbar)=\tilde{y}(z,b) = \tilde{y}\left(\hbar^{-\frac{2}{d+2}} q,\hbar^{\frac{2(a-d)}{d+2}} u\right)
\end{equation}
Noticing that the Symanzik rotation \refeq{eq:SimanzikRot} are just a rotation of $\hbar$, i.e. that
\begin{equation}
\omega^{-k} z = (e^{i \pi k} \hbar)^{-\frac{2}{d+2}}~,\quad \omega^{(a-d)k} b_a = (e^{i \pi k} \hbar)^{\frac{2(a-d)}{d+2}}
\end{equation}
one can extend the solution $y(q,u,\hbar)$ to other sectors of the complex plane using
\begin{equation}
	y_k(q,u,\hbar) = y(q,u,e^{i \pi k}\hbar)
\end{equation}
 The associated Wronskian, $W_{k_1+1,k_2+1}(u,\hbar) = \tilde{W}_{k_1+1,k_2+1}(b)$, is then
 \begin{equation}
 W_{k_1+1,k_2+1}(u,\hbar) = \hbar^{-\frac{2}{d+2}}\, W(y_{k_1}(q,u,\hbar),y_{k_2}(q,u,\hbar))
 \end{equation}
 and we can define the $Y$-functions as
 \begin{align}\label{eq:Yfunction}
 \begin{split}
 Y_{2n}(u,\hbar)&=\left(\frac{W_{-n,n} W_{-n-1,n+1}}{W_{n,n+1} W_{-n-1,-n}}\right)(u,\hbar)\\
 Y_{2n+1}(u,e^{-i \pi/2}\hbar)&=\left(\frac{W_{-n-1,n}W_{-n-2,n+1}}{W_{n,n+1}W_{-n-2,-n-1}}\right)(u,\hbar)
 \end{split}
 \end{align}
 such that they satisfy the functional equations
 \begin{equation} \label{eq:Ysys}
Y_{a}(u,e^{i \pi/2}\hbar)Y_{a}(u,e^{-i \pi/2}\hbar) =  \left(1+Y_{a+1}(u,\hbar)\right)\left(1+Y_{a-1}(u,\hbar)\right)
 \end{equation}
 with $Y_{0}(u,\hbar)=Y_{d}(u,\hbar)=0$ as in the $Y$-system \eqref{eq:YsysTilde}. This time, unlike \eqref{eq:YsysTilde}, the $Y$-system \eqref{eq:Ysys} is involving the same $u_k$ parameters on both sides.\\
 
The ultimate goal of this section is to derive the TBA equations \eqref{eq:TBAsysMinC}. Now that we found the $Y$-system \eqref{eq:Ysys}, we need to convert it into a TBA system. To achieve this, we still need to make the $\epsilon$-functions and masses \refeq{eq:masses} appear. Let us look at the low $\hbar$ behaviour of the $Y$-functions. In a sense we shall make very precise soon, the small $\hbar$ regime in this section corresponds to the large $\theta$ regime in the previous section. We should then have an equivalent formulation of the asymptotic property \eqref{eq:asymptoticPropEps} containing the masses and $\epsilon$-functions. In order to evaluate the asymptotic behaviour of the $Y$-functions, we need the Wronskian which means we need the asymptotic solutions. Proceeding to the WKB expansion of $y_k$ at small $\hbar$, one gets
\begin{equation}
y_k(q,u,\hbar) \sim \frac{(-1)^{\frac{1}{2}k}}{\sqrt{2i}} \hbar^{\frac{d}{2(d+2)}} \exp\left(\mp (-1)^k\frac{ i}{\hbar}\int_{s_k}^{q}P(\bar{q})d\bar{q}\right)
\end{equation}
where $\mp$ denotes in which Riemann sheet $q$ lives and $s_k\in \mathcal{S}_k$. Using this form of $y_k(q,u,\hbar)$, one can evaluate the Wronskian $W_{k_1+1,k_2+1}(u,\hbar)$ then the $Y$-functions using \eqref{eq:Yfunction}. Their asymptotic behavior for $\hbar\to 0$, valid for $\abs{\arg(\hbar)}<\pi$, is
\begin{align}\label{eq:Ysmallhbar}
	\begin{split}
	\log Y_{2k}(u,\hbar) &\sim -\frac{i}{\hbar} \oint_{\gamma_{d-2k,d-2k+1}} p(q)dq = -\frac{1}{\hbar}m_{d-2k,d-2k+1}\\
	\log Y_{2k+1}(u,\hbar) &\sim -\frac{1}{\hbar} \oint_{\gamma_{d-1-2k,d-2k}} p(q)dq = -\frac{1}{\hbar} m_{d-1-2k,d-2k}
	\end{split}
\end{align}
In \eqref{eq:Ysmallhbar}, the cycles thus the masses have confusing indices due to the definition \eqref{eq:Yfunction}. To make the equations nicer, let us relabel the $Y$-functions as $Y_k \mapsto Y_{d-k}$. This relabeling leaves \eqref{eq:Ysys} invariant. Now, let us define the analytic functions $\ell_{a,a+1}(\hbar)$ in $\abs{\arg(\hbar)}\leq \pi/2$ as
\begin{equation}
	\ell_{a,a+1}(\hbar) = \log Y_{a}(\hbar) + \frac{m_{a,a+1}}{\hbar}
\end{equation}
where we omitted the dependencies on the moduli $u_k$. The $Y$-system translates to
 \begin{align}\label{eq:Ysysell}
\begin{split}
\ell_{a,a+1}(e^{i \pi/2}\hbar)+\ell_{a,a+1}(e^{-i \pi/2}\hbar) &=  \log\left(1+Y_{a-1}(\hbar)\right)+\log\left(1+Y_{a+1}(\hbar)\right)\\
&=L_{a-1,a}(\hbar)+L_{a+1,a+2}(\hbar)
\end{split}
\end{align}
We can conclude by setting
\begin{equation} \label{eq:ODEIMthetaEpsilon}
	\hbar = e^{-\theta}~,\quad Y_a = e^{-\epsilon_{a,a+1}(\theta)}
\end{equation}
which is coherent with $L_{a,b}(\theta)= \log\left(1+\exp\left(-\epsilon_{a,b}(\theta)\right)\right)$ as it has been already defined previously. Convoluting \eqref{eq:Ysysell} with the kernel $K(\theta)$ defined in \eqref{eq:CoshKernel} and using the analyticity of $\ell$ yields the TBA system \eqref{eq:TBAsysMinC} and complete our derivation.

\section{Wall Crossing}
\label{chap:wallCrossing}
\subsection{Preamble on notations}
In order to make the equations in the present section more readable, we will group the couple of indices into one ``edge'' index: quantities of the form $Q_{i,j}$ will sometime be written as $Q_{(a)}$, where $(a)$ is to be understood as the ordered couple $(i,j)$ with $j>i$, corresponding to a cycle encircling the turning points $q_i$ and $q_j$. The orderedness of $(a)$ is due to the fact that the quantities $Q_{i,j}$ are typically antisymmetric since they are involving integration over cycles, i.e. $Q_{i,j}=-Q_{j,i}$. Later, when developing the theory of TBA graphs, the indices $(a)$ will be representing oriented edges. The natural way to represent such edges, because of their antisymmetric nature, is the structure we just mentioned. In other words, we have a one to one map between quantities $Q_{(i,j)}$ and $Q_{(j,i)}$, hence then can be used interchangeably and we make the choice to use the canonical form where $j>i$. \\

In previous sections (where we were in the context of the minimal chamber), we used notation of the form $Q_{i,i+1}$, notation that seems more cumbersome than it should be since one could use a unique index $i$ instead. But later, when we lose the ordering property of the turning points \refeq{eq:orderingTP} (because some of them are sent into the complex plane), we will find out that new relevant cycles are appearing into TBA systems and having two indices will be helpful in order to link together arbitrary turning points instead of limiting ourselves to consecutive ones. Additionally, we introduce this edge index notation $Q_{(a)}$ because it will be useful for writing more general TBA system in a concise way, especially when summing (so we can sum over the appropriate set of edges). For example, the TBA system \refeq{eq:TBAsysMinC} can be written as
\begin{equation}
\label{eq:TBAsysMinC2}
\epsilon_{(a)}(\theta) = m_{(a)}\exp(\theta) -\sum_{(b) \in s_d^{(a)}} K\star L_{(b)}(\theta)
\end{equation}
where $(a)=(i,i+1)$ and $i\in \{1,\ldots,d-1\}$ are labeling the $d-1$ relevant cycles in the minimal chamber and where $s_d^{(a)}=\{(i-1,i),(i+1,i+2)\}=\{(a-1),(a+1)\}$ is the set of the edges ``tangent'' to the edge $(a)$ and where we defined addition on the element $(a)$, $(a+n)$, as a short hand for $(i+n,j+n)$. This notation will be especially useful when the kernel depends on two distinct cycles (thus will have two edges indices). As we shall soon see, we can write these kind of TBA in the following compact form: 
\begin{equation} \label{eq:TBAnotation}
\epsilon_{(a)}(\theta) = \abs{\Pi_{(a)}}\exp(\theta) +\sum_{(b) \in S_d} K_{(a),(b)}\star L_{(b)}(\theta)
\end{equation}
where $ K_{(a),(b)}$ is a carefully selected kernel, $(a)=(i,i+k)$, $i\in \{1,\ldots,d-k\}$, $k\in \{1,\ldots,{d-1}\}$ and $S_d$ the set containing the  $d(d-1)/2$ possible nonequivalent oriented edges. When there is no kernels or sums and that the two indices are explicitly specified, we will prefer $Q_{i,j}$ to $Q_{(i,j)}$ for shortness of notation.

Of course, one could use four indices, writting $K_{ijkl}$ instead of $K_{(a),(b)} = K_{(i,j),(k,l)}$, but when we leave the edges unspecified (as mute variable in sums for example) the edge index notation is a little bit more compact, natural and reminiscent of a traditional vector-matrix multiplication. Alternatively, as in \cite{Ito:2018eon}, one could use only one index and write quantities $Q_{(a)}$ as $Q_a$, with $a\in \{1,\ldots,d(d-1)/2\}$, which is even more compact, but at the cost of transparency (it is harder to see which turning points are involved once we are grouping the edges $(a)$ consisting of pairs into only one larger list). In our opinion, the ordered pair $(a)=(i,j)$ with $j>i$ is the most natural way to describe more complicated TBA equations and related quantities, especially in the so-called \emph{maximal chamber} regime, for which the TBA system is the most intricate.

\subsection{Analytical continuation of two TBA equations in the mass representation}
\label{chap:analcontmasses}
The equations \refeq{eq:TBAsysMinC2} hold as long as we are in this special region of the moduli space where all the masses are real -- or equivalently where all the turning points of $V(q)$ are on the real axis -- i.e. in the minimal chamber. We can analytically continue the TBA system outside this region of the moduli space. Doing so, the $d-1$ masses are acquiring a complex part, and we can decompose them into their usual polar form:
\be
m_{(a)}=\abs{m_{(a)}} e^{\phi_{(a)}}~,~~~(a)=(i,i+1)~\text{and}~i\in \{1,\ldots,d-1\}
\ee
We also introduce the shifted function
\be
\tilde{f}_{(a)}(\theta) = f_{(a)}(\theta- i \phi_{(a)})
\ee
such that $\tilde{\epsilon}_{(a)}(\theta) \sim \abs{m_{(a)}}$ in the large $\theta$ limit. Rewriting \refeq{eq:TBAsysMinC2} with the shifted functions yields
\begin{equation}
	\label{eq:TBAsysMinC3}
	\tilde{\epsilon}_{(a)}(\theta) = \abs{m_{(a)}}\exp(\theta) -\sum_{(b) \in s_r^{(a)}} K_{(a),(b)}\star \tilde{L}_{(b)}(\theta)
\end{equation}
where we are convoluting with the kernel
\begin{equation}
	 K_{(a),(b)}(\theta) = K(\theta+i(\phi_{(b)}-\phi_{(a)})) = \frac{1}{2\pi}\frac{1}{\cosh(\theta+i(\phi_{(b)}-\phi_{(a)}))}
\end{equation}
As long as $\abs{\phi_{(a)}-\phi_{(a\pm 1)}}<\pi/2$, the modified TBA system \refeq{eq:TBAsysMinC3} holds and the analytical continuation is trivial. However, notice that we are hitting a pole once we cross $\abs{\phi_{(a)}-\phi_{(a\pm 1)}}=\abs{\phi_{(a),(a\pm 1)}}=\pi/2$ and we have to modify the TBA system in order to incorporate this non-trivial contribution.\\

Let us proceed to the analytical continuation of two consecutive TBA equations as an exercise. Let's say the  $\tilde{\epsilon}_{(a)}$ and $\tilde{\epsilon}_{(a+1)}$ equations, with $(a)=(1,2)$. As $\phi_{(a),(a+1)}$ crosses $\pi/2$ (hence $\phi_{(a+1),(a)}$ crosses $-\pi/2$), we pick the pole contribution from the kernel and modify the $\tilde{\epsilon}_{(a)}$ and $\tilde{\epsilon}_{(a+1)}$ accordingly:
\begin{align}
\label{eq:sheps12}
	\tilde{\epsilon}_{1,2}(\theta) =& ~\abs{m_{1,2}}\exp(\theta) - K_{(1,2),(2,3)}\star \tilde{L}_{2,3}(\theta)-\tilde{L}_{2,3}^-\left(\theta+i\phi_{(2,3),(1,2)}+i\delta\right)\\
\label{eq:sheps23}
	\tilde{\epsilon}_{2,3}(\theta) =& ~\abs{m_{2,3}}\exp(\theta) -\sum_{(b) \in \{(1,2),(3,4)\}} K_{(2,3),(b)}\star \tilde{L}_{(b)}(\theta)-\tilde{L}_{1,2}^+\left(\theta+i\phi_{(1,2),(2,3)}-i\delta\right)
\end{align}
where the subscript $\pm$ indicates a shift of $\pm i\pi/2$ in the argument, i.e.
\be
f^\pm(\theta) = f^\pm\left(\theta\pm\frac{i \pi}{2}\right)
\ee
The resulting system of $d-1$ TBA equations is not closed anymore. On could close the TBA system by adding the two additional equations corresponding to $\tilde{\epsilon}_{1,2}^+\left(\theta+i\phi_{(1,2),(2,3)}-i\delta\right)$ and $\tilde{\epsilon}_{2,3}^-\left(\theta+i\phi_{(2,3),(1,2)}+i\delta\right)$, thus obtaining a TBA system with $d+1$ equations. But there is a clever way to close this system using $d$ equations instead, as it was done in \cite{Gaiotto:2008cd}. By redefining a new set of $Y$-functions (related to the $\epsilon$-functions through $Y_{(a)}=e^{-\epsilon_{(a)}}$, see \eqref{eq:ODEIMthetaEpsilon}) one can absorb the new source terms in the LHS. This set of new $Y$-functions is
\begin{equation}
\label{eq:newYfcosh}
Y_{1,2}^n=\frac{Y_{1,2}}{1+Y_{2,3}^-},~~~~Y_{2,3}^n=\frac{Y_{2,3}}{1+Y_{1,2}^+},~~~~Y_{1,3}^n=\frac{Y_{1,2}Y_{2,3}^-}{1+Y_{1,2}+Y_{2,3}^-}
\end{equation}
Rewriting \refeq{eq:TBAsysMinC3} with the new functions, then omitting the superscript $n$, one find that the two concerned TBA and their neighbor (there is only one in this case: $(3,4)$) are modified:
\begin{align}
\begin{split}
\label{eq:TBAnewEps}
\tilde{\epsilon}_{1,2}(\theta) =& ~\abs{m_{1,2}}\exp(\theta) - K_{(1,2),(2,3)}\star \tilde{L}_{2,3}(\theta)-K_{(1,2),(1,3)}^+\star \tilde{L}_{1,3}(\theta)\\
\tilde{\epsilon}_{2,3}(\theta) =& ~\abs{m_{2,3}}\exp(\theta) -\sum_{(b) \in \{(1,2),(3,4),(1,3)\}} K_{(2,3),(b)}\star \tilde{L}_{(b)}(\theta)\\
\tilde{\epsilon}_{3,4}(\theta) =& ~\abs{m_{3,4}}\exp(\theta) - \sum_{(b) \in \{(2,3),(4,5)\}}K_{(3,4),(b)}\star \tilde{L}_{(b)}(\theta)-K_{(3,4),(1,3)}^+\star \tilde{L}_{1,3}(\theta)
\end{split}
\end{align}
The rest of the system stays the same. In order to get the additional TBA equation, one need to consider the sum \refeq{eq:sheps12}$+$\refeq{eq:sheps23}, evaluated at $\theta \mapsto \theta + i \phi_{(1,3),(1,2)}$ and  $\theta\mapsto\theta+i\phi_{(1,3),(2,3)}-i\pi/2$ respectively, yielding
\begin{equation}
	\tilde{\epsilon}_{1,3}(\theta) = ~\abs{m_{1,3}}\exp(\theta) - \sum_{(b) \in \{(1,2),(3,4)\}}K_{(1,3),(b)}^-\star \tilde{L}_{(b)}(\theta) -K_{(1,3),(2,3)}\star \tilde{L}_{2,3}(\theta) 
\end{equation}
where $\abs{m_{1,3}} e^{\phi_{1,3}}=m_{1,3}=m_{1,2}-i m_{2,3}$. This last definition for the mass is better understood in the period representation where $\Pi_{1,3} = \Pi_{1,2}+\Pi_{2,3}$.\\

As an application, let's write the TBA system for the cubic potential ($d=3$) in the maximal chamber:
\begin{align*}
\tilde{\epsilon}_{1,2}(\theta) =& ~\abs{m_{1,2}}\exp(\theta) - K_{(1,2),(2,3)}\star \tilde{L}_{2,3}(\theta)-K_{(1,2),(1,3)}^+\star \tilde{L}_{1,3}(\theta)\\
\tilde{\epsilon}_{2,3}(\theta) =& ~\abs{m_{2,3}}\exp(\theta) - K_{(2,1),(1,2)}\star \tilde{L}_{1,2}(\theta) - K_{(2,1),(1,3)}\star \tilde{L}_{1,3}(\theta)\\
\tilde{\epsilon}_{1,3}(\theta) =& ~\abs{m_{1,3}}\exp(\theta) -  K_{(1,3),(1,2)}^-\star \tilde{L}_{1,2}(\theta) - K_{(1,3),(2,3)}\star \tilde{L}_{2,3}(\theta)
\end{align*}

The mass representation has the advantage that it is convenient to use in the minimal chamber, and it is easy to define the branch cuts and orientations prescription: all the masses are real and positive by definition in the minimal chamber. Furthermore, the kernel takes a particular simple form in this regime. But it also have some drawbacks: the masses are pairing in an unnatural way. Worse, we have these shifted kernels ${K}^\pm$ after wall-crossing, which require additional data in the form of the matrix $\langle (a),(b)\rangle_\pm$. Notice that one can encode all the TBA system in any chamber writing the TBA equations in the compact form
\begin{equation}
\epsilon_{(a)}(\theta) = \abs{m_{(a)}}\exp(\theta) -\sum_{(b) \in S_d} K_{(a),(b)}\star \tilde{L}_{(b)}(\theta)
\end{equation}
by defining the kernel to be
\begin{equation}
K_{(a),(b)}(\theta) = \frac{1}{2\pi}\frac{\langle (a),(b)\rangle_I}{\cosh\left(\theta+i\phi_{(b),(a)}+\langle (a),(b)\rangle_\pm\right)}
\end{equation}
All the data of the TBA system is contained in two matrices. On one hand the matrix $\langle (a),(b)\rangle_I$, encoding the intersection of cycles; it is symmetric and can take the values 0, 1 or 2. On the other hand, $\langle (a),(b)\rangle_\pm$, encoding the shifts on the argument of the kernel ${K}^\pm$; it is antisymmetric and can take the values $\pm i \pi/2$ or 0. By using the period representation, one can encode any TBA system using a unique antisymmetric matrix $\langle (a),(b)\rangle$ with values $\pm 2$, $\pm 1$ or 0. For this reason, we shall now depart from our present conventions (the one in \cite{Ito:2018eon,Alday:2010vh}) in order to adopt the period formulation (used in \cite{ToledoJonathan2016}).

\subsection{Analytical continuation of two TBA equations in the period representation}
\label{chap:analcontperiods}

Instead of the masses \refeq{eq:masses}, we are now going to use the classical periods
\begin{equation}
		\Pi_{(a)} = \oint_{(a)}p(q)dq
\end{equation}
where we omitted the superscript $(0)$ and where $\oint_{(a)}$ denotes an integration along the cycle encircling the two turning points contained in the edge index $(a)$. We slightly change our conventions for the contours orientations and branches prescriptions: we chose them such that the periods corresponding to cycles encircling the allowed (resp. forbidden) region is real (resp. imaginary) and positive. Equivalently, it simply comes down to multiply the even masses of the form $m_{2k,2k+1}$ by $i$ in our previous representation. We want to write our complex periods in the polar coordinates:
\begin{equation}
\Pi_{(a)} = \abs{\Pi_{(a)}} e^{i \varphi_{(a)}}
\end{equation}
In the minimal chamber, the argument $\varphi_{(a)}$ is either $0$ or $\pi/2$ with the prescription adopted above, such that $\varphi_{(a)}=\phi_{(a)}$ or $\varphi_{(a)}=\phi_{(a)}+\pi/2$ depending if $(a)$ is selecting a classically allowed or forbidden region. In order to take into account the fact that the periods are already complex in the minimal chamber, we have to define an index-dependent kernel from the start:
\begin{equation}
\label{eq:kernelSinh}
K_{(a),(b)}(\theta) = \frac{1}{2\pi i}\frac{\langle (a),(b)\rangle}{\sinh\left(\theta+i\varphi_{(a),(b)}\right)}
\end{equation}
which is obviously equivalent to the kernel \refeq{eq:CoshKernel} in the minimal chamber\footnote{Since $\sinh(x-i\pi/2)=-i\cosh(x)$.} with the appropriate choice of sign for the intersection (antisymmetric) matrix $\langle (a),(b)\rangle$; for now it can only take the value $\pm 1$ (the cycles are intersecting) or 0 (no intersection of cycles). If we rewrite the TBA equations for a system in the minimal chamber in the following way
\begin{equation}
\label{eq:TBAsysMinC4}
\tilde{\epsilon}_{(a)}(\theta) = \abs{\Pi_{(a)}}\exp(\theta) +\sum_{(b) \in s_d} K_{(a),(b)}\star \tilde{L}_{(b)}(\theta)
\end{equation}
where $s_d=\{(i,i+1)\}$ with $i \in \{1,\ldots,d-1\}$ (i.e. the set of all the relevant couplings in the minimal chamber), then the system \refeq{eq:TBAsysMinC4} is equivalent to the system in \refeq{eq:TBAsysMinC3} if the intersection matrix is a tridiagonal antisymmetric matrix ${\langle (a+k),(a+k+1)\rangle} = (-1)^k = -\langle (a+k+1),(a+k)\rangle$; ${\langle(a),(b)\rangle}=0$ otherwise. A graphical representation of such matrices can be fond for $d \in \{3,\ldots,8\}$ in figure \ref{fig:TBAmaxC}: the $(d-1)\times (d-1)$ matrix in the uper left corner (delimited by the black lines) is $(-1)^{d-1} {\langle(a),(b)\rangle}$. The sign of the intersection matrix ${\langle(a),(b)\rangle}$ is correlated to the sign of $\varphi_{(a),(b)}$.\\

We just reformulated our minimal chamber TBA system using the classical periods instead of the masses.  Let us now proceed to its analytical continuation. This time, because of the modified kernel \refeq{eq:kernelSinh}, we hit a pole when $\varphi_{(a),(a+1)}$ crosses $0$, i.e. when the phases of the periods align (this fact will be important for the diagrammatic procedure in section \ref{chap:wallCrossingDiagram}). As for the masses case, we will focus on the $(a)=(1,2)$ case that is coupling $\tilde{\epsilon}_{1,2}$ and $\tilde{\epsilon}_{2,3}$ together. Appart from the small convention modifications, we can reproduce the story of the previous section verbatim. Deforming the contour and picking the pole leads to additional source terms.
\begin{align}
\label{eq:sinhsheps12}
\tilde{\epsilon}_{1,2}(\theta) =& ~\abs{\Pi_{1,2}}\exp(\theta) + K_{(1,2),(2,3)}\star \tilde{L}_{2,3}(\theta)+\tilde{L}_{2,3}\left(\theta+i\varphi_{(1,2),(2,3)}+i\delta\right)\\
\label{eq:sinhsheps23}
\tilde{\epsilon}_{2,3}(\theta) =& ~\abs{\Pi_{2,3}}\exp(\theta) + \sum_{(b)\in \{(1,2),(3,4)\}} K_{(2,3),(b)}\star \tilde{L}_{(b)}(\theta)+\tilde{L}_{1,2}\left(\theta+i\varphi_{(2,3),(1,2)}-i\delta\right)
\end{align}
As before, they can be absorbed by defining new $Y$-functions (hence new $\epsilon$-functions through $Y_{(a)}=e^{-\epsilon_{(a)}}$):
\begin{equation}
\label{eq:newYfsinh}
\tilde{Y}_{1,2}^n=\frac{\tilde{Y}_{1,2}}{1+\tilde{Y}_{2,3}^+},~~~~\tilde{Y}_{2,3}^n=\frac{\tilde{Y}_{2,3}}{1+\tilde{Y}_{1,2}^-},~~~~\tilde{Y}_{1,3}^n=\frac{\tilde{Y}_{1,2}\tilde{Y}_{2,3}^-}{1+\tilde{Y}_{1,2}+\tilde{Y}_{2,3}^+}
\end{equation}
except that the $\pm$ superscrip in \refeq{eq:newYfsinh} is no longer describing a shift by $\pm i \pi/2$ as in \refeq{eq:newYfcosh} but a shift by $\pm i \varphi_{(1,2),(2,3)}= \mp i \varphi_{(2,3),(1,2)}$. Besides, one can notice that in the additional source terms in \refeq{eq:sinhsheps12} and \refeq{eq:sinhsheps23} as well as in the new $Y$-functions definition \refeq{eq:newYfsinh}, one does not need this $\pm i \pi/2$ shift anymore. Plugging \refeq{eq:newYfsinh} into \refeq{eq:sinhsheps12} and omitting the superscript $n$, one gets the reformulated version of \refeq{eq:TBAnewEps}
\begin{align}
\begin{split}
\label{eq:TBAnewEpsSinh}
\tilde{\epsilon}_{1,2}(\theta) =& ~\abs{\Pi_{1,2}}\exp(\theta)+\sum_{(b) \in \{(2,3),(1,3)\}} K_{(1,2),(b)}\star \tilde{L}_{(b)}(\theta) \\
\tilde{\epsilon}_{2,3}(\theta) =& ~\abs{\Pi_{2,3}}\exp(\theta) +\sum_{(b) \in \{(1,2),(3,4),(1,3)\}} K_{(2,3),(b)}\star \tilde{L}_{(b)}(\theta)\\
\tilde{\epsilon}_{3,4}(\theta) =& ~\abs{\Pi_{3,4}}\exp(\theta) + \sum_{(b) \in \{(2,3),(4,5),(1,3)\}}K_{(3,4),(b)}\star \tilde{L}_{(b)}(\theta)
\end{split}
\end{align}
and the new TBA equation involving $\tilde{\epsilon}_{1,3}$ and the additional period $\Pi_{1,3}=\Pi_{1,2}+\Pi_{2,3}$ is obtained by adding $\tilde{\epsilon}_{1,2}$ and $\tilde{\epsilon}_{2,3}$, with the appropriate shifts. One finds
\begin{equation}
\tilde{\epsilon}_{1,3}(\theta) = ~\abs{\Pi_{1,3}}\exp(\theta) + \sum_{(b) \in \{(1,2),(2,3),(3,4)\}}K_{(1,3),(b)}\star \tilde{L}_{(b)}(\theta)
\end{equation}

In fact, the lack of $\pm i \pi/2$ shifts allow us to rewrite the general analytic continuation of the couple $(\tilde{\epsilon}_{(i,i+1)},\tilde{\epsilon}_{(i+1,i+2)})$ leading to the additionnal TBA equation $\tilde{\epsilon}_{(i,i+2)}=\tilde{\epsilon}_{(c)}$ in a nice and compact manner:
\begin{align}
\begin{split}
\label{eq:TBAnewEpsSinhCompact}
\tilde{\epsilon}_{(a)}(\theta) =& ~\abs{\Pi_{(a)}}\exp(\theta)+\sum_{(b) \in s_d^{(a)}\cup(c) } K_{(a),(b)}\star \tilde{L}_{(b)}(\theta) \\
\tilde{\epsilon}_{(c)}(\theta) =& ~\abs{\Pi_{i,i+2}}\exp(\theta) +\sum_{(b) \in A}K_{(a),(b)}\star \tilde{L}_{(b)}(\theta)
\end{split}
\end{align}
where the $(a) \in A$ and $A$ is the set including the four couples $(i+k,i+k+1)$ with $k \in \{-1,0,1,2\}$. The rest of the TBA system is unchanged. As a final simplifying remark, let's notice that we only selected the explicitly non-zero kernels in the system above. If we allow for null kernels (through the intersection matrix $\langle (a),(b)\rangle$), one can simply defines the set of the $r$ relevant coupling in the minimal chamber: $s_d=\{(j,j+1)\}$ with $j \in \{1,\ldots,d\}$, and extend it with the new pair $(c)=(i,i+2)$. The system \refeq{eq:TBAnewEpsSinhCompact} reduces to
\begin{equation}
\label{eq:TBAsysCompact}
\tilde{\epsilon}_{(a)}(\theta) = \abs{\Pi_{(a)}}\exp(\theta) +\sum_{(b) \in s_d\cup(c)} K_{(a),(b)}\star \tilde{L}_{(b)}(\theta)
\end{equation}
where $(a)\in s_d\cup(c)$ such that it is unifying the $d$ TBA equations into the same expression. All the terms that are not appearing explicitly in \refeq{eq:TBAnewEpsSinhCompact} are put to 0 through the intersection matrix $\langle (a),(b)\rangle$ included into the definition of the kernel \eqref{eq:kernelSinh}. In the next section (section \ref{chap:NumberOfTBA}), we will prove that we have at most $d(d-1)/2$ TBA equation for the maximally involved TBA system describing the resumed WKB periods of a degree $d$ polynomial. In other words, it means that one can encode any TBA system in the (at most) $d(d-1)/2 \times d(d-1)/2$ dimensional intersection matrix. We will explain how to find this matrix graphically in the section \ref{chap:wallCrossingDiagram}, then we will proceed to write it for an arbitrary degree $d$ polynomial.

\subsection{Number of TBA equation and the associated region in the moduli space}
\label{chap:NumberOfTBA}

In the minimal chamber, for a polynomial with $d$ roots, they are only $d-1$ relevant pairings of these turning points, corresponding to the $d-1$ couples of successive turning points i.e. of the form $(i,i+1)$. This is the a priori minimal number of TBA equation\footnote{i.e. before simplifications due to symmetries; see section \ref{chap:simplifyTBA} for examples of TBA system reduced because of symmetries.} one need to solve in order to resum the quantum periods, hence the ``\emph{minimal}'' in minimal chamber. One can see the wall-crossing procedure presented above as a way to form (TBA equations that are involving) the desired additional pair of turning points, hence period. By repeating the process, one can continue to produces the desired periods until one obtains the appropriate TBA system. For a number of TBA equation $>d-1$, we will designate the corresponding region in the moduli space as the \emph{intermediate chamber}. However, this process can't continue forever. Indeed, after the $d-1$ pairs of the form $(i,i+1)$, one can add the pairs of the form $(i,i+2)$, $(i,i+3)$ etc. until one reaches the last possible pair: $(1,d)$. Thus, all the possible non-equivalent pairs can be labeled by $(i,i+k)$ with $i\in\{1,\ldots,d-k\}$, $k\in\{1,\ldots,d-1\}$. We call this set $S_d$, and it has exactly $\sum_{n=1}^{d-1} n = d(d-1)/2$ elements. We will call a TBA system with exactly this number of TBA equation in the \emph{maximal chamber}.

We explain in section \ref{chap:analcontperiods} that we can encode any TBA system using a (at most) $d(d-1)/2 \times d(d-1)/2$ dimensional matrix. Indeed, it is always possible to write any TBA system in the form 
\begin{equation}
\label{eq:ultimateTBA}
\tilde{\epsilon}_{(a)}(\theta) = \abs{\Pi_{(a)}}\exp(\theta) +\sum_{(b) \in S_d} K_{(a),(b)}\star \tilde{L}_{(b)}(\theta)
\end{equation}
and $(a) \in S_d$. Of course, if the system is not in the maximal chamber, we can select the appropriate subset of $S_d$ and the system \refeq{eq:ultimateTBA} reduces to a simpler one.

\subsection{Wall Crossing as a diagrammatic procedure}
\label{chap:wallCrossingDiagram}

In the previous sections (\ref{chap:analcontperiods} and \ref{chap:analcontmasses}), we derived in great detail the standard procedure to follow in order to analytically continue the TBA equations outside the minimal chamber. However, this process can be quite cumbersome to carry on multiple times using only algebra, especially when the degree of the polynomial of interest is large (the number of times we have to wall-cross is increasing with the square of $d$ for the maximal chamber). For example, if one was supposed to analytically continue the TBA system for the sextic potential from the minimal chamber to the maximal chamber, one would need to repeat the wall-crossing process ten times! In order to make this process simple, tractable and less prone to errors, it is a good idea to abstract it into a diagrammatic procedure. The purpose of this section is to provide simple rules that accomplish just that, inspired by Jonathan Toledo's thesis \cite{ToledoJonathan2016}.
The diagrammatic rules could in principle accommodate the mass formulation of the TBA equations. Yet, we found the diagrammatic process easier to carry on using the period formulation. We shall then use it in the following.

\begin{figure}
	\center
	\includegraphics[width=0.7\textwidth]{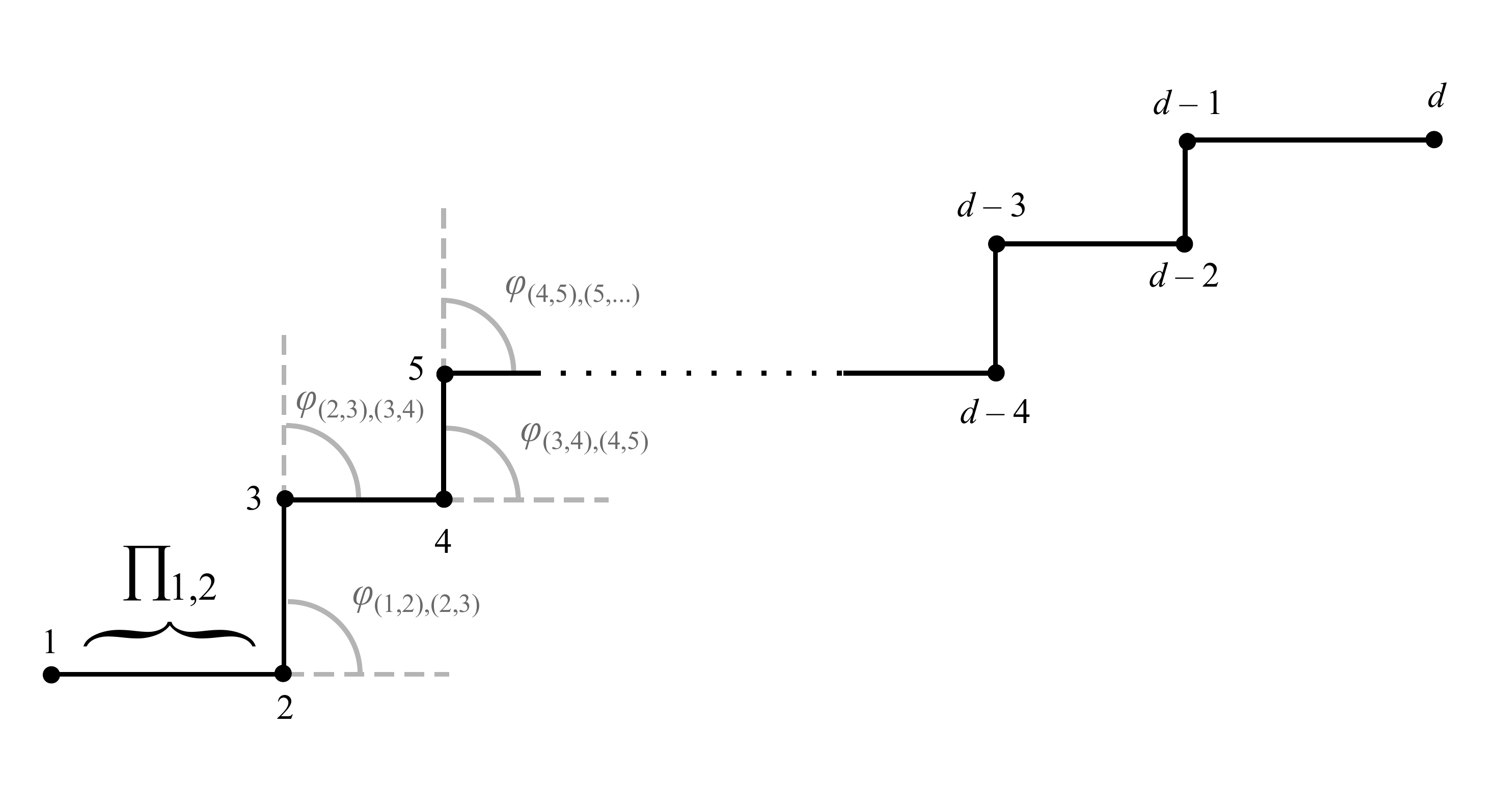}
	\caption{TBA graph corresponding to a polynomial of degree $d$ in the minimal chamber.}
	\label{fig:MinCdiagram}
\end{figure}

First, let us define how to associate a TBA system with a diagram. The starting point is Toledo's idea consisting of drawing the successive periods in the minimal chamber configuration as two dimensional vectors $(\Re \Pi_{(a)},\Im \Pi_{(a)})$, starting with $\Pi_{1,2}$ placed at some origin point, then arranging the next vectors of periods tail to tip, i.e. the endpoint of a given period vector is the starting point of the next. In the minimal chamber, this construction produces some kind of ``stair'' diagram, because we alternate between classically allowed and forbidden regions, such that if $\varphi_{(a)} =0$, then $\varphi_{(a+1)} =\pi/2$ etc. Therefore, the angle difference $\varphi_{(a),(a+1)}=\pm \pi/2$ $\forall a$ in the minimal chamber. The diagram should look like the figure \ref{fig:MinCdiagram} at this point.\\

We can abstract this construction a little bit further. Indeed, the length $\abs{\Pi_{(a)}}$, if intervening into the associated TBA equation through the $\abs{\Pi_{(a)}}\exp(\theta)$ term, will not be relevant to derive the TBA system (the length of the periods vector is always confined in the ``length'' term, the structure of which stays unchanged whatever the TBA system is). All we need to find using this diagrammatic procedure is the intersection matrix $\langle(a),(b) \rangle$. Furthermore, the global orientation is inconsequential, and one can even  ``locally'' transform the angles $\varphi_{(a),(a+1)}$, but at a \emph{crucial condition}. As we have already stated, as we are deforming the integral contour, the kernel \refeq{eq:kernelSinh} hits a pole when $\varphi_{(a),(b)}=0$, and we want to keep that property of paramount importance. Therefore, we are allowed to transform these angles as long as the transformed difference of angles $f_{(a),(b)}(\varphi_{(a),(b)})$ is 0 exactly when the untouched difference of angles is 0 and nowhere else\footnote{i.e. the functions $f_{(a),(b)}$ must satisfy $f_{(a),(b)}(x)=0$ if and only if $x=0$.}. We can throw all these transformed diagrams into the same class of equivalence, which defines a new, more abstract, diagrammatic object. The later is a lot more flexible but just as useful at the task of deriving the morphed TBA equations. If this diagrammatic object has a little bit more structure than a graph (because of the angle property), we will call it the TBA graph nonetheless, by abuse of notation and with the purpose of using some graph lingo, like \emph{edge} and \emph{vertex}. In this regard, each turning point is symbolized by a vertex $q_i$ and each TBA equation is symbolized by an edge $(a)=\{q_i,q_j\}=(i,j)$. With this in mind, it is time to state the very simple first diagrammatic rule: \emph{the connection rule}.
\begin{enumerate}
\item  \emph{Connection rule: the intersection matrix} $\langle(a),(b) \rangle = \pm 1$ \emph{if and only if the edges} $(a)$ \emph{and} $(b)$ \emph{are connected (i.e. they share a common vertex).}
\end{enumerate}
\begin{figure}
	\center	\includegraphics[width=0.8\textwidth]{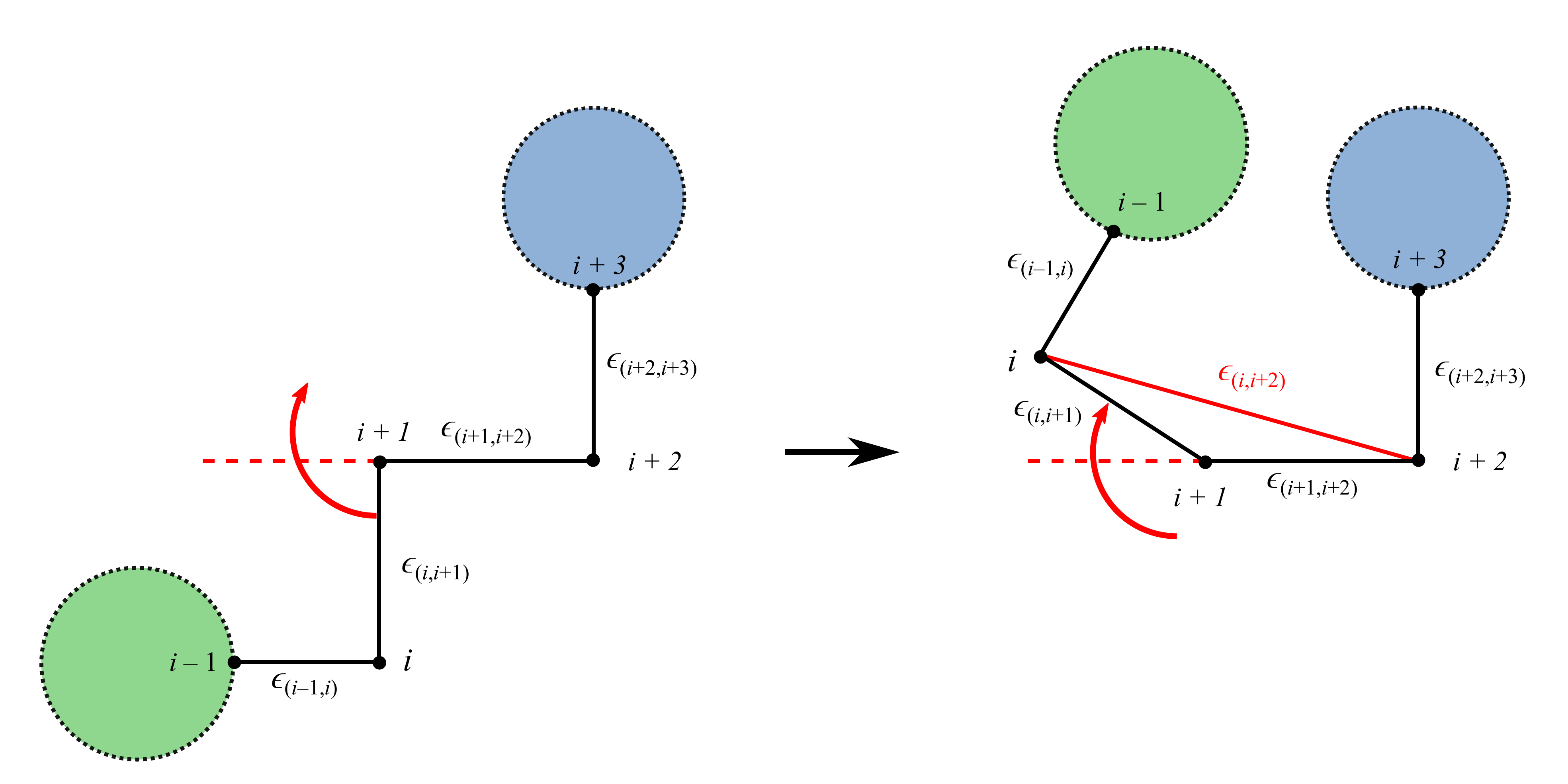}
	\caption{Four TBA equations in the "stair" configuration with arbitrary TBA graph attached at the end points vertices $i-1$ and $i+3$ (green and blue disks). We are analytically continuing from the left TBA graph, where $\varphi_{(i,i+1),(i+1,i+2)}\neq 0$, then we cross the 0 angle, where the periods are aligning (red dashed line); appling the Wall-crossing rule, we end up with an additional TBA equation for $\epsilon_{(i,i+2)}$ (red full line) in the TBA graph on the right. If the discs are other "stairs", this is corresponding to the morphing from the minimal chamber as presented in section \ref{chap:analcontperiods}.}
	\label{fig:AnalcontDiagram}
\end{figure}
The precise sign depends on the sign of $\varphi_{(a),(b)}$. One can check that the TBA graph in figure \ref{fig:MinCdiagram} is indeed reproducing the TBA system in the minimal chamber: since the edge $(a)$ is connected to the edges $(a\pm 1)$, we find this tridiagonal antisymmetric structure we found previously (the anti-symmetry of the intersection matrix is a direct result of the anti-symmetry of $\varphi_{(a),(b)}$).\\

Now that we are armed with the TBA graph and its first rule, let us show how and under what additional rules it is reproducing the wall-crossing procedure. As stated before,  nothing really dramatic happens as long as $\varphi_{(a),(b)}$ is not crossing 0. But when it does, we get an additional $\epsilon$ function. To be precise, when $\varphi_{(i,i+1),(i+1,i+2)}$ cross 0, we get an additional TBA equation involving the new $\epsilon_{(i,i+2)}$. This new pseudoenergy is coupled with four of the ``neighbour'' TBA, as defined in  \refeq{eq:TBAnewEpsSinhCompact}. That's the whole story of section \ref{chap:analcontperiods}. Let us reproduce this procedure by stating the second diagrammatic rule: \emph{the wall-crossing rule}.
\begin{enumerate}
	\setcounter{enumi}{1}
	\item  \emph{Wall-crossing rule: when the edges $(i,j)$ and $(j,k)$ cross their common alignment line, one add the new edge $(i,k)$ to the TBA graph.}
\end{enumerate}
\begin{figure}
	\center
	\includegraphics[width=\textwidth]{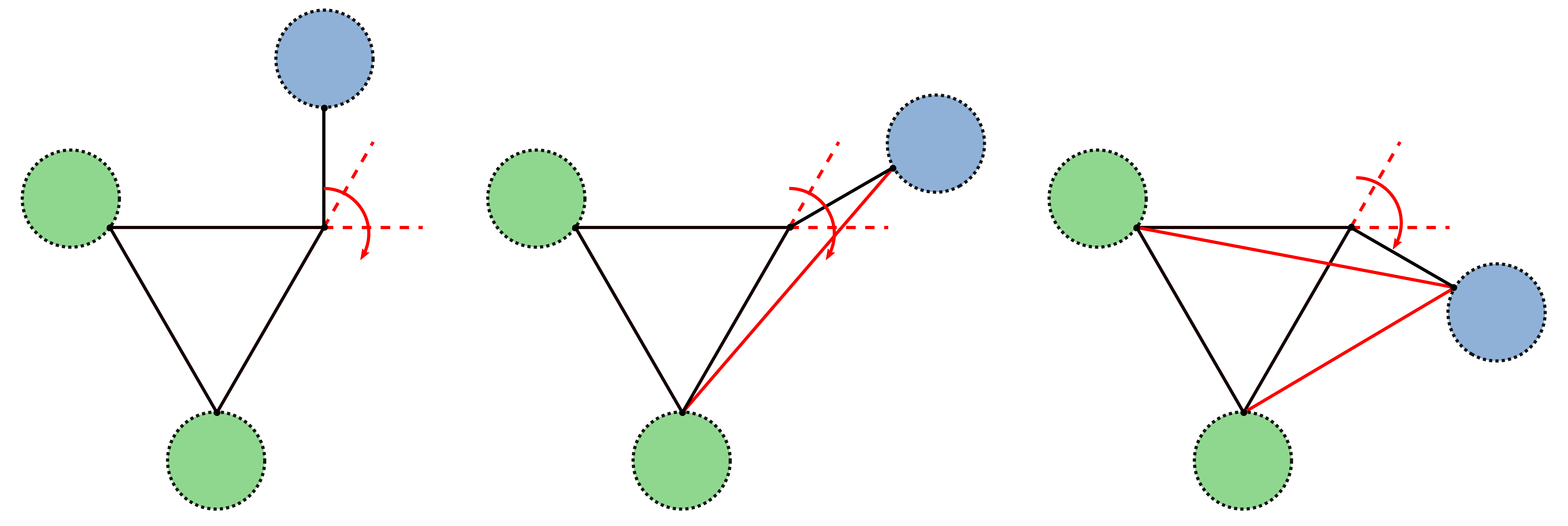}
	\caption{Analytical continuation crossing two consecutive period alignments around the same ``pivot'' vertex. The blue and green discs are arbitrary TBA graphs. Crossing the second alignment angle produces an intersection from which we can extract an additional entry of the intersection matrix through the intersection rule. Notice that we could have generalized this diagram, having $k$ green discs (linked or not) instead of only two. One would have obtained $k$ additional TBA equations after morphing the TBA equation attached to the blue disc around the ``pivot'' vertex, $k-1$ of which would be intersecting the further right TBA equation with linked green disc, $k-2$ would be intersecting the second TBA from the right etc.}
	\label{fig:AnalContCoef2}
\end{figure}
Of course, $(i,j)$ and $(j,k)$ are aligning exactly when $\varphi_{(i,j),(j,k)}=0$. This means that, with the addition of this rule, the TBA graph is at least modified at the right time. Now we can use the connection rule to verify if we are reproducing the correct TBA. Before applying the wall-crossing rule, we are in the following situation: two connected edges, $(i,j)$ and $(j,k)$, are aligning. They are themselves possibly connected to other edges through the vertex $i$ and $k$, but let's assume for now that the edges $(i,j)$ and $(j,k)$ have only one neighbor each, $(m,i)$ and $(k,n)$ (this is the case in the minimal chamber for example, where $m=i-1,j=1+1,k=1+2$ and $n=i+3$). Applying the wall-crossing rule, we draw an additional edge $(i,k)$. Now, we can read the new TBA system by applying the connection rule hence updating the intersection matrix. The TBA equation associated to the new edge is coupling the four neighbor together and no other. This is the case in \refeq{eq:TBAnewEpsSinhCompact}. Only the four neighbor are morphed by this wall-crossing. This is still the case in \refeq{eq:TBAnewEpsSinhCompact}. The whole story of section \ref{chap:analcontperiods} is in fact reproduced by the diagrammatic process presented in figure~\ref{fig:AnalcontDiagram}.\\

We stated a ``construction rule'' (the wall-crossing rule), that is allowing us to analytically continue a TBA graph, and a ``reading rule'' (the connection rule) that provides a way to read a TBA graph and make it corresponds to a TBA system. However, we still miss an ingredient in the reading side to complete the story and have a correct correspondence between TBA graphs and TBA systems. The third and last diagrammatic rule is taking care of this missing element and complete the translation. We purposefully put an important detail under the rug until now: there are some $\pm2$ elements appearing in the intersection matrix for some TBA systems. This is a new effect that is not predicted by our current set of rules. As an exercise, one can start by working out the simpler case where this new effect takes place: the quartic potential ($d=4$). After the first wall-crossing, we are in the situation described diagrammatically by a triangle with an unique leg, corresponding to the TBA system \refeq{eq:TBAnewEpsSinhCompact} and containing $3+1=4$ equations (this is the starting configuration of figure \ref{fig:AnalContCoef2} with empty discs). By repeating the wall-crossing procedure algebraically two more times, one obtains the following TBA system corresponding to a quartic potential in the maximal chamber:

\footnotesize
\begin{align}
\begin{split}
\label{eq:quarticMaxChamber}
 \epsilon_{1,2}(\theta) =& \abs{\Pi_{1,2}} \exp(\theta)+K_{(1,2),(2,3)}\star L_{2,3}+K_{(1,2),(1,3)}\star L_{1,3}+K_{(1,2),(2,4)}\star L_{2,4}+K_{(1,2),(1,4)}\star L_{1,4}\\\epsilon_{2,3}(\theta) =& \abs{\Pi_{2,3}} \exp(\theta)-K_{(2,3),(1,2)}\star L_{1,2}-K_{(2,3),(3,4)}\star L_{3,4}-K_{(2,3),(1,3)}\star L_{1,3}-K_{(2,3),(2,4)}\star L_{2,4}\textcolor{red}{-2K_{(2,3),(1,4)}\star L_{1,4}}\\\epsilon_{3,4}(\theta) =& \abs{\Pi_{3,4}} \exp(\theta)+K_{(3,4),(2,3)}\star L_{2,3}+K_{(3,4),(1,3)}\star L_{1,3}+K_{(3,4),(2,4)}\star L_{2,4}+K_{(3,4),(1,4)}\star L_{1,4}\\\epsilon_{1,3}(\theta) =& \abs{\Pi_{1,3}} \exp(\theta)-K_{(1,3),(1,2)}\star L_{1,2}+K_{(1,3),(2,3)}\star L_{2,3}-K_{(1,3),(3,4)}\star L_{3,4}-K_{(1,3),(1,4)}\star L_{1,4}\\\epsilon_{2,4}(\theta) =& \abs{\Pi_{2,4}} \exp(\theta)-K_{(2,4),(1,2)}\star L_{1,2}+K_{(2,4),(2,3)}\star L_{2,3}-K_{(2,4),(3,4)}\star L_{3,4}-K_{(2,4),(1,4)}\star L_{1,4}\\\epsilon_{1,4}(\theta) =& \abs{\Pi_{1,4}} \exp(\theta)-K_{(1,4),(1,2)}\star L_{1,2}\textcolor{red}{+2K_{(1,4),(2,3)}\star L_{2,3}}-K_{(1,4),(3,4)}\star L_{3,4}+K_{(1,4),(1,3)}\star L_{1,3}+K_{(1,4),(2,4)}\star L_{2,4}
 \end{split}
\end{align}\normalsize
where we omitted the $\theta$ dependence on the convolutions and the tildes for shortness, and where we factored out the intersection matrix $\langle(a),(b) \rangle$ in the definition \refeq{eq:kernelSinh} for clarity. Here, all the terms are correctly predicted by our two diagrammatic rules, except two terms with coefficient $\pm2$ (in red). If we repeat this exercise with other TBA systems, the same phenomenon occurs each time we have to wall-cross two (or more) successive edges attached to the same vertex, thus creating intersecting graphs. To take these coefficient 2 terms into account, one can add the third and final diagrammatic rule to our current set: \emph{the intersection rule}.
\begin{enumerate}
	\setcounter{enumi}{2}
	\item  \emph{Intersection rule: the intersection matrix} $\langle(a),(b) \rangle = \pm 2$ \emph{if and only if the edges} $(a)$ \emph{and} $(b)$ \emph{are intersecting.}
\end{enumerate}
\begin{figure}
	\center
	\includegraphics[width=\textwidth]{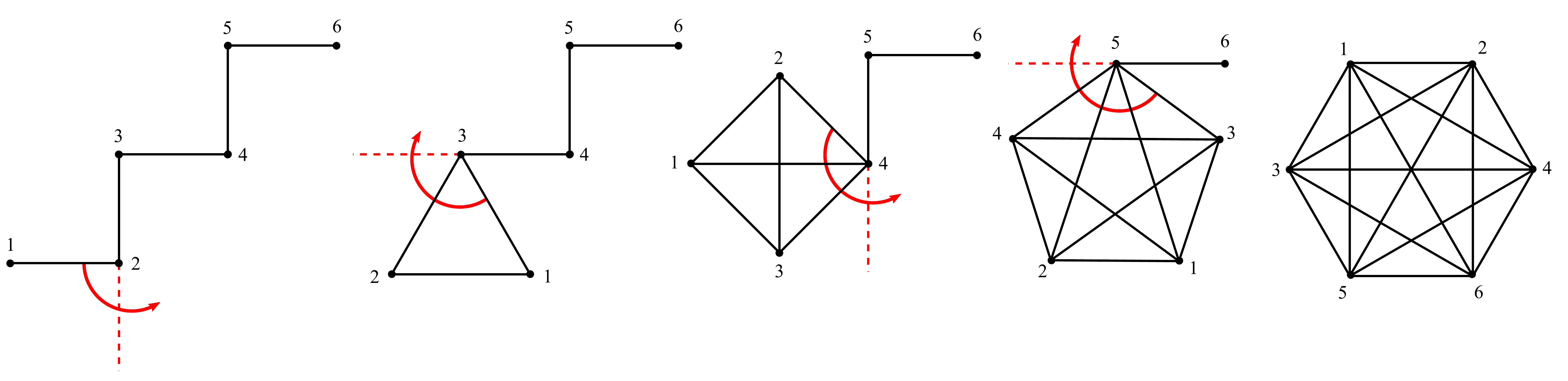}
	\caption{Analytical continuation of the TBA system corresponding to a sextic ($d=6$) potential, from the minimal chamber (far left) to the maximal chamber (far right). There are multiple possible analytical continuations that lead to the maximal chamber; in this specific example, we chose to analytically continue around the ``pivot vertex'' $2$, i.e. when $\varphi_{(1,2),(2,3)}=0$, then the ``pivot vertex'' $3$ etc. Because of the wall-crossing rule, we are obtaining one additional equation at step one, two at step two etc. After four of these steps, we are finally in the maximal chamber. The TBA system is then consisting of 15 TBA equations, corresponding to the maximal graph with 6 vertices (graph on the far right).}
	\label{fig:FromMinToMaxSextic}
\end{figure}

Let us review the diagrammatic process explained and justified above in a compact paragraph. The rule two -- the wall-crossing rule -- is a ``construction rule''. One could start with any TBA graph and apply it as much as needed in order to end up with the required morphed TBA graph. Once this process is finished and the final TBA graph is obtained, one can read it using the ``reading rules'' one and three -- the connection and intersection rules -- encoding the intersection matrix thus encoding the full TBA system.

As an example, let's apply this procedure on some polynomials potential: from the cubic up to the octic potential and from the minimal chamber all the way to the maximal chamber. The first step is to find the associated TBA graph. A specific example is provided for the sextic polynomial in figure \ref{fig:FromMinToMaxSextic}, and we end up with the maximal graph with 6 vertices. This is always the case for any polynomial of degree $d$, with one exception if we are talking about stricto sensu graphs: the quartic. In proper graph theory, the maximal graph formed by 4 vertex is planar, hence having no intersecting edges. This is not possible with our TBA graphs defined above because of the alignment property. Otherwise, the TBA graph of a degree $d$ polynomial in the maximal chamber is equivalent to a complete graph with $d$ vertices.

\begin{figure}
	\center
	\includegraphics[width=\textwidth]{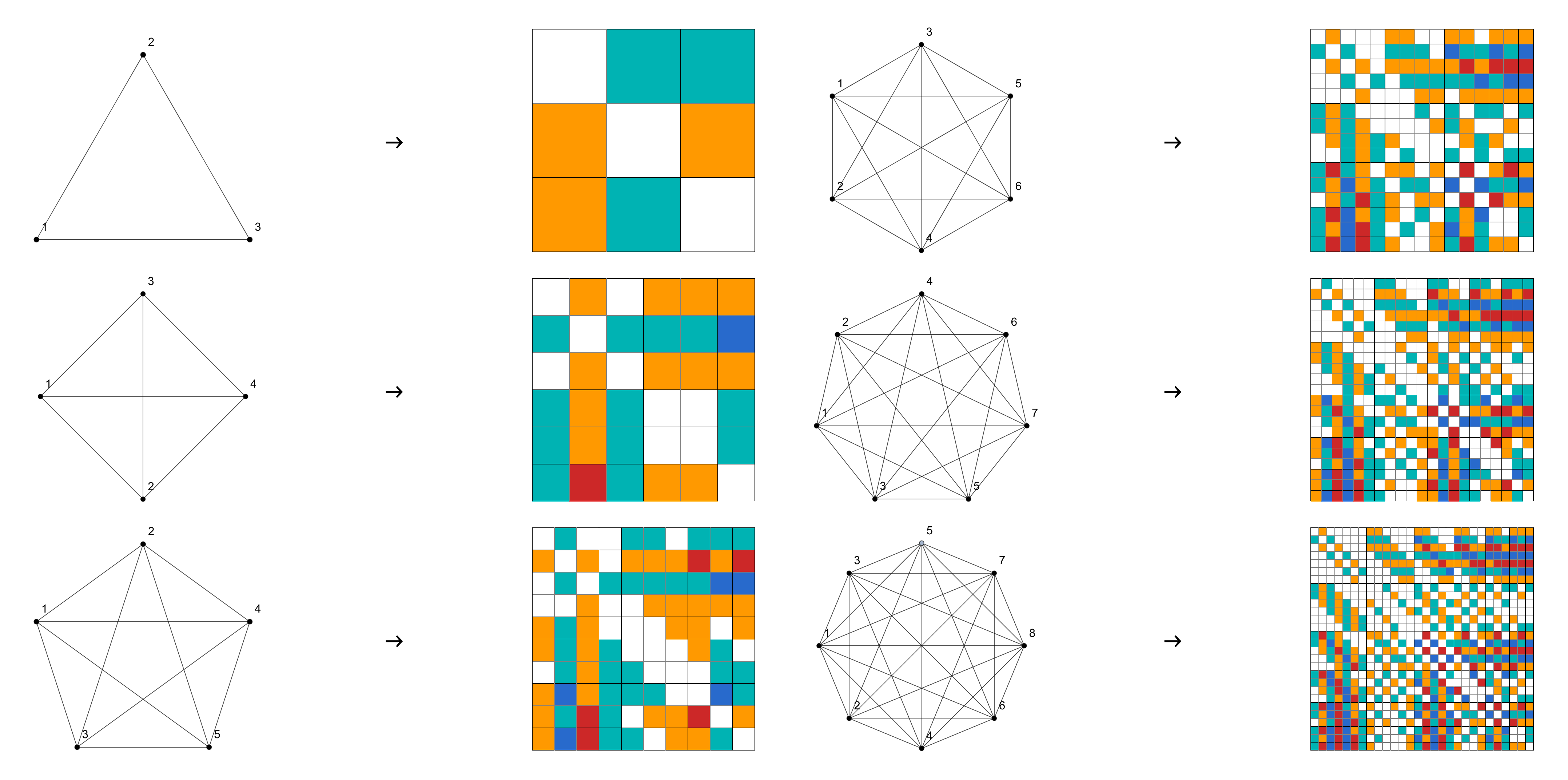}
	\caption{The TBA graph and intersection matrices $\langle (a),(b) \rangle$ for the cubic up to the octic ($d\in\{3,\ldots,8\}$) in the maximal chamber. The intersection matrix is read from the TBA graph according to the diagrammatic rules of section \ref{chap:wallCrossingDiagram}. Orange, red, teal, blue and white are color coding respectively $+1$, $+2$, $-1$, $-2$ and $0$. The intersection matrices are organized such that the $d(d-1)/2$ entries $(a)$ are, from top to bottom (resp. $(b)$, from left to right) $(i,i+k)$, $i\in \{1,\ldots,d-k\}$ and $k\in \{1,\ldots,d-1\}$; we denote an increment of $k$ by a black line. For example, the 6 entries for the quartic reads in this order: $(1,2),(2,3),(3,4),(1,3),(2,4),(1,4)$, and the element $\langle (2,3),(1,4) \rangle = -2$.}
	\label{fig:TBAmaxC}
\end{figure}

Now that we have obtained the TBA graph, we can read the TBA system encoded in it by applying the connection and intersection rule. Remember that the intersection matrix is antisymmetric, therefore one only has to find a little bit less than half of the matrix elements\footnote{Since the matrix is a $N\times N$ antisymmetric matrix with null diagonal and $N=d(d-1)/2$, one need to identify exactly $d(d-1)(d^2-d-2)/8$ independent matrix elements.}. First, let us use the connection rule to set all the matrix elements that are equal to $\pm$1. Read row after row (or column after column) -- starting from the colomn (or row) $k+1$ if you are considering the row (or column) $k$, i.e. one square after the diagonal in case you want to save time by computing only the independents elements -- if there is only one shared vertex in the associated column (resp. row) entry, put a 1 \footnote{If both are the same, this is obviously a diagonal element, which is 0.}. In the sextic case, if we are considering the row $(1,2)$ for example, look for every edge that is containing either $1$ or $2$ (but not both) in the corresponding column, i.e. $(2,3)$, $(1,3)$, $(2,4)$, $(1,4)$, $(2,5)$, $(1,5)$, $(2,6)$ and $(1,6)$. This specific example is contained in figure \ref{fig:TBAmaxC} (first line/row of the sextic potential). At this point you are already done with $d$ TBA equation, since the concave hull of the polygon formed by the $d$ vertices provides necessarily non-intersecting edges\footnote{The is why the first non trivial intersecting case is the quartic in the maximal chamber: the cubic has only $d=3=d(d-1)/2$ TBA equations.}. Furthermore, every row (column) should have exactly $2(d-2)$ entries set to one at this point: since every vertex is shared between $d-1$ edges in the maximal chamber, each edge formed by a given couple of vertices will be connected to $d-2$ different edges through each of its two vertices. With the explication above, one can see that the matrix elements for the signless connection matrix (the $\pm 1$ part of the full intersection matrix without taking into account the sign) is then simply given by:
\begin{equation}\label{eq:connectionMatrix}
\abs{\langle (i,i+k),(j,j+l) \rangle^\text{connection}} =\delta _{i,j}+\delta _{i,j+l}+\delta _{i+k,j}+\delta _{i+k,j+l}-2 \delta _{i,j} \delta _{i+k,j+l}
\end{equation}
with $i \in \{1,\ldots,d-k\} $, $j \in \{1,\ldots,d-l\} $, and $k,l \in \{1,\ldots,d-1\} $. An example of this structure for $d\in\{18,19\}$ can be observed in figure \ref{fig:fullintMat} by looking at the teal and orange entries as $1$, $0$ otherwise.

In order to take care of the $\pm 2$ elements, chose a row (or column) associated to an edge inside the convex polynomial hull but not on the boundary (hence with intersection), then check all the columns (or rows)  -- after the diagonal and not already set to one. Put a two if the column edge and the row edge are intersecting in the TBA graph. Let's consider the sextic example once more: if one start with the row corresponding to the edge $(2,5)$, one have to put a $2$ in the column corresponding to the edges $(3,4)$, $(1,5)$, $(2,6)$ and $(1,6)$. Finding the exact number of intersection for any edges, i.e. the number of elements set to $\pm 2$ in a given row/column, is a little bit more involved than finding the number of matrix elements set to $\pm 1$. Let us open a parenthesis about it in the next paragraph. 

First, note that because of the $\mathbb{Z}_d$ symmetry, this problem is not affected by which particular vertex we chose as the starting point for an edge, such that there are only $d-1$ non-equivalent edges in the resulting equivalence class. We will denote them as $[k] = (v_0,v_{0}+k)$ with $k \in \{1,\ldots,d-1\}$ and where $v_0$ is inconsequential. For our labeling purpose, the non-equivalent edge $[k]$ is obtained by connecting the starting vertex with the vertex we reach after $k$ ``jumps'' along the convex polygonal hull formed by all the vertices (or equivalently, this jump can be seen as a rotation of $2\pi i/d$ with respect to the center of the polygon). Furthermore, because of the reflection symmetry, we will ultimately only need ``half'' of them: $\ceil{\frac{d-1}{2}}$ to be precise (we could also have started directly with the $D_d$ dihedral symmetry). This new equivalent class is obtained by identifying a ``jump'' in the clockwise and anticlockwise direction, and we will still denote the elements of this class by $[k]$, but this time with $k \in \{1,\ldots,\ceil{\frac{d-1}{2}}\}$. Each of them have a unique number of intersections. A simple recursive observation allows us to fully count them: if we call $I_{k,d}$ the number of edges intersecting with $[k]$ for a maximally connected polygonal containing $d$ vertex, then $I_{k,d+1}=I_{k,d}+k-1$ and $I_{k+1,d}=I_{k,d}+1$ the first time a new $k$ is appearing\footnote{It also happens to be a perfect square: $I_{k+1,d}=k^2$ the first time $k+1$ is appearing.} (i.e. each time $\ceil{\frac{d-1}{2}}$ increases), starting with $I_{2,4}=1$. Of course, $I_{1,d}=0$ since $[1]$ is always on the convex polygonal hull. These recursion relations imply that $I_{k,d} = (k-1)(d-k-1)$. As an example, this formula predicts that for the TBA system corresponding to an octic potential in the maximal chamber we have either 0, 5, 8 or 9 terms with coefficient $\pm 2$ in a given TBA equation (corresponding to $[1]$,$[2]$,$[3]$ and $[4]$ resp.). This is what is observed in figure \ref{fig:TBAmaxC}.\\

We know how to count them, but we still want to determine the exact matrix elements of this signless pure intersection matrix (\emph{i.e.} without the connection part described in \refeq{eq:connectionMatrix}). In order to achieve that, let us fix a labeling for concreteness. Let's say for simplicity that all of our vertices are the $d$th roots of the unity. Instead of labeling the vertices by complex numbers in the complex plane, we label them by the $n$ in $e^{2\pi n/d}$. One can then define $[k]^{v} = (v,v+k)$ in order to label the class of $d-1$ edges containing $v_0$ among its vertices. Because edges are non ordered pairs, \emph{i.e.} $(v,v+k) = (v+k,v)$, we have the identifications $[k]^{v} = [d-k]^{v+k}$. Another useful property is that the pure signless intersection matrix  $\abs{\langle [n]^{v},[m]^{v+k} \rangle^\text{intersection}}$ is transposed under the reflection $k\mapsto d-k$, \emph{i.e.} $\abs{\langle [n]^{v},[m]^{v+d-k} \rangle^\text{intersection}}=\abs{\langle [m]^{v},[v]^{v+k} \rangle^\text{intersection}}$ (we used that fact during the counting of the crossings in the previous paragraph). Now, let's fix the matrix elements. The edges of the form $[k]^{v}$ will never intersect among each other $\forall k \in \{1,\ldots,d-1\}$ since they share the same starting point but are ending on a different endpoint. Furthermore, any edge on the convex polygonal hull, \emph{i.e.} such that it can be written as $[1]^{v}$ or $[d-1]^{v}$ will never intersect with any edge. As a result, the pure signless intersection matrix will have a factor $\abs{\langle [n]^{v},[m]^{v+k} \rangle^\text{intersection}} \propto~ (1-\delta_{n,1}-\delta_{n,d-1})(1-\delta_{m,1}-\delta_{m,d-1})$. Additionally, because of the non ordered properties, two equivalent edges (of the form $[k]^{v} = [d-k]^{v+k}$) will never intersect since they are the same edge. The pure signless intersection matrix hence takes a factor $\abs{\langle [n]^{v},[m]^{v+k} \rangle^\text{intersection}} \propto (1-\delta_{n,k})(1-\delta_{m,d-k})$. The procedure is splitting the $(d-1)\times(d-1)$ matrix $\abs{\langle [n]^{v},[m]^{v+k} \rangle^\text{intersection}}$ (at $k$ fixed and arbitrary $v$) in four sub-matrices, delimited by the cross of zeroes introduced by the factor $(1-\delta_{n,k})(1-\delta_{m,d-k})$:
\begin{equation}
\abs{\langle [n]^{v},[m]^{v+k} \rangle^\text{intersection}}=
\begin{pmatrix}
0 & 0 & 0 & 0 & 0\\
0 & R^{(1)}_{(d-k-2)\times(k-2)} & 0 & L_{(k-2)\times(k-2)} & 0\\
0 & 0 & 0 & 0 & 0\\
0 & U_{(d-k-2)\times(d-k-2)} & 0 & R^{(2)}_{(k-2)\times (d-k-2)} & 0\\
0 & 0 & 0 & 0 & 0
\end{pmatrix}
\end{equation}
where the subscript is indicating the size of the sub-matrix (of course, if one of the component is of size $\leq 0$, the ``cross'' is not really splitting our matrix into 4 sub-matrices but only one square sub-matrix of size $d-1$ if $k \in\{1,d-1\}$ or $d-2$ if $k \in\{2,d-2\}$). By studying this geometric problem (using recursions for example), one can convince himself that the rectangular matrices $R^{(i)}=0$, when $U$ is the upper triangular matrix and $L$ the lower triangular matrix. Putting all these results together, we get the matrix elements
\begin{align}
\begin{split}
\abs{\langle [n]^{v},[m]^{v+k} \rangle^\text{intersection}}=& (1-\delta_{n,1}-\delta_{n,d-1})(1-\delta_{m,1}-\delta_{m,d-1})(1-\delta_{n,k})(1-\delta_{m,d-k})\\
&\left(\sum_{\substack{l_x \leq l_y\\l_y=0}}^{d-k-3} \delta_{n,k+1+l_y} \delta_{m,2+l_x} + \sum_{\substack{l_y \leq l_x\\l_x=0}}^{k-3} \delta_{n,2+l_y} \delta_{m,d-k+1+l_x}\right)
\end{split}
\end{align}
or, going back to the pair of vertices notation, themselves denoted by the $d$th root of the unity :
\begin{align}\label{eq:intersectionMatrix}
\begin{split}
\abs{\langle (v,v+n),(w,w+m) \rangle^\text{int.}}=&(1-\delta_{v,w})(1-\delta_{n,1}-\delta_{n,d-1})(1-\delta_{m,1}-\delta_{m,d-1})\\
&(1-\delta_{n,((w-v) \bmod d)})(1-\delta_{m,d-((w-v) \bmod d)})\\
&\Bigg(\sum_{\substack{l_x \leq l_y\\l_y=0}}^{d-((w-v) \bmod d)-3} \delta_{n,((w-v) \bmod d)+1+l_y} \delta_{m,2+l_x}\\
& + \sum_{\substack{l_y \leq l_x\\l_x=0}}^{((w-v) \bmod d)-3} \delta_{n,2+l_y} \delta_{m,d-((w-v) \bmod d)+1+l_x}\Bigg)
\end{split}
\end{align}
For the following, it will be useful to relabel our vertices, still considering them as the $d$th roots of the unity, but instead sorting them according to their real part then imaginary part. Doing so, $0 \leftrightarrow d$, $d-1 \leftrightarrow d-1$, $1 \leftrightarrow d-2$ etc. and we get the following dictionary, given by the bijection $v\leftrightarrow i$
\begin{align}\label{eq:bij}
\begin{split}
\{0,\ldots d-1\} \ni~~ v &= \left(\frac{1}{4} \left(2 d+1+(-1)^{d+i}(2 i-1)\right)\right) \bmod d\\
\{1,\ldots d\} \ni~~ i &=  \begin{cases}
d & v=0 \\
d-2 v+1 & 2 v\leq d \\
2 v-d & 2 v>d
\end{cases}
\end{split}
\end{align}
An example of this structure for $d\in\{18,19\}$ and the labeling defined above can be observed in figure \ref{fig:fullintMat}.\\

The remaining elements (i.e. not set to $\pm 1$ or $\pm 2$ after application of the connection and intersection rules) of the triangular part of the intersection matrix $\langle(a),(b)\rangle$ are set to $0$. In order to determine the signs, you can multiply every element in this resulting triangular matrix with $\text{sign}(\varphi_{(a),(b)})$. Once this is done, copy in $\langle (b),(a) \rangle$ the opposite of the result in $\langle (a),(b) \rangle$ and the full intersection matrix $\langle (a),(b) \rangle$ is finally 
\begin{equation}\label{eq:fullintersectionMatrix}
	\langle(a),(b)\rangle = \text{sign}\left(\varphi_{(a),(b)}\right)\left(\abs{\langle(a),(b)\rangle^\text{connection}}+2 \abs{\langle(a),(b)\rangle^\text{intersection}}\right)
\end{equation}
as stated earlier by the connection and intersection rules. The associated TBA system is given by \refeq{eq:ultimateTBA}. This process is carried on for polynomials of degree $d\in \{3,\ldots,8\}$ in the maximal chamber and the resulting intersection matrices can be found in figure \ref{fig:TBAmaxC}. In order to have a broader view of the general structure for arbitrary $d$, one can also look at the matrices depicted in figures \ref{fig:signMat} and \ref{fig:fullintMat} for larger $d$ ($d\in\{18,19\}$ in that case).

\subsection{Simplifying a TBA system using symmetries}
\label{chap:simplifyTBA}
\begin{figure}
	\center
	\includegraphics[width=0.8\textwidth]{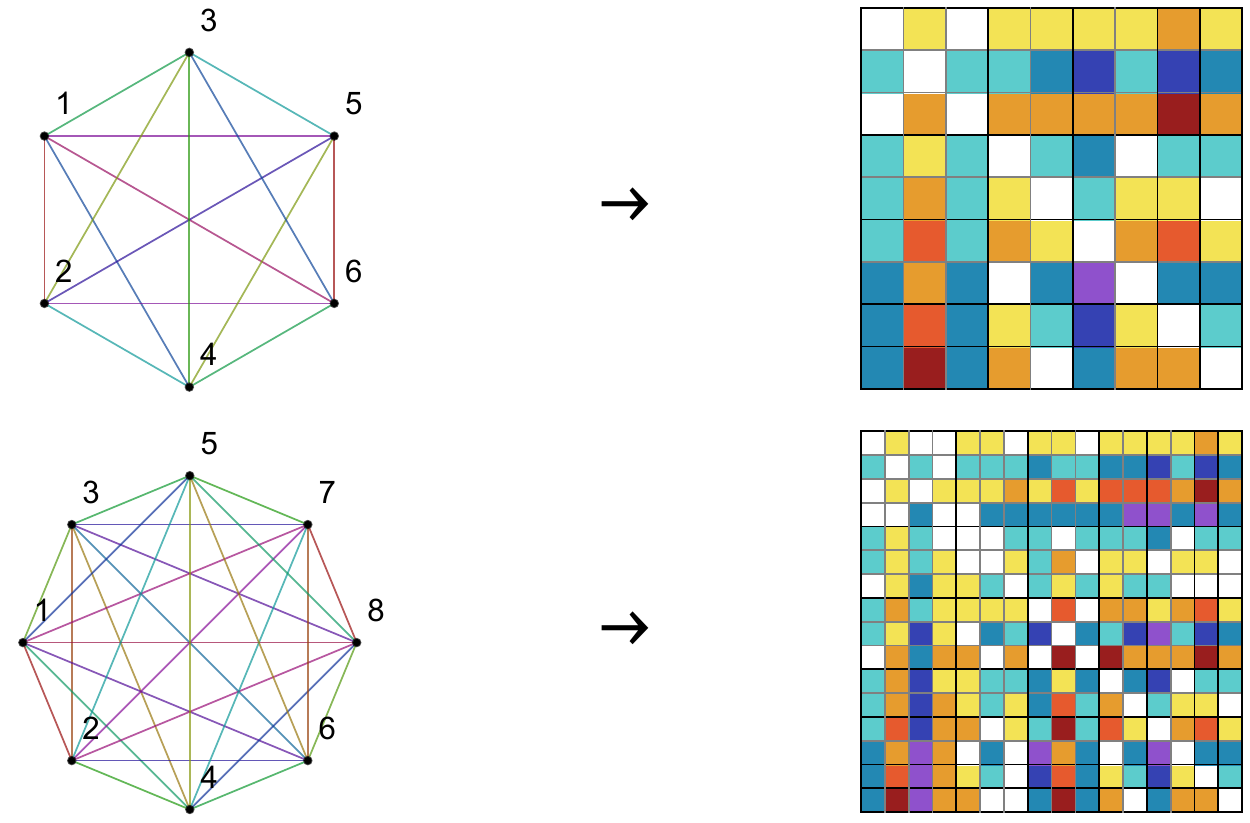}
	\caption{The color codded TBA graph and reduced intersection matrices $\langle (a),(b) \rangle$ for the symmetric sextic and octic potentials in the maximal chamber. We paired the equivalent TBA equations through \refeq{eq:symRedundancies}. This identification is denoted in the TBA graph by edges of the same color, resulting in $\frac{1}{2} \left(\left\lceil \frac{3}{2}-\frac{d}{2}\right\rceil -3\right) \left\lceil
		\frac{3}{2}-\frac{d}{2}\right\rceil +\frac{1}{2} \left\lfloor \frac{d}{2}\right\rfloor 
		\left(\left\lfloor \frac{d}{2}\right\rfloor +1\right)+1$ independent TBA equations in the maximal chamber case (parallel lines). Concerning the intersection matrix: yellow, orange, orange-red, dark red, light teal, cerulean, dark blue, purple and white are color coding respectively $+1$, $+2$, $+3$, $+4$, $-1$, $-2$, $-3$, $-4$ and $0$. The intersection matrices are organized in the same way as figure \ref{fig:TBAmaxC}, with the exception of the redundant $\epsilon$-functions that are deleted.}
	\label{fig:TBAmaxCsym}
\end{figure}

In the precedent sections, we assumed our polynomials of interest were completely general. Let us now assume they have some sort of symmetry. This symmetry will relates some $\epsilon$-functions together and reduces the TBA system to a simpler one, with less TBA equations. The tilde on the  $\epsilon$ or $L$-functions is omitted in the following.\\
Let us consider an important class of polynomials in order to illustrate this fact: the symmetric polynomials (i.e. with the $q\mapsto-q$ symmetry). In that case, because the turning points/classical periods/masses on one side are the same than on the other side, we can pair the $\epsilon$-functions using
\be
\label{eq:symRedundancies}
\epsilon_{i,i+k}(\theta) = \epsilon_{d-i-k+1,d-i+k}(\theta)
\ee
By substituting the redundant $\epsilon$-functions into the TBA system, one gets a simpler and reduced TBA system. Alternatively, one can simplify the TBA system by considering the reduced intersection matrix obtained from the full intersection matrix by deleting every row and column corresponding to the same equivalent TBA equation except from one, adding the value of of each deleted matrix element to the corresponding still existing matrix element in the reduced matrix. This intersection matrix can also be read from the TBA graph by identifying the equivalent TBA equations together. We provide two examples of such computation for the symmetric sextic and symmetric octic in the figure \ref{fig:TBAmaxCsym}.

Let us explicitly write the resulting reduced TBA system for the quartic potential. Starting from \refeq{eq:quarticMaxChamber}, we identify $\epsilon_{1,2}=\epsilon_{3,4}$ and $\epsilon_{1,3}=\epsilon_{2,4}$ since the corresponding periods/masses are also equal. We can simply substitute these identifications into \refeq{eq:quarticMaxChamber} and obtain
\begin{align}
\begin{split}
\label{eq:SymQuarticMaxChamber}
\epsilon_{1,2}(\theta) =& \abs{\Pi_{1,2}} \exp(\theta)+K_{(1,2),(2,3)}\star L_{2,3}+2 K_{(1,2),(1,3)}\star L_{1,3}+K_{(1,2),(1,4)}\star L_{1,4}\\
\epsilon_{2,3}(\theta) =& \abs{\Pi_{2,3}} \exp(\theta)-2 K_{(2,3),(1,2)}\star L_{1,2}-2 K_{(2,3),(1,3)}\star L_{1,3}-2K_{(2,3),(1,4)}\star L_{1,4}\\
\epsilon_{1,3}(\theta) =& \abs{\Pi_{1,3}} \exp(\theta)-2 K_{(1,3),(1,2)}\star L_{1,2}+K_{(1,3),(2,3)}\star L_{2,3}-K_{(1,3),(1,4)}\star L_{1,4}\\
\epsilon_{1,4}(\theta) =& \abs{\Pi_{1,4}} \exp(\theta)-2K_{(1,4),(1,2)}\star L_{1,2}+2K_{(1,4),(2,3)}\star L_{2,3}+2K_{(1,4),(1,3)}\star L_{1,3}
\end{split}
\end{align}
where we factored out the intersection matrix $\langle(a),(b) \rangle$ coefficients as in \refeq{eq:quarticMaxChamber}, i.e. the kernel is \refeq{eq:kernelSinh} without the intersection matrix or
\begin{equation}
\label{eq:kernelSinhwoIM}
K_{(a),(b)}(\theta) = \frac{1}{2\pi i}\frac{1}{\sinh\left(\theta+i\varphi_{(a),(b)}\right)}
\end{equation}
for clarity. Notice that in this symmetric reduction procedure, the periods/masses are exactly equal, such that the kernels are also equal, which allows us to keep that compact intersection matrix formulation of the TBA system. This is not always the case for other symmetries as we shall see.

Similar statements (with appropriate modifications on the rules linking the pseudo-energies together) can be made for other symmetries. Important examples are antisymmetry, potentials leading to a PT-symmetric Hamiltonian or -- the most restrictive one and the subject of the rest of the next section -- the $D_d$ dihedral symmetry of pure potentials.

\subsection{TBA equations and pure potentials}\label{chap:purePot}
\subsubsection{Geometric and preliminary observations}
\label{chap:Geometric and preliminary observations}
\begin{figure}
	\center
	\includegraphics[width=0.8\textwidth]{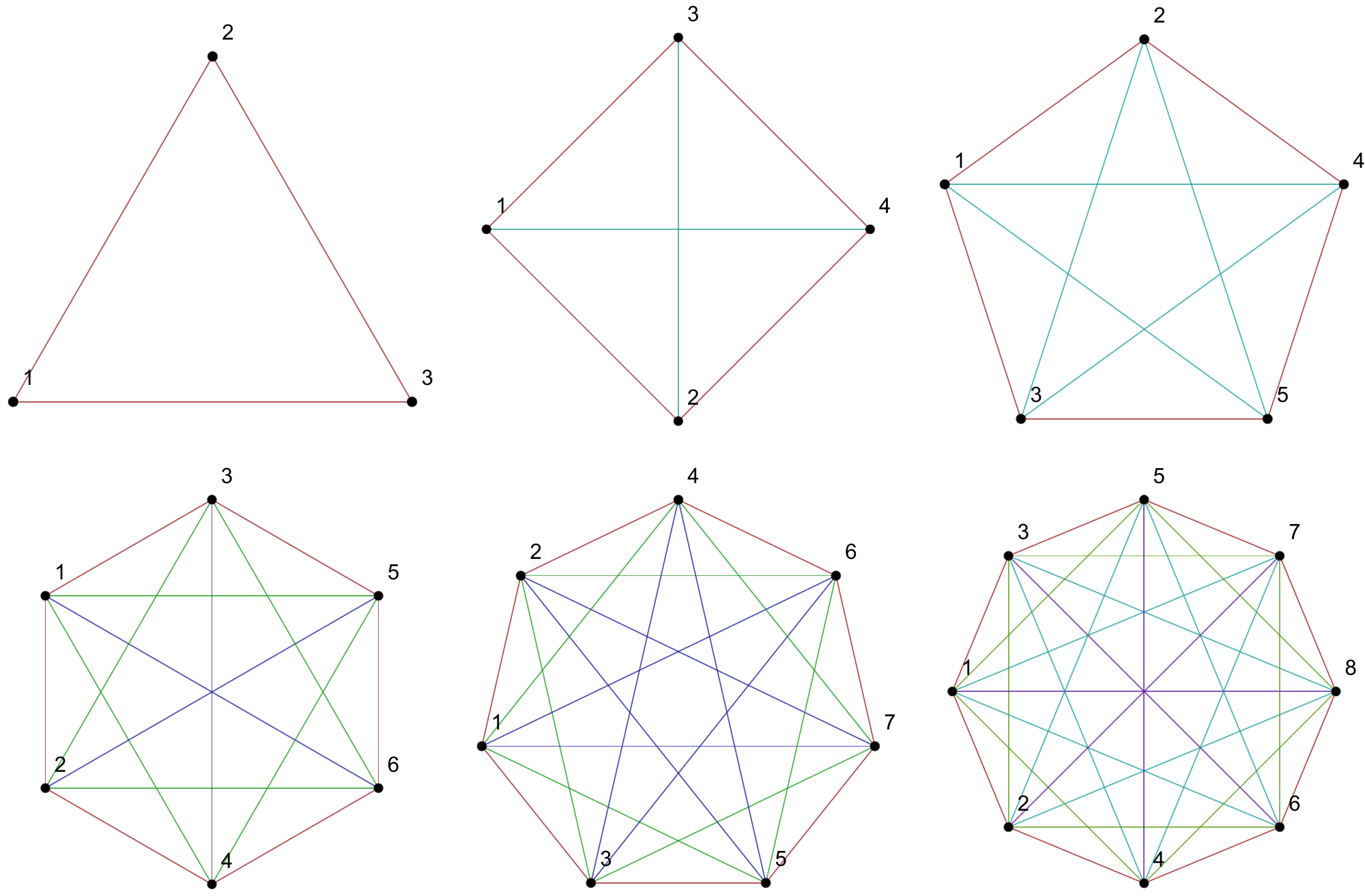}
	\caption{The color codded TBA graph (taking into account the dihedral symmetry) resulting in the reduced TBA system for a monic potential of degree $d$ for $d\in\{3,\ldots,8\}$. One can observe the $\ceil{\frac{d-1}{2}}$ different colors corresponding to the independent $\epsilon$-functions. If one does not take into account the symmetry under reflection and only consider the $\mathbb{Z}_d$ symmetry, one has $d-1$ colors instead, corresponding to the equations in \cite{Dorey:1998pt}.}
	\label{fig:TBAmaxCmonic}
\end{figure}

In order to link our result to Dorey and Tateo equations -- which is next section's goal -- we still have to investigate how the TBA system simplifies when the potential is pure, \emph{i.e.} in the form $V(q) = a q^d$. In that case, we are always in the maximal chamber (as long as $d>2$) since the turning points are forming a regular convex $d$-gon in the complex plane. If one draw all the possible cycles between the turning points of the $d$-gon, the resulting graph is a complete graph like the one in figure \ref{fig:TBAmaxC}
. We can apply the same arguments presented in section \ref{chap:wallCrossingDiagram} when we were computing the number of intersections of a given edge in a complete graph on the absolute value of the periods: because of the dihedral symmetry of the problem, we can group the $d(d-1)/2$ absolute values of the periods in $\ceil{\frac{d-1}{2}}$ non-equivalent elements, denoted by the equivalent classes of edge $[k]$ with $k \in \{1,\ldots,\ceil{\frac{d-1}{2}}\}$, resulting in the independent pseudo-energies denoted by $\epsilon_{[k]}$\footnote{The $\mathbb{Z}_d$ symmetry leaves $d-1$ independent TBA equations, equivalent to the equations in \cite{Dorey:1998pt} as we shall see. The reflection reduces this number to $\ceil{\frac{d-1}{2}}$.}. The $\epsilon_{[k]}$ can be read from the reduced TBA graph with $\ceil{\frac{d-1}{2}}$ color, like the ones in figure \ref{fig:TBAmaxCmonic}.
Unlike the symmetric case, the periods/masses are not exactly equal: their absolute values are, but their phases are not the same. For this reason, we cannot simply reduce the intersection matrix as described in the symmetric case since some kernels will not have the same complex shift. We can however group the sum of the kernels having same $L$-functions as a common factor into a new kernel. Doing so, the TBA system for a pure potential takes the form
\begin{equation}
\label{eq:TBAmonic}
\epsilon_{[k]}(\theta) = \abs{\Pi_{[k]}}\exp(\theta) +\sum_{l =1}^{\ceil{\frac{d-1}{2}}} \mathcal{K}_{[k],[l]}\star L_{[l]}(\theta)
\end{equation}
where $\mathcal{K}_{[k],[l]}$ is the sum of kernels with common factor $L_{[l]}$, i.e.
\begin{equation}
\label{eq:TBAmonicKernels}
\mathcal{K}_{[k],[l]}(\theta) = \sum_{b\in[l]} K_{(k),(b)}(\theta)
\end{equation}
Where the sum $\sum_{b\in[l]}$ is to be understood as a sum over every edge $(b)$ that is in the equivalence class $[l]$ (same color in the TBA graph). In other words, instead of a $d(d-1)/2 \times d(d-1)/2$ intersection matrix of coefficients, we can encode the reduced TBA system with a $\ceil{\frac{d-1}{2}}\times\ceil{\frac{d-1}{2}}$ matrix of kernels $\mathcal{K}_{[k],[l]}(\theta)$. Since the $\varphi_{[k],[l]}$ are always some multiple of $\pi/d$ in the pure case (see the next subsection for an exact statement), the resulting sum of kernels simplifies (in particular, it is always a real function).

Let us give a concrete example by computing the kernels for the pure quartic. Starting with the TBA system associated with the symmetric quartic \refeq{eq:SymQuarticMaxChamber}, we have two more identifications to do in order to end up with the TBA system for a pure quartic: $\epsilon_{1,2}=\epsilon_{1,3}=\epsilon_{[1]}$ (red TBA in figure \ref{fig:TBAmaxCmonic}) and $\epsilon_{2,3}=\epsilon_{1,4}=\epsilon_{[2]}$ (cyan TBA). The reduced TBA system is 
\begin{align}
\begin{split}
\label{eq:PureQuarticMaxChamber}
\epsilon_{1,2}(\theta) =& \abs{\Pi_{1,2}} \exp(\theta)+2 K_{(1,2),(1,3)}\star L_{1,2}+(K_{(1,2),(2,3)}+K_{(1,2),(1,4)})\star L_{2,3}\\
\epsilon_{2,3}(\theta) =& \abs{\Pi_{2,3}} \exp(\theta)-2 (K_{(2,3),(1,2)}+K_{(2,3),(1,3)})\star L_{1,2}-2 K_{(2,3),(1,4)}\star L_{2,3}
\end{split}
\end{align}
which simplifies in the system \refeq{eq:TBAmonic} with the following matrix of kernels
\begin{equation}
\label{eq:KmonicQuartic}
\mathcal{K}(\theta)=
\begin{pmatrix}
	\frac{1}{\pi} \frac{1}{\cosh(\theta)} & \frac{\sqrt{2}}{\pi} \frac{\cosh(\theta)}{\cosh(2\theta)} \\
	\frac{2\sqrt{2}}{\pi} \frac{\cosh(\theta)}{\cosh(2\theta)} & \frac{1}{\pi} \frac{1}{\cosh(\theta)}
\end{pmatrix}
\end{equation}
or, explicitly written:
\begin{align}
 \begin{split}
 \label{eq:PureQuarticMaxChamberExpl}
 \epsilon_{[1]}(\theta) =& \abs{\Pi_{[1]}} \exp(\theta)+\frac{1}{\pi} \int_{\mathbb{R}}  \frac{L_{[1]}(\bar{\theta})~d\bar{\theta}}{\cosh(\theta-\bar{\theta})}+\frac{\sqrt{2}}{\pi} \int_{\mathbb{R}}  \frac{\cosh(\theta-\bar{\theta})}{\cosh(2(\theta-\bar{\theta}))}L_{[2]}(\bar{\theta})~d\bar{\theta}\\
 \epsilon_{[2]}(\theta) =& \abs{\Pi_{[2]}} \exp(\theta)+\frac{2\sqrt{2}}{\pi} \int_{\mathbb{R}}  \frac{\cosh(\theta-\bar{\theta})}{\cosh(2(\theta-\bar{\theta}))}L_{[1]}(\bar{\theta})~d\bar{\theta}+\frac{1}{\pi} \int_{\mathbb{R}}  \frac{L_{[2]}(\bar{\theta})~d\bar{\theta}}{\cosh(\theta-\bar{\theta})}
 \end{split}
\end{align}
As an additional example, the matrix of kernels for the pure sextic (3 colors) yields
\begin{equation}
\label{eq:KmonicSextic}
\mathcal{K}(\theta)=
\begin{pmatrix}
\frac{\sqrt{3} \sinh(2\theta )}{\pi \sinh(3\theta )} & \frac{3 \cosh (2 \theta )}{\pi \cosh(3 \theta )} & \frac{\sqrt{3} \sinh(2\theta )}{\pi \sinh(3\theta )} \\
\frac{3 \cosh (2 \theta )}{\pi \cosh(3 \theta )}  & \frac{3\sqrt{3} \sinh(2\theta )}{\pi \sinh(3\theta )} & \frac{3 \cosh (2 \theta )}{\pi \cosh(3 \theta )}  \\
\frac{2 \sqrt{3} \sinh(2\theta )}{\pi \sinh(3\theta )} & \frac{6 \cosh (2 \theta )}{\pi \cosh(3 \theta )}  & \frac{2 \sqrt{3} \sinh(2\theta )}{\pi \sinh(3\theta )} \\
\end{pmatrix}
\end{equation}
where $\epsilon_{[1]}=\epsilon_{1,2}=\epsilon_{5,6}=\epsilon_{1,3}=\epsilon_{2,4}=\epsilon_{3,5}=\epsilon_{4,6}$ (red), $\epsilon_{[2]}=\epsilon_{2,3}=\epsilon_{4,5}=\epsilon_{1,4}=\epsilon_{3,6}=\epsilon_{1,5}=\epsilon_{2,6}$ (green) and $\epsilon_{[3]}=\epsilon_{3,4}=\epsilon_{2,5}=\epsilon_{1,6}$ (blue) as can be read from figure \ref{fig:TBAmaxCmonic}.

\subsubsection{Exact classical periods for pure potentials}\label{chap:ExactClassPerPurePot}
A pure potential $V(x)=a q^d$ with energy $E>0$ has a momentum given by \refeq{eq:WKBcurve}, i.e. $\sqrt{2(E-a q^d)}$. By the rescaling $q\mapsto \left(\frac{E}{a}\right)^{\frac{1}{d}} q$ one can rewrite the classical periods in a friendly manner,
\begin{equation}\label{eq:classPmonic1}
\Pi_\gamma =  \sqrt{2E} \left(\frac{E}{a}\right)^{\frac{1}{d}} \int_{\gamma} \sqrt{1-q^d} dq
\end{equation}
where $\gamma$ denotes a path encircling two turning points. With the previous scaling, the turning points are simply listed by the $d$th roots of unity, i.e. $q_{\text{tp}}^{k} = e^{2\pi i k/d}$, with $k\in\mathbb{Z}$, such that the previous integral \refeq{eq:classPmonic1} resolves to
\begin{equation}\label{eq:classPmonic2}
2 q_{\text{tp}}^{k_2} \, _2F_1\left(-\frac{1}{2},\frac{1}{d};1+\frac{1}{d};\left(q_{\text{tp}}^{k_2}\right)^d\right)-2 q_{\text{tp}}^{k_1} \, _2F_1\left(-\frac{1}{2},\frac{1}{d};1+\frac{1}{d};\left(q_{\text{tp}}^{k_1}\right)^d\right)
\end{equation}
We can factorize the only building block we need,
\begin{equation}\label{eq:classPmonicHypergeomToGamma}
2 ~ _2F_1\left(-\frac{1}{2},\frac{1}{d};1+\frac{1}{d};1\right)=\sqrt{\pi } ~\frac{\Gamma \left(1+\frac{1}{d}\right)}{\Gamma \left(\frac{3}{2}+\frac{1}{d}\right)}
\end{equation}
in order to express all of our classical periods as
\begin{equation}\label{eq:classPmonic3}
\Pi_{q_{\text{tp}}^{k_1},q_{\text{tp}}^{k_2}} = \sqrt{2\pi E} \left(\frac{E}{a}\right)^{\frac{1}{d}} \frac{\Gamma\left(1+\frac{1}{d}\right)}{\Gamma\left(\frac{3}{2}+\frac{1}{d}\right)}~ (q_{\text{tp}}^{k_2} -  q_{\text{tp}}^{k_1})
\end{equation}
For the following, let us fix the labeling of the turning points for concreteness: the indices $1\ldots d$ are listing the sorted turning points, ordered from the smallest to largest real part first, then imaginary part. We want to relate this labeling with the $d$th root of unity $q_{\text{tp}}^{k} = e^{2\pi i k/d}$. The one with largest real part, $q_d$, is obviously $q_{\text{tp}}^{0}=1$. Then we can identify $q_{d-2n}=q_{\text{tp}}^{n}$ and $q_{d-2n-1}=q_{\text{tp}}^{-n}$ until we run out of turning point after $d-1$ labeling if $d$ is odd. If $d$ is even, the smallest turning point is $q_1=q_{\text{tp}}^{d/2}=-1$. Grouping all these results together, we get the dictionary
\begin{equation}\label{eq:dicoTurningPoints}
q_n = \exp \left(\frac{i \pi}{2 d} \left(2 d+1+(-1)^{d+n}(2 n-1)\right)\right)
\end{equation}
relating $q_n$ and $q_{\text{tp}}^{k}$ through the bijection \refeq{eq:bij}. With this labeling, we can rewrite our exact classical periods as
\begin{equation}\label{eq:classPmonic4}
\Pi_{n,m} = \sqrt{2\pi E} \left(\frac{E}{a}\right)^{\frac{1}{d}} \frac{\Gamma\left(1+\frac{1}{d}\right)}{\Gamma\left(\frac{3}{2}+\frac{1}{d}\right)}~ (q_{m} -  q_{n})
\end{equation}
The last result is analytically explaining the geometric observation that there are $\ceil{\frac{d-1}{2}}$ differently colored $\epsilon$-functions in figure \ref{fig:TBAmaxCmonic}: since the absolute value of the periods is given by
\begin{align}\label{eq:absclassPmonic}
\abs{\Pi_{q_{\text{tp}}^{k_1},q_{\text{tp}}^{k_2}}} &= \sqrt{8\pi E} \left(\frac{E}{a}\right)^{\frac{1}{d}} \frac{\Gamma\left(1+\frac{1}{d}\right)}{\Gamma\left(\frac{3}{2}+\frac{1}{d}\right)}~  \abs{\sin\left(\frac{\pi  \left(k_1-k_2\right)}{d}\right)}\\
\abs{\Pi_{n,m}} &= \sqrt{8\pi E} \left(\frac{E}{a}\right)^{\frac{1}{d}} \frac{\Gamma\left(1+\frac{1}{d}\right)}{\Gamma\left(\frac{3}{2}+\frac{1}{d}\right)}~ \abs{\sin\left(\frac{\pi}{4d} \left((-1)^m(2m-1)-(-1)^n(2n-1)\right)\right)} \nonumber\\
\abs{\Pi_{n,n+k}} &= \sqrt{8\pi E} \left(\frac{E}{a}\right)^{\frac{1}{d}} \frac{\Gamma\left(1+\frac{1}{d}\right)}{\Gamma\left(\frac{3}{2}+\frac{1}{d}\right)}~ \abs{\sin\left(\frac{\pi}{4d} \left(2 k+2 n-1+(-1)^k(1-2 n)\right)\right)}\nonumber
\end{align}
and, in particular, realizing that we can identify the equivalence classes as $\epsilon_{[n]}=\epsilon_{n,n+1}$  $\forall n \in \{1\ldots \ceil{\frac{d-1}{2}}\}$ (or, alternatively, that $\epsilon_{[k]}=\epsilon_{n,n+2k}$ $\forall n \in \{1\ldots d\}$ and $\forall k \in \{1\ldots \ceil{\frac{d-1}{2}}\}$)  intervening in \refeq{eq:TBAmonic},
\begin{equation}\label{eq:absclassPmonicEqClasses}
\abs{\Pi_{[n]}} =\abs{\Pi_{n,n+1}} = \sqrt{8\pi E} \left(\frac{E}{a}\right)^{\frac{1}{d}} \frac{\Gamma\left(1+\frac{1}{d}\right)}{\Gamma\left(\frac{3}{2}+\frac{1}{d}\right)}~ \abs{\sin\left(\frac{n \pi}{d}\right)}
\end{equation} 
is indeed describing $\ceil{\frac{d-1}{2}}$ different periods. With \refeq{eq:classPmonic4}, we entirely described all the exact classical periods $\Pi_{(a)}$ intervening in the TBA graph for arbitrary pure potentials of degree $d$. With \refeq{eq:absclassPmonic} and \refeq{eq:absclassPmonicEqClasses}, we gave precision on the exact form of their absolute values. To complete the picture, let us work their explicit arguments, i.e. the angles $\varphi_{(a)}$:
\begin{align}\label{eq:argclassPmonic}
\varphi_{n,m} =&\frac{(-1)^d\pi}{4 d} \left(2(-1)^d+ \left(2d \left(2+(-1)^m\right)+(-1)^{n}(2 n-1) +(-1)^{m}(2 m-1)\right)\right)\\
\varphi_{n,n+k} =&\frac{(-1)^d\pi }{4 d} \left((-1)^{k+n} (2 d+2 k+2 n-1)+4 d+2 (-1)^d+(-1)^n (2 n-1)\right)
\end{align}

such that the exact periods can be written in the polar form as
\begin{align}\label{eq:Pmonic}
\begin{split}
\Pi_{n,n+k} =& \sqrt{8\pi E} \left(\frac{E}{a}\right)^{\frac{1}{d}} \frac{\Gamma\left(1+\frac{1}{d}\right)}{\Gamma\left(\frac{3}{2}+\frac{1}{d}\right)}~ \abs{\sin\left(\frac{\pi}{4d} \left(2 k+2 n-1+(-1)^k(1-2 n)\right)\right)} \\
&e^{\frac{i \pi }{4 d} (-1)^d \left((-1)^{k+n} (2 d+2 k+2 n-1)+4 d+2 (-1)^d+(-1)^n (2 n-1)\right)}
\end{split}
\end{align}
\subsubsection{Intersection matrix for general $d$}
\begin{figure}
	\center
	\includegraphics[width=0.495\textwidth]{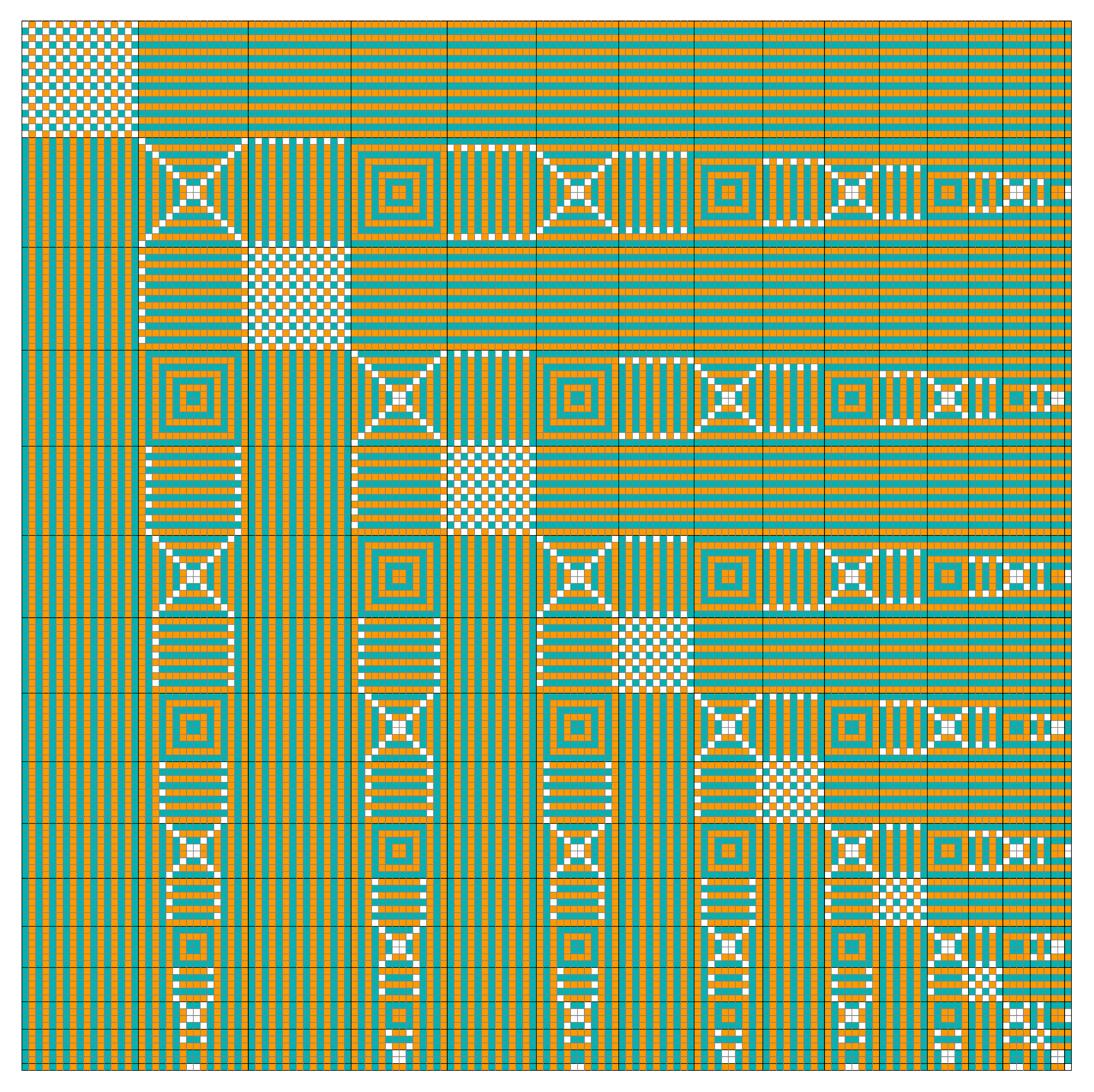}
	\includegraphics[width=0.495\textwidth]{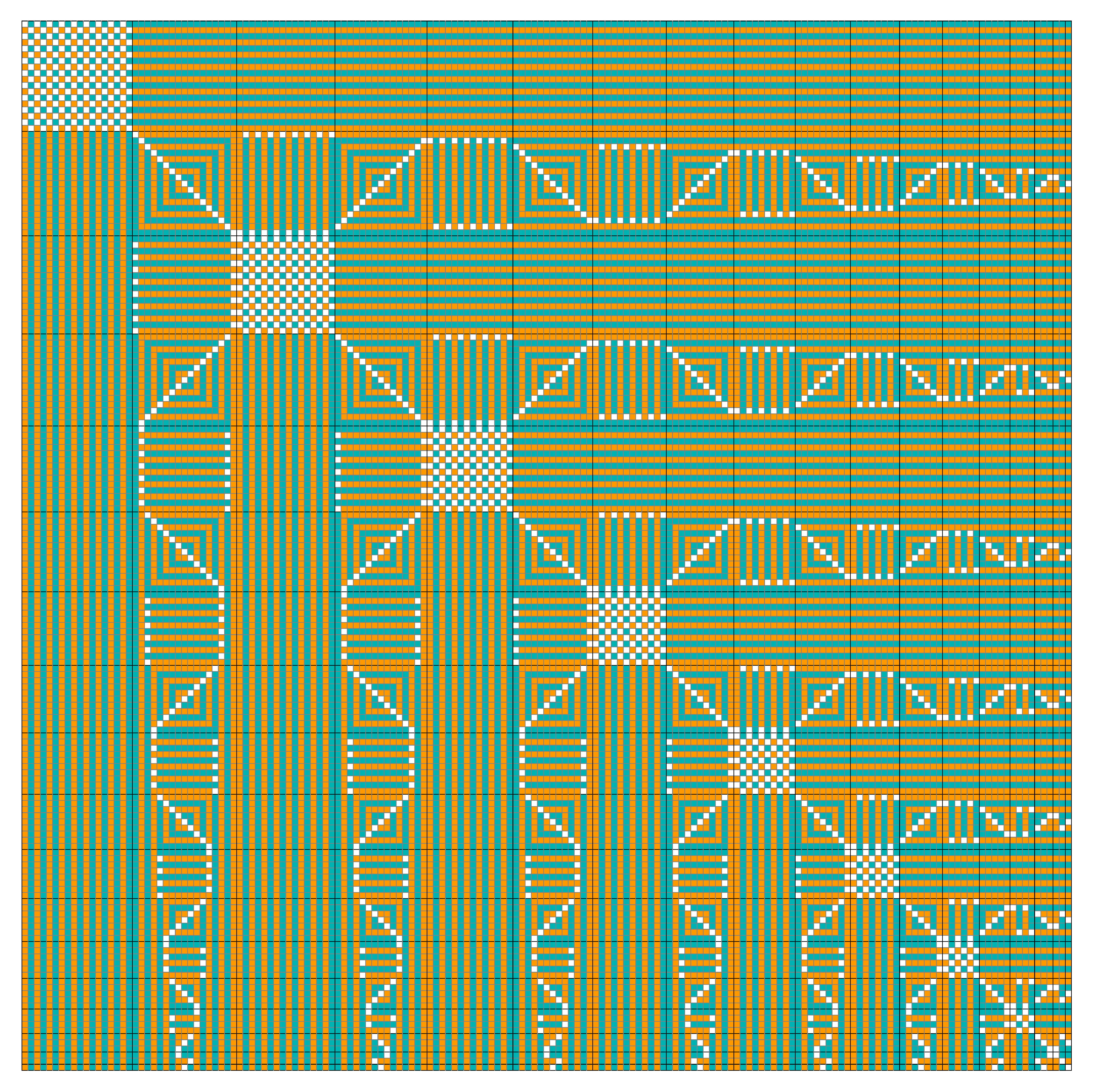}
	\caption{Structure of the sign matrix for polynomials of degree $18$ and $19$ respectively. Orange, teal and white are color coding  $+1$, $-1$ and $0$ respectively. The sign matrix is organized such that the $d(d-1)/2$ entries $(a)$ are, from top to bottom (resp. $(b)$, from left to right) $(i,i+k)$, $i\in \{1,\ldots,d-k\}$ and $k\in \{1,\ldots,d-1\}$; we denote an increment of $k$ by a black line.}
	\label{fig:signMat}
\end{figure}

\begin{figure}
	\center
	\includegraphics[width=0.495\textwidth]{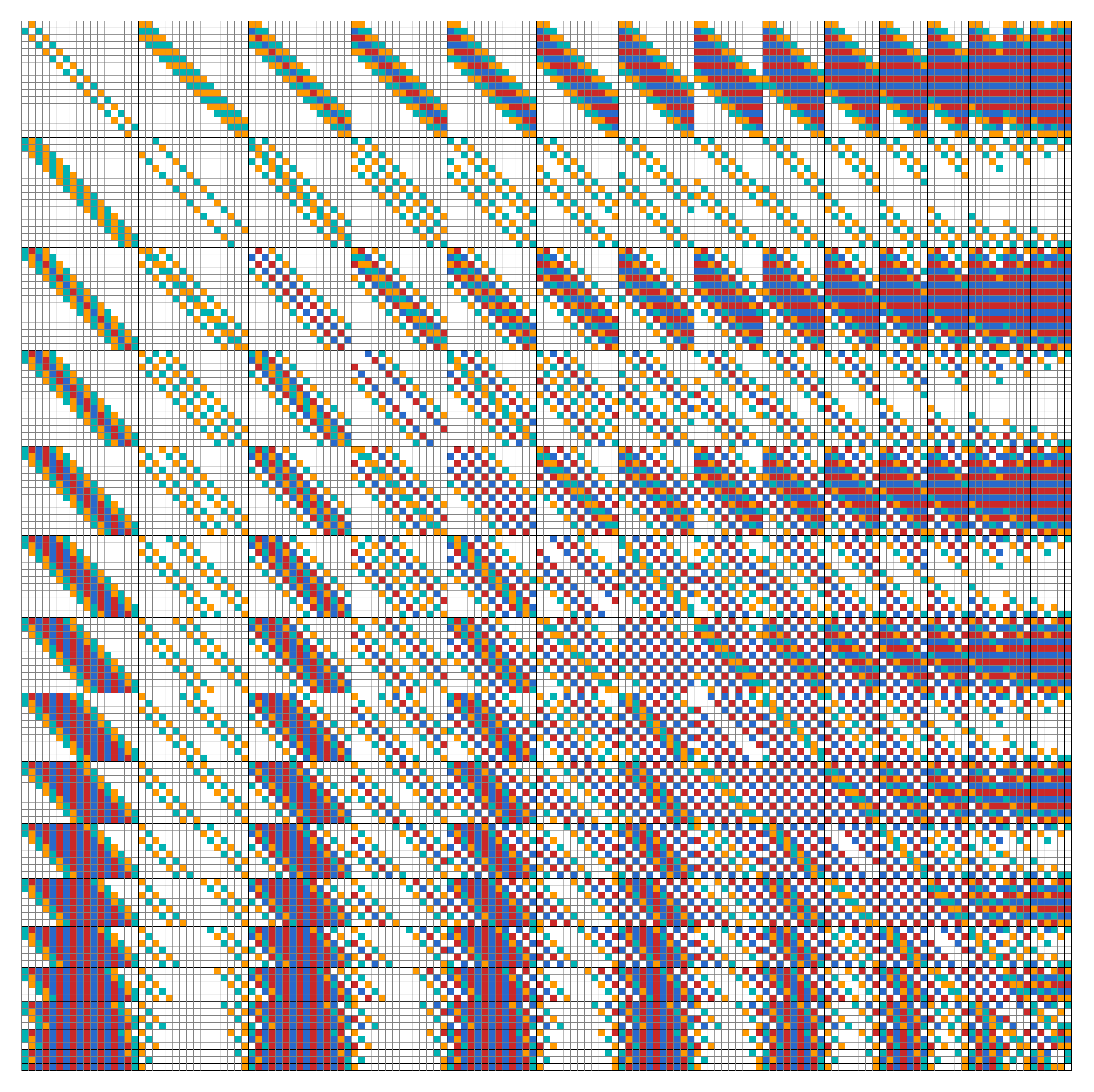}
	\includegraphics[width=0.495\textwidth]{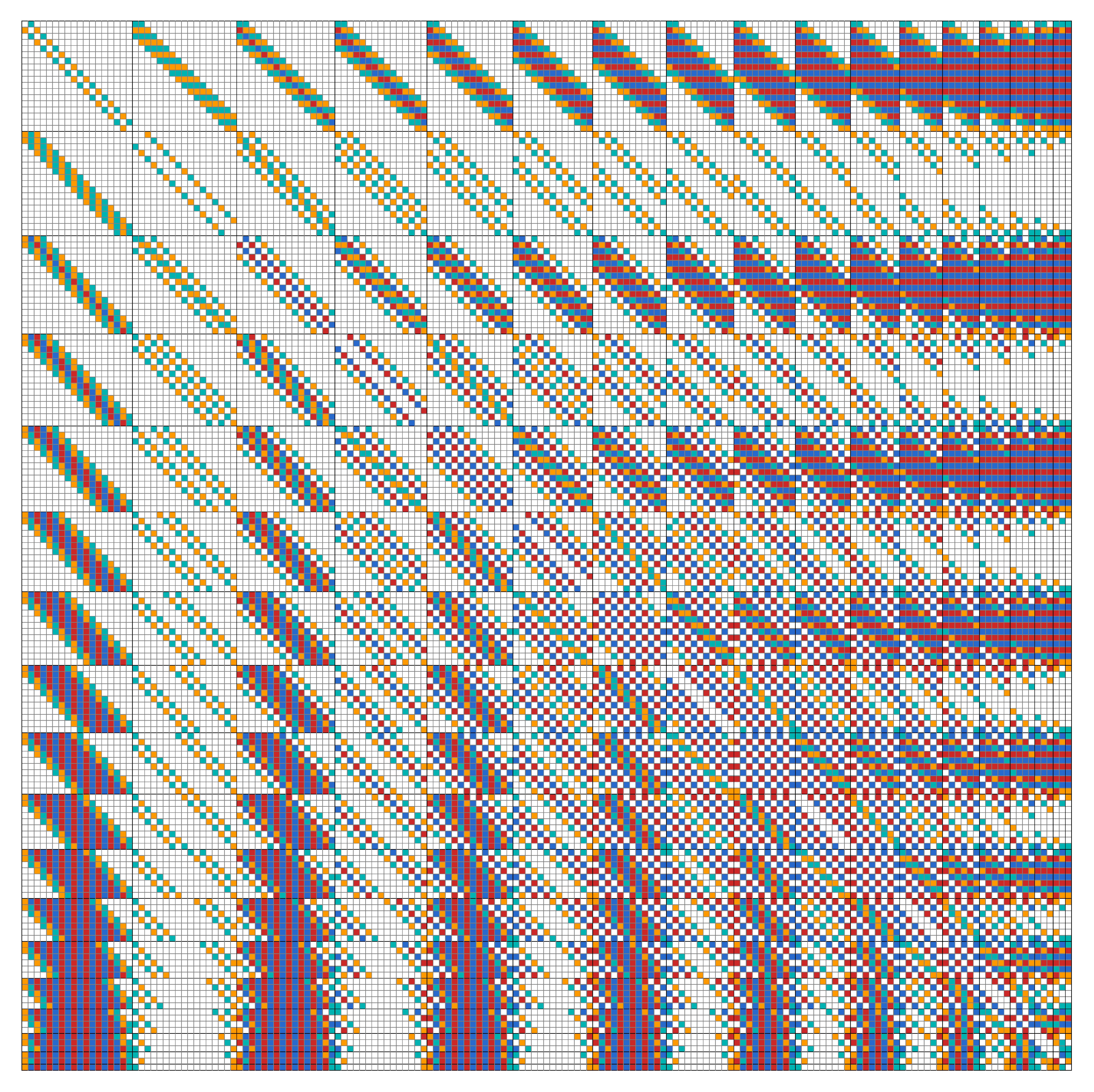}
	\caption{Structure of the (full, signed) intersection matrix \refeq{eq:fullintersectionMatrix} for polynomials of degree $18$ and $19$ respectively. Orange, red, teal, blue and white are color coding respectively $+1$, $+2$, $-1$, $-2$ and $0$. Orange and teal (resp. red and blue) entries are the non-null entries of the signless connection (resp. signless intersection) matrix \eqref{eq:connectionMatrix} (resp. \eqref{eq:intersectionMatrix}). We used the same organization conventions as in figure \ref{fig:signMat}.
	}
	\label{fig:fullintMat}
\end{figure}

Thanks to the computations above, we know the exact sign matrix $\text{sign}(\varphi_{(a),(b)})$ for arbitrary $d$. On can write it as
\begin{equation} \label{eq:signMat1}
\text{sign}(\varphi_{(i,i+k),(j,j+l)}) = \text{sign}\left(i \log \left(e^{-\frac{i \pi}{d} v_{ijkl}}\right)\right)
\end{equation}
in the real/imaginary labeling, where $v_{ijkl}=v_{ji}+(-1)^d (v_{j+l+d,i+k+d})$, $v_{ij} = v_i-v_j$ and $v_{i} = (2d + 1 +(-1)^{d+i}(2i-1))/4$, the map found in \refeq{eq:bij}, i.e. $v_i/d$ is simply the argument of the root $\varphi_i$ in the $d$th root of unity labeling for a monic potential. We left it in the form $i \log(\exp(-i x))$ since one has to pay attention to the periodicity of this function. An alternative non ambiguous form would be
\begin{equation} \label{eq:signMat2}
\text{sign}(\varphi_{(i,i+k),(j,j+l)}) = \text{sign}\left(
v_{ijkl}
-2 d \left\lfloor \frac{v_{ijkl}}{2 d}+\frac{1}{2}\right\rfloor
\right)
\end{equation}
For example of this sign matrix at large $d$ (so the repeating structure can be observed), one can look at figure \ref{fig:signMat}. In order to get the full exact intersection matrix \refeq{eq:fullintersectionMatrix} for polynomials in the maximal chamber, one just need to put together the the signless connection matrix \refeq{eq:connectionMatrix} and signless intersection matrix \refeq{eq:intersectionMatrix}.
For explicitness' sake, let us rewrite them in our labeling system:
\begin{align} \label{eq:explicitConnectionPart}
\abs{\langle (i,i+k),(j,j+l) \rangle^\text{connection}} =&\delta _{i,j}+\delta _{i,j+l}+\delta _{i+k,j}+\delta _{i+k,j+l}-2 \delta _{i,j} \delta _{i+k,j+l}\\
\label{eq:explicitIntersectionPart} \abs{\langle (i,i+k),(j,j+l) \rangle^\text{int.}}=&(1-\delta _{v_i,v_j}) (A_{ijkl}+A_{jilk})
\end{align}
where, after simplification,
\begin{align}
\begin{split}
A_{ijkl} =& \delta_{2,\mu_{i+k,i}} \delta_{1+\mu_{ij},\mu_{j+l,j}} \theta_{d>3+\mu_{ij}}+\theta_{\mu_{i+k,i}>1} \delta_{d,2+\mu _{j+l,j}} \delta_{d,1+\mu_{ij}+\mu_{i+k,i}}\\
&+\theta_{d>2+\mu _{j+l,j}} \theta_{\mu _{j+l,j}<\mu_{ij}+\mu _{i+k,i}} \theta_{d>\mu_{ij}+\mu_{i+k,i}} \theta_{d<\mu_{i+k,i}+\mu_{j+l,j}}\\
&+\theta_{\mu _{i+k,i}>2} \theta_{\mu _{ij}<\mu _{j+l,j}} \theta_{d>2+\mu_{j+l,j}} \theta_{\mu_{j+l,j}<\mu_{ij}+\mu_{i+k,i}} \theta_{d+1>\mu_{i+k,i}+\mu_{j+l,j}}
\end{split}
\end{align}
$\mu_{ij}= v_{ij} \bmod d$ and $\theta_b = 1$ if $b$ is true, $0$ otherwise.
As an important note, if we initially derived the formulas \refeq{eq:signMat1} or \refeq{eq:signMat2} for the specific case of pure potentials, let us remark than it also applies to a larger class of polynomials. As long as one is in presence of some set of roots of $E-V(q)$ ``not to far'' from the $d$th roots of the unity, in the sense that as long as identifying the two sets of roots is not modifying the sign of \refeq{eq:signMat1}, the intersection matrix is invariant under the deformation from the pure potential case to the more difficult case of interest. By using the appropriate classical periods and labeling system, on can solve TBA systems of the form \refeq{eq:ultimateTBA} with the same matrix of intersection we derived above in the context of pure potential. For example, $d/2$-wells problems with energy above the maximas. Using \refeq{eq:fullintersectionMatrix} with the sign matrix derived here, we are indeed reproducing the structures found in figure \ref{fig:TBAmaxC}. 

%

\subsubsection{Restricting the TBA equations to pure polynomial potentials}

Putting \refeq{eq:fullintersectionMatrix} and \refeq{eq:absclassPmonicEqClasses} into \refeq{eq:ultimateTBA}, one finds, in our real/imaginary labeling prescription,
\begin{align}
\label{eq:TBAforMonic}
\begin{split}
\tilde{\epsilon}_{(i,i+k)}(\theta) =& \sqrt{8\pi E} \left(\frac{E}{a}\right)^{\frac{1}{d}} \frac{\Gamma\left(1+\frac{1}{d}\right)}{\Gamma\left(\frac{3}{2}+\frac{1}{d}\right)}~ \abs{\sin\left(\frac{\pi}{d} v_{i+k,i}\right)} \exp(\theta) \\
&+\sum_{\substack{j \leq d-l\\l\leq d-1}} K_{(i,i+k),(j,j+l)}\star \tilde{L}_{(j,j+l)}(\theta)
\end{split}
\end{align}
where
\begin{align}
\label{eq:KernelforMonic}
\begin{split}
K_{(i,i+k),(j,j+l)}(\theta) = \frac{\langle (i,i+k),(j,j+l) \rangle}{2\pi i \sinh\left(\theta + \log \left(e^{-\frac{i \pi}{d} v_{ijkl}}\right)\right)}
\end{split}
\end{align}
and where the full intersection matrix $\langle (i,i+k),(j,j+l) \rangle$ is given by \refeq{eq:fullintersectionMatrix}, with the intersection and connection part given explicitly in our prescription by \refeq{eq:explicitConnectionPart} and \refeq{eq:explicitIntersectionPart}. The TBA system \refeq{eq:TBAforMonic} is containing $d(d+1)/2$ equations, but most of them are redundant : as explained in section \ref{chap:Geometric and preliminary observations}, we can associate a TBA graph to this pure polynomial potential system, which is the fully connected polygonal with $d$ vertices, and the identifications are given by the edges with same lengths. In the $d$th-root of unity prescription, it means that the edge (or TBA equation or $\epsilon$-function) $(v,v+k) = [k]$ is in the equivalence class $[k]$ $\forall v \in \{1,\ldots,d\}$. We can go back to our real/imaginary prescription using the bijection \refeq{eq:bij}: in this prescription, the equivalence class is given by the relation $(i,j) = [\abs{v_{ji}}]$ (with $v_{ij} = v_i-v_j$ and $v_{i} = (2d + 1 +(-1)^{d+i}(2i-1))/4$). Equivalently, one can also look at $(i,i+k)$ directly in the real imaginary prescription: if $k$ is odd, then it is a member of $\left[i+\frac{k-1}{2}\right]$, if $k$ is even then it is a member of $[k/2]$. In any prescription, this leaves us with $d-1$ TBA equations that are equivalent to Dorey and Tateo equations after simplification. However, these equations can still be paired with the reflected edge of same length, such that we can enlarge the equivalence class adding the relation $[d-k]=[k]$. At the end of the day, we are left with $\ceil{\frac{d-1}{2}}$ independent TBA equations
\begin{align}
\label{eq:TBAforMonicKilledRedundancies}
\begin{split}
\tilde{\epsilon}_{(k,k+1)}(\theta) = \epsilon_{[k]}(\theta) =& \sqrt{8\pi E} \left(\frac{E}{a}\right)^{\frac{1}{d}} \frac{\Gamma\left(1+\frac{1}{d}\right)}{\Gamma\left(\frac{3}{2}+\frac{1}{d}\right)}~ \abs{\sin\left(\frac{\pi}{d} v_{k+1,k}\right)} \exp(\theta) \\
&+\sum_{\substack{j \leq d-l\\l\leq d-1}} K_{(k,k+1),(j,j+l)}\star \tilde{L}_{(j,j+l)}(\theta)
\end{split}
\end{align}
in the real/imaginary prescription, with $k\in \{1,\ldots,\ceil{\frac{d-1}{2}}\}$. The expression \refeq{eq:TBAforMonicKilledRedundancies} can be further simplified realizing than the sum is involving equivalent $L$-functions (of course, since the $\epsilon_{[k]}$ are members of the same equivalence class, so do the $L_{[k]} = \log\left(1+\exp(-\epsilon_{[k]})\right)$). We can use the linearity of the integration and factorize these equivalent $L$-functions into a common factor of a new kernel, sum of the old ones, as we already outlined in \refeq{eq:TBAmonicKernels}. Doing so, we get the following simplified TBA system, involving only member of the equivalence class, as expected then exemplified from \refeq{eq:TBAmonic},
\begin{equation}
\label{eq:TBAforMonicSimplified}
\epsilon_{[k]}(\theta) = \sqrt{8\pi E} \left(\frac{E}{a}\right)^{\frac{1}{d}} \frac{\Gamma\left(1+\frac{1}{d}\right)}{\Gamma\left(\frac{3}{2}+\frac{1}{d}\right)}~ \abs{\sin\left(\frac{\pi}{d} v_{k+1,k}\right)} \exp(\theta) +\sum_{l=1}^{\ceil{\frac{d-1}{2}}} \mathcal{K}_{[k],[l]}\star L_{[l]}(\theta)
\end{equation}
where the matrix of kernels is given by
\begin{equation}
\label{eq:KernelMatrixMonic}
\mathcal{K}_{[k],[l]}(\theta) = \sum_{(a,b) \in [l]} \frac{\langle (k,k+1),(a,b) \rangle}{2\pi i \sinh\left(\theta + \log \left(e^{-\frac{i \pi}{d} v_{k,a,k+1,b}}\right)\right)}
\end{equation}
If one want to rewrite \refeq{eq:KernelMatrixMonic} in a more concrete fashion, one has to explicit the sum over the edges $(a,b)$ which are members of the equivalent class $[l]$, i.e. find how to generate the set $[l]$ over which we will be summing. Let us work before the reflection pairings $[k]=[d-k]$ for now. The rules above, telling in which equivalence class an arbitrary edge $(i,i+k)$ is, need to be inverted. Using these rules, it is easy to see that the equivalence class $[1]$ is the set,
$$[1] = \{(1,2)\} \cup \left(\bigcup_{i=1}^{d-2} \{(i,i+2)\}\right)$$
in our prescription. Likewise,
$$[2] = \{(2,3)\} \cup \{(1,4)\} \cup \left(\bigcup_{i=1}^{d-4} \{(i,i+4)\}\right),$$
etc. and it follows that the sum spawn over
$$[l] =\left(\bigcup_{i=1}^{\min(l,d-l)} \{(l+1-i,l+i)\}\right)  \cup \left(\bigcup_{i=1}^{d-2l} \{(i,i+2l)\}\right)$$
and one can write the matrix of kernels \refeq{eq:KernelMatrixMonic} involved in the TBA system \refeq{eq:TBAforMonicSimplified} as
\begin{equation}
\label{eq:kerIMS}
\mathcal{K}_{[k],[l]}(\theta) = 
\begin{cases} 
\phi_{k,l}^{\text{IMS}}(\theta) & d-l=l\\
\phi_{k,l}^{\text{IMS}}(\theta)+\phi_{k,d-l}^{\text{IMS}}(\theta) & \text{otherwise} 
\end{cases}
\end{equation}
thus taking into account the reflection symmetry $[d-k]=[k]$ we glossed out until now, and with $\phi$ the matrix of kernels one would obtain considering $d-1$ TBA equations,
\begin{align}
\begin{split}
\label{eq:KernelMatrixMonic2}
\phi_{k,l}^{\text{IMS}}(\theta) =& \sum_{n=1}^{\min(l,d-l)} \frac{\langle (k,k+1),(l+1-n,l+n) \rangle}{2\pi i \sinh\left(\theta + \log \left(e^{-\frac{i \pi}{d} v_{k,l+1-n,k+1,l+n}}\right)\right)} \\
&+ \sum_{n=1}^{d-2l} \frac{\langle (k,k+1),(n,n+2l) \rangle}{2\pi i \sinh\left(\theta + \log \left(e^{-\frac{i \pi}{d} v_{k,n,k+1,n+2l}}\right)\right)}
\end{split}
\end{align}
We purposefully called it $\phi$ since it analogous to the matrix of kernels \refeq{eq:DTkernel} found in Dorey and Tateo equations \refeq{eq:DTeq}. However, notice that $\phi_{k,l}^{\text{IMS}}(\theta) \neq \phi_{k,l}(\theta)$; it is only when we are applying the reflection symmetry that we have $\mathcal{K}_{[k],[l]}(\theta) =\mathcal{K}_{[k],[l]}^{\text{DT}}(\theta)$ element-wise. For example, one can notice that $\phi_{d-1,d-1}^{\text{IMS}}(\theta) = 0$ and $\phi_{k,l}^{\text{IMS}}(\theta) = 0$ $\forall k \in \{1,\cdots,2l-d-1\}$ and $\forall l \in \{\ceil{\frac{d}{2}}+1,\cdots,d-1\}$ (a triangle of zeroes of height $d-3$ and length $\ceil{\frac{d-3}{2}}$), when $\phi$ is symmetric and non zero $\forall k,l \in \{1,\cdots,d-1\}$ (excepted for some specific values of $\theta$) as one can see from the definition \refeq{eq:DTkernel}. Nonetheless, we expect $\phi_{k,l}^{\text{IMS}}(\theta)+\phi_{k,d-l}^{\text{IMS}}(\theta) = \phi_{k,l}(\theta)+\phi_{k,d-l}(\theta)$ element-wise, since it is equivalent to the statement $\mathcal{K}_{[k],[l]}(\theta) =\mathcal{K}_{[k],[l]}^{\text{DT}}(\theta)$.
\subsection{Dorey and Tateo equations as a special case}\label{chap:DTasSpecialCase}
Let us present a lightning review of the main result of \cite{Dorey:1998pt} (a more recent and complete review about the ODE/IM correspondence can be found in \cite{Dorey:2007zx}). The starting point is to consider an integrable massive quantum field theory associated with the $A_{d-1}$ Lie algebra and $d-1$ massive particle species. The scattering theory is factorisable with two particle S-matrix elements:
\begin{equation}
	S_{ab}(\theta)=\prod_{\substack{\abs{a-b}+1 \\ \text{step 2}}}^{a+b+1} (p-1)(p+1),~~(x)=\frac{\sinh\left(\frac{\theta}{2}+\frac{i\pi}{2 d}x\right)}{\sinh\left(\frac{\theta}{2}-\frac{i\pi}{2 d}x\right)}~~\text{and}~~a,b \in \{1,\ldots,d-1\}
\end{equation} 
where $\theta$ is the rapidity. Using TBA techniques, one can find the following system consisting of $d-1$ pseudo-energies $\epsilon_{a}(\theta)$ that are solving
\begin{equation}
\label{eq:DTeq}
\epsilon_a(\theta) = m_a r \cosh(\theta) + \sum_{b=1}^{d-1} \phi_{ab} \star L_b (\theta),~~a \in \{1,\ldots,d-1\}
\end{equation}
where $r$ is linked with the finite-size scaling of the system of interest and with kernel
\begin{equation}
\label{eq:DTkernel}
\phi_{ab}(\theta) = i ~\partial_\theta \log(S_{ab}(\theta))
\end{equation}
Dorey and Tateo's conjecture is that the $T$-functions (cousins of the $Y$-functions which are the exponentiated pseudo-energies) coincide with the spectral determinant of the quantum system of interest -- a pure potential of the form $q^{d}$ -- after taking the ``conformal limit'' of \refeq{eq:DTeq}. For all of our purposes, we can write \refeq{eq:DTeq} in this limit as
\begin{equation}
\label{eq:DTeqConf}
\epsilon_a(\theta) = \abs{\Pi_a} \exp({\theta}) + \sum_{b=1}^{d-1} \phi_{ab} \star L_b (\theta)
\end{equation}
This system is very reminiscent of the TBA system \refeq{eq:TBAmonic} we wrote earlier for a pure potential. In fact, this is the ``$A_{d-1}$ reduction'' of our maximal system. Furthermore, ``half'' of the $\epsilon$-functions are redundant, because of the ``reflection'' identification $\epsilon_{a}=\epsilon_{d-a}$. This identification is actually the missing reflection symmetry we need to quotient out in order to end up with the reduced system \refeq{eq:TBAmonic}. Indeed, by applying this identification on \refeq{eq:DTeqConf}, one find 
\begin{equation}
\label{eq:DTeqIMS}
\epsilon_{[k]}(\theta) = \abs{\Pi_{[k]}} \exp({\theta}) + \sum_{l=1}^{\ceil{\frac{d-1}{2}}} \mathcal{K}_{[k],[l]}^{\text{DT}} \star L_{[l]} (\theta)
\end{equation}
where 
\begin{equation}
\label{eq:kerDT}
\mathcal{K}_{[k],[l]}^{\text{DT}}(\theta) = 
\begin{cases} 
\phi_{k,l}(\theta) & d-l=l\\
\phi_{k,l}(\theta)+\phi_{k,d-l}(\theta) & \text{otherwise} 
\end{cases}
\end{equation}
and $\phi_{a,b}$ defined in \refeq{eq:DTkernel}. As we already stated, the matrix of kernels \refeq{eq:kerDT} should be equal element-wise to \refeq{eq:kerIMS} $\forall \theta$. Let us work out the monic cubic in detail as a simple concrete example. In that simple case, the scattering matrix and matrix of kernels for the $A_{2}$ equations are 
\begin{equation*}
S(\theta)=
\begin{pmatrix}
-\frac{\cosh\left(\frac{\theta}{2}-\frac{i\pi}{6}\right)}{\cosh\left(-\frac{\theta}{2}+\frac{i\pi}{6}\right)} & \frac{\sinh\left(\frac{-\theta}{2}-\frac{i\pi}{6}\right)}{\sinh\left(\frac{\theta}{2}+\frac{i\pi}{6}\right)} \\
\frac{\sinh\left(\frac{-\theta}{2}-\frac{i\pi}{6}\right)}{\sinh\left(\frac{\theta}{2}+\frac{i\pi}{6}\right)} & -\frac{\cosh\left(\frac{\theta}{2}-\frac{i\pi}{6}\right)}{\cosh\left(-\frac{\theta}{2}+\frac{i\pi}{6}\right)}
\end{pmatrix}
~~\Rightarrow~~
\phi(\theta)=
\begin{pmatrix}
\frac{\sqrt{3}}{4 \pi  \cosh (\theta )+2 \pi } & \frac{\sqrt{3}}{4 \pi  \cosh (\theta )-2 \pi } \\
\frac{\sqrt{3}}{4 \pi  \cosh (\theta )-2 \pi } & \frac{\sqrt{3}}{4 \pi  \cosh (\theta )+2 \pi } \\
\end{pmatrix}
\end{equation*}
Applying \refeq{eq:kerDT}, we find
\begin{equation}
	\mathcal{K}^{\text{DT}}(\theta)  =\phi_{1,1}(\theta)+\phi_{1,2}(\theta)= \frac{\sqrt{3}\sinh(2\theta)}{\pi  \sinh(3\theta )}
\end{equation}
By reading the monochromatic TBA graph of the monic cubic in figure \ref{fig:TBAmaxCmonic} (or simply computing \refeq{eq:kerIMS}, given explicitly), one indeed finds the following TBA equation:
\begin{equation}
\epsilon (\theta) = \abs{\Pi} \exp(\theta) + \frac{\sqrt{3}}{\pi}\int_{\mathbb{R}} \frac{\sinh(2(\theta-\bar{\theta}))}{\sinh(3(\theta-\bar{\theta}))} L(\bar{\theta}) d\bar{\theta}
\end{equation}
As an exercise, an interested reader can compute $\mathcal{K}^{\text{DT}}$ for the quartic (resp. sextic) and find that it is indeed matching the matrix provided in \refeq{eq:KmonicQuartic} (resp. \refeq{eq:KmonicSextic}). We wrote a program that is computing the exact matrices of kernels $\mathcal{K}$ and $\mathcal{K}^{\text{DT}}$ for arbitrary $d$. We were able to prove a symbolic and exact equality for $d\in\{3,\ldots,14\}$. We also computed systematically their numerical differences $\abs{\mathcal{K}(\theta)-\mathcal{K}^{\text{DT}}(\theta)}$. Taking the maximum of this numerical difference $\forall \theta$ and $\forall$ matrices entries, we were able to verify that it is at most of the order $10^{-1000}$ for $d\in\{1,\ldots,230\}$. In order to achieve the general proof, the only remaining step is to demonstrate the identity, $\mathcal{K}_{[k],[l];d}(\theta) =\mathcal{K}^{\text{DT}}_{[k],[l];d}(\theta)$ which we will not provide in the present work. This identity is purely mathematical at this point, and we know the l.h.s. and r.h.s. for arbitrary parameters $l$, $k$ and $d$.

\subsection{Computing the WKB periods using the $\epsilon$-functions.}
\label{chap:WKBperiods}
The pseudo-energies are encoding the all order WKB periods. Thus, one should in principle be able to extract the quantum corrections to the classical periods $\Pi^{(n)}$ for any $n$ from the pseudo-energies. For simplicity, let us start working in the minimal chamber, in the mass representation. We know that the asymptotic behavior of the $\epsilon$-functions at large $\theta$ is given in this case by $\epsilon_{(a)}(\theta) \sim m_{(a)} \exp(\theta)$. In fact, the all-orders expansion of \refeq{eq:TBAsysMinC} at large $\theta$ yields
\begin{equation}
	\epsilon_{(a)}(\theta) \sim m_{(a)} e^{\theta} + \sum_{n \geq 1}  m_{(a)}^{(n)}e^{(1-2n)\theta}
\end{equation}
where the coefficients $m_{(a)}^{(n)}$ are
\begin{equation}
m_{(a)}^{(n)} = \frac{(-1)^n}{\pi} \int_{\mathbb{R}} e^{(2n-1)\bar{\theta}} \left(L_{(a-1)}(\bar{\theta})+L_{(a+1)}(\bar{\theta})\right) d\bar{\theta}
\end{equation}
and are related to the periods through relations similar to \refeq{eq:masses}:
\begin{align}
\label{eq:massescorr}
\begin{split}
m_{2k-1,2k}^{(n)} &= (-1)^n \Pi^{(n)}_{2k-1,2k}\\
m_{2k,2k+1}^{(n)} &= i~\Pi^{(n)}_{2k,2k+1}
\end{split}
\end{align}
This is due to the large $\theta$ identity
\begin{equation}
\frac{1}{\cosh\left(\theta-\bar{\theta}\right)} = 2 e^{-\theta}e^{\bar{\theta}} \sum_{n \in \mathbb{N}} (-1)^n e^{-2n\theta}e^{2n\bar{\theta}}
\end{equation}
The same derivation can be reproduced mutatis mutandis for any given TBA system. In the general form \refeq{eq:ultimateTBA} with kernel \refeq{eq:kernelSinh}, a similar computation yields the following quantum correction to the periods:
\begin{equation}
\label{eq:WKBTBA}
	\Pi^{(n)}_{(a)} = i \sum_{(b) \in S_d}  \langle(a),(b)\rangle e^{i(1-2n)\varphi_{(b)}}~ \frac{(-1)^{n}}{\pi} \int_{\mathbb{R}} e^{(2n-1)\bar{\theta}}\tilde{L}_{(b)}(\bar{\theta}) d\bar{\theta}
\end{equation}

\section{Examples}
\label{chap:res}
In the following, we apply the TBA machinery developed in the previous sections in order to solve the Schrödinger spectral problem \refeq{eq:Schro} for various potentials, computing numerically their WKB periods at arbitrary order and Voros spectra.\\

The two first examples (cubic and quartic potentials) are already covered in \cite{Ito:2018eon} with an important and interesting exception: we computed resonant (thus complex) Voros' spectra for cubic potentials when only the periods are computed in \cite{Ito:2018eon}. Anyway, these examples are to be thought of as warm-up exercises and trivial checks for the analytical continuation procedure described above. The next example, a general complex quintic potential, is an application going beyond genus one. We compute all the WKB periods for this generic unphysical potential in order to check if our method still holds with this complicated and non-symmetric case. In the sextic example, we are solving the triple well problem for energies below (resp. above) the wells, corresponding to a sextic potential in the minimal (resp. maximal) chamber configuration.
\subsection{The cubic potential}
The first natural example beyond the harmonic oscillator is the cubic oscillator of the form
\begin{equation}
\label{eq:cubicOscPot}
V(q)= \frac{q^2}{2}-g q^3
\end{equation}
which has been extensively studied for this precise reason. However, because the potential is odd, a lot of subtleties are arising. Let us study them in the point of view of the TBA equations and resurgence.

As long as $0<E<\frac{1}{54 g^2}$, we are in the minimal chamber (\emph{i.e.} we are in presence of three real turning points). A particular cases of which will be studied in the next section. In the last section dedicated to cubic potentials, we will basically promote the coupling constant $g$ to a complex number, hence forcing the turning points to rotate in the complex plane, outside the real line. In fact, we will focus on PT-symmetric cubic potential, since it is a beautiful and interesting example, well studied in the literature and relatively easy to study in the context of TBA equations and resurgence. Of course, the methods developed here can accommodate more general cubic polynomials.

This particular exercise was already covered (partially\footnote{In the minimal chamber, they found the correct periods but did not compute the resonant Voros spectrum.}) in \cite{Ito:2018eon}. In the recent \cite{Hollands:2019wbr}, they are covering the cubic case from the point of view of abelianization.

\subsubsection{The cubic oscillator in the minimal chamber}\label{chap:cubicOsc}
The cubic oscillator \refeq{eq:cubicOscPot} with $g$ and $E$ $\in \mathbb{R}$ respecting $0<E<\frac{1}{54 g^2}$ seems to be the simplest non-trivial example of a polynomial potential in the minimal chamber, with its only two very simple TBA equations:
\begin{align}
\begin{split}
\label{eq:TBAcubicOsc}
\epsilon_{1,2}(\theta) =& \abs{\Pi_{1,2}} \exp(\theta)-\frac{1}{2\pi} \int_{\mathbb{R}}  \frac{L_{2,3}(\bar{\theta})~d\bar{\theta}}{\cosh(\theta-\bar{\theta})}\\
\epsilon_{2,3}(\theta) =& \abs{\Pi_{2,3}} \exp(\theta)-\frac{1}{2\pi} \int_{\mathbb{R}}  \frac{L_{1,2}(\bar{\theta})~d\bar{\theta}}{\cosh(\theta-\bar{\theta})}
\end{split}
\end{align}
However, the Hamiltonian of the cubic oscillator \refeq{eq:cubicOscPot} for $g$ real is clearly unbounded and the resulting spectrum is \emph{resonant} (see appendix \ref{app:CDilres} for a brief description). Let us focus on two examples: $g=1$ with $E=1/200$, in order to connect with the results of \cite{Ito:2018eon}, and $g=1/10$ with $E=1$, because in this case the non-perturbative effects are stronger and it will be easier to show, within our precision, that the corresponding level $E_n$ of the standard spectrum (i.e. spectrum evaluated at $\hbar_n = \exp(-\theta_n)$) is purely real, even if the spectrum of such resonant state is a priori complex, as discussed in \ref{app:CDilres}.\\

Solving the TBA system \refeq{eq:TBAcubicOsc} numerically, using the results of appendix \ref{app:num}, one can then compute the WKB periods using \refeq{eq:WKBTBA}. Our results can be found in the table \ref{tab:QperiodsCubicOsc} and is compared with the WKB periods computed with standard quantum mechanical techniques described in appendix \ref{app:qperiods}.\\

\begin{figure}
	\center
	\includegraphics[width=0.86\textwidth]{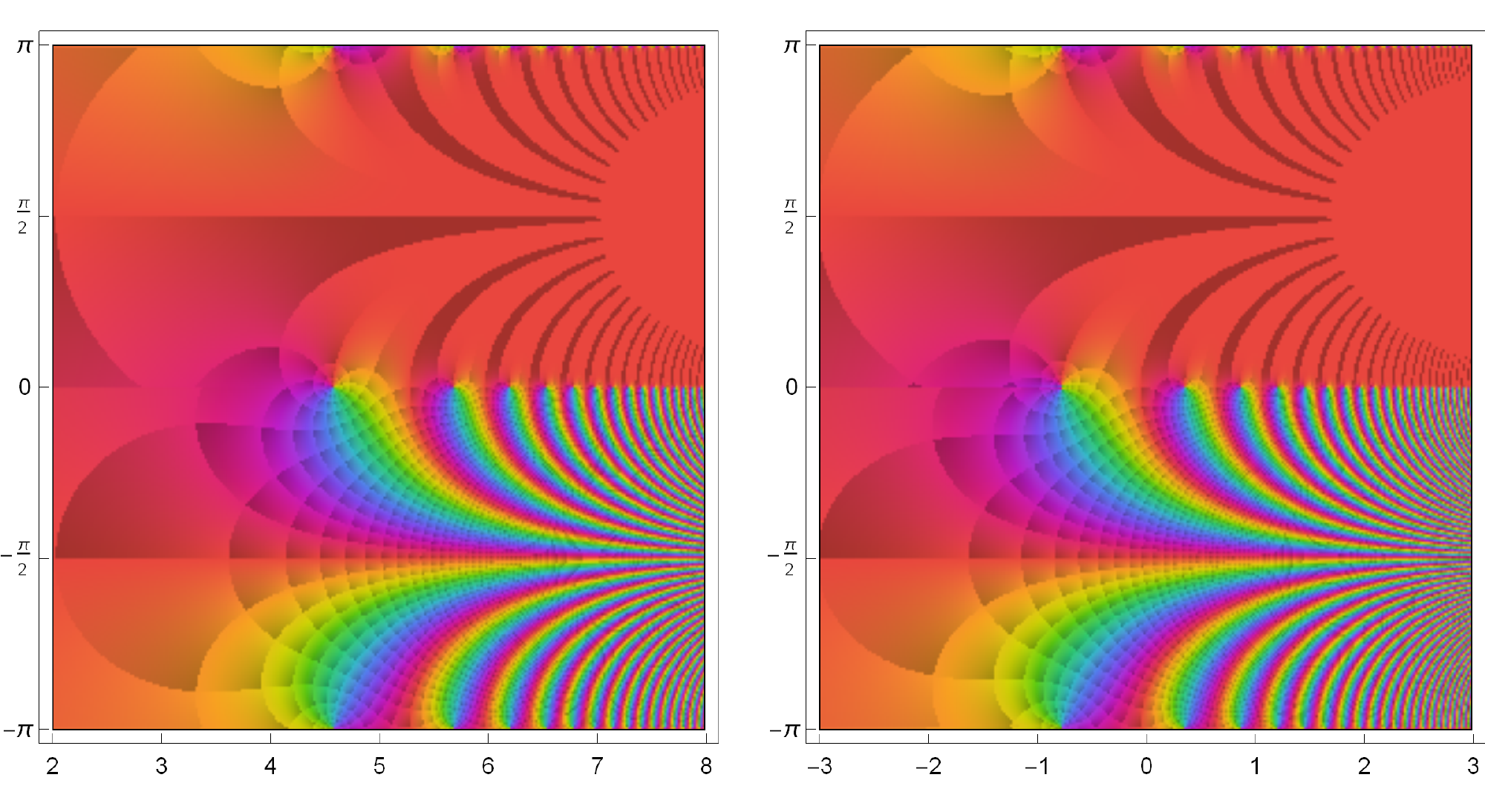}
	\includegraphics[width=0.094\textwidth]{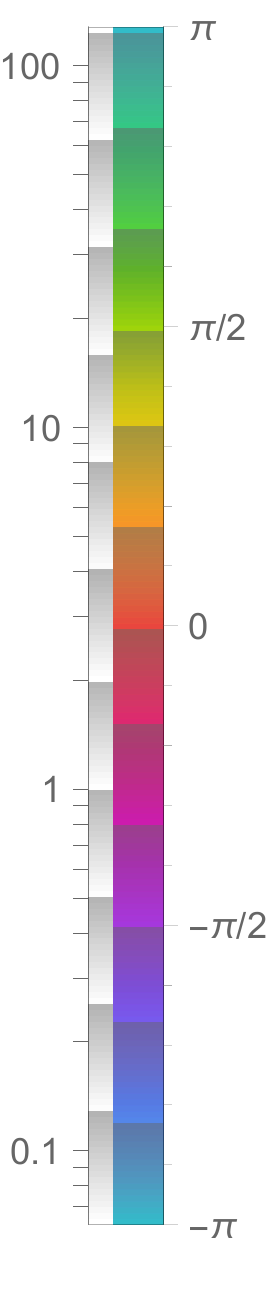}
	\caption{Inverse of the EQC \refeq{eq:EQCcubicOsc} plotted in the complex $\theta$-plane for $g=1$ with $E=1/200$ (left) and $g=1/10$ with $E=1$ (right) respectively. The color is encoding the argument of this function, the poles of which are the zeroes of the EQC, thus are selecting the resonant Voros spectrum for the associated WKB curve parameters. The imaginary part of the zeroes are non-perturbatively close to $i k \pi$. When $k$ is even, they are related to the same $\hbar$, with positive real part and non-perturbatively suppressed imaginary part. When $k$ is odd, it corresponds to the same  $\hbar$ than in the previous case except it has a negative real part. However, one can recover the appropriate control spectrum in table \ref{tab:SpectrumCubicMinC} by changing the dilatation angle appropriately (taking it close to the real negative axis instead of the real positive axis).}
	\label{fig:EQCcubicCplane}
\end{figure}

Using Voros-Silverstone connection formulae, developped in appendix \ref{app:EQC}, one can extract the exact quantization conditions associated to this problem. Depending on the choice of lateral Borel resummation, one finds
\begin{equation}
\label{eq:EQCcubicOsc}
1+e^{\pm\frac{i}{\hbar}\mathcal{B}_{\pm}\left(\Pi_{1,2}\right)(\theta)} +\frac{1}{2}(1\pm 1) e^{-\frac{i}{\hbar}\mathcal{B}\left(\Pi_{2,3}\right)(\theta)} = 0
\end{equation}
where the $\mathcal{B}_{\pm}$ are the lateral borel resumations defined in \refeq{eq:LateralBorelResum} (the sign $\pm$ is encoding the choice of lateral resummation). We can go from one EQC to the other by the use of the Stockes automorphism, using Delabaere-Pham formula, as explained in section \ref{chap:WKB&ResurgentQM}. In that case, it is very easy to show, so let's do it explicitly as an exercise. Rewriting $\mathcal{V}_\pm = e^{\pm\frac{i}{\hbar}\mathcal{B}_{\pm}\left(\Pi_{1,2}\right)(\theta)}$ and $\mathcal{V}_{\text{np}}=e^{-\frac{i}{\hbar}\mathcal{B}\left(\Pi_{2,3}\right)(\theta)}$, the EQCs are given by
\begin{equation}
 1+\mathcal{V}_\pm+\frac{1}{2}(1\pm 1 )\mathcal{V}_{\text{np}} =0
\end{equation}
In this notation, Delabaere-Pham formula takes the form
\begin{equation}
	\frac{\mathcal{V}_+}{\mathcal{V}_-} = 1+\mathcal{V}_{\text{np}}
\end{equation}
such that
\begin{align*}
	0&=1+\mathcal{V}_+ +\mathcal{V}_{\text{np}} &&\mapsto 1+\mathcal{V}_-(1+\mathcal{V}_{\text{np}}) +\mathcal{V}_{\text{np}} = (1+\mathcal{V}_{\text{np}})(1+\mathcal{V}_-)&&\Leftrightarrow 1+\mathcal{V}_- = 0\\
	0&=1+\mathcal{V}_- &&\mapsto 1+\frac{\mathcal{V}_-}{1+\mathcal{V}_{\text{np}}} = \frac{1}{1+\mathcal{V}_{\text{np}}}(1+\mathcal{V}_+ +\mathcal{V}_{\text{np}}) &&\Leftrightarrow 1+\mathcal{V}_+ +\mathcal{V}_{\text{np}} = 0
\end{align*}
This general procedure provides a non-trivial check that our EQC are working.\\
In order to relate the all order resummed periods involved in the EQC \refeq{eq:EQCcubicOsc} with the $\epsilon$-functions, we simply use \refeq{eq:regEpsilon}. The analytical structure of the EQS's inverse can be observed in figure \ref{fig:EQCcubicCplane}, the poles of which are corresponding to the zeroes of the EQC. Solving \refeq{eq:EQCcubicOsc} using the appropriate lateral Borel resummation prescription yields the Voros spectrum, which can be checked numerically with great accuracy using the complex dilatation method described in appendix \ref{app:CDil}. The result are summarized in the table \ref{tab:SpectrumCubicMinC}.

\begin{table}[tb]
	\begin{center}
		\begin{tabular}{|c|l|l|l|}
			\multicolumn{4}{l}{\textbf{WKB curve parameters:} $g=1$, $E=1/200$}\\ \hline
			\multicolumn{1}{|c}{$\gamma$} & \multicolumn{1}{|c}{$\Pi_\gamma^{(1)}$ (TBA)}  &  \multicolumn{1}{|c|}{$\Pi_\gamma^{(2)}$ (TBA)}&  \multicolumn{1}{|c|}{$\Pi_\gamma^{(3)}$ (TBA)} \\ \hline
			$(1,2)$ & $3.657475832644622$ & $948.7944867184156$ & $1.368408366637427\times 10^6$ \\
			$(2,3)$ & $ -9.193962022855980~i $ & $19138.83173033597~i$& $-2.280904640847162\times 10^8~i$ \\
			\hline
			\multicolumn{1}{|c}{$\gamma$} & \multicolumn{1}{|c}{$\Pi_\gamma^{(1)}$ (Diff. Op.)}  &  \multicolumn{1}{|c|}{$\Pi_\gamma^{(2)}$ (Diff. Op.)}&  \multicolumn{1}{|c|}{$\Pi_\gamma^{(3)}$ (Diff. Op.)} \\ \hline
			$(1,2)$ & $3.657475832642637$ & $948.7944867169863$ & $1.368408366655313\times 10^6$\\
			$(2,3)$ & $ -9.193962022850793~i $ & $19138.83173030450~i$& $-2.280904640893138\times 10^8~i$ \\
			\hline
			\multicolumn{4}{l}{}\\
			\multicolumn{4}{l}{\textbf{WKB curve parameters:} $g=1/10$, $E=1$}\\ \hline
			\multicolumn{1}{|c}{$\gamma$} & \multicolumn{1}{|c}{$\Pi_\gamma^{(1)}$ (TBA)}  &  \multicolumn{1}{|c|}{$\Pi_\gamma^{(2)}$ (TBA)}&  \multicolumn{1}{|c|}{$\Pi_\gamma^{(3)}$ (TBA)} \\ \hline
			$(1,2)$ & $0.05591075945249572$ & $0.003835061419185378$ & $0.001501382556091332$ \\
			$(2,3)$ & $ -0.04821582740725232~i $ & $0.002362302662884354~i$& $-0.0006569433399559175~i$ \\
			\hline
			\multicolumn{1}{|c}{$\gamma$} & \multicolumn{1}{|c}{$\Pi_\gamma^{(1)}$ (Diff. Op.)}  &  \multicolumn{1}{|c|}{$\Pi_\gamma^{(2)}$ (Diff. Op.)}&  \multicolumn{1}{|c|}{$\Pi_\gamma^{(3)}$ (Diff. Op.)} \\ \hline
			$(1,2)$ & $0.05591075945221964$ & $0.003835061419129100$ & $0.001501382556077192$\\
			$(2,3)$ & $ -0.04821582740701447~i $ & $0.002362302662849769~i$& $-0.0006569433399535173~i$ \\
			\hline 
				
		\end{tabular}
		
		\caption{The three first quantum corrections to the periods for the two examples of cubic potential \refeq{eq:cubicOscPot} in the minimal chamber, \emph{i.e.} with $g=1$, $E=1/200$ and with $g=1/10$, $E=1$ respectively, computed using the $\epsilon$-functions obtained by solving the appropriate TBA system numerically using the iterative integration method with $16000$ and $8000$ Gaussian distributed points respectively, and cutoff $\in [-75,22]$ and $\in [-75,18]$ respectively, as described in appendix \ref{app:num} (TBA). These values are compared with exact values obtained using the differential operator method described in appendix \ref{app:qperiods} (Diff. Op.).}
		\label{tab:QperiodsCubicOsc}
	\end{center}
\end{table}

\begin{table}[tb]
	\begin{center}
		\begin{tabular}{|l|c|c|}
			\multicolumn{3}{l}{\textbf{WKB curve parameters:} $g=1$, $E=1/200$}\\ \hline
			&$\theta_1=4.5720059951096 + 0.0000262355612~i$ & $\theta_2=5.68181569531799 - 1.31\times 10^{-13}~i$ \\
			\hline
			$E_1$&$1.0000000000000072- 1.87\times 10^{-16}~i$ & $0.3374194868285303+ 4.38\times 10^{-14}~i$  \\
			$E_2$&$2.836066671194304 + 0.008972793092311~i$ & $1.0000000000000007 + 0  \times 10^{-17}~i$  \\
			$E_3$&$4.308166892233317 + 0.253226263150458~i$ & $1.640969751176588 - 1.08768\times10^{-10}~i$  \\
			$E_4$&$5.785710072491928 + 0.973307366723122~i$ & $2.255952635641602 - 5.0473565\times10^{-8}~i$  \\
			\hline
			&$\theta_3=6.193492295534587 + 0 \times 10^{-19}~i$& $\theta_4=6.530198014144337 + 0\times10^{-25}~i$\\
			\hline
			$E_1$&$0.2030946908164711+ 0  \times 10^{-17}~i$ & $0.1452772223687511+ 0  \times 10^{-17}~i$ \\
			$E_2$& $0.6050732019463855+ 0  \times 10^{-17}~i$ & $0.4337227134778471+ 0  \times 10^{-17}~i$ \\
			$E_3$& $1.0000000000000002+ 0  \times 10^{-17}~i$ & $0.7187015508690309+ 0  \times 10^{-17}~i$ \\
			$E_4$&$1.3872112810554597+ 0  \times 10^{-17}~i$ & $1.0000000000000001+ 0  \times 10^{-17}~i$ \\
			\hline
		\end{tabular}
	\begin{tabular}{|l|c|c|}
		\multicolumn{3}{l}{\textbf{WKB curve parameters:} $g=1/10$, $E=1$}\\ \hline
		&$\theta_1=-0.7833886178219 + 0.0142637449697~i$ & $\theta_2=0.3544576265167 + 0.0000199830828~i$ \\
		\hline
		$E_1$&$1.0000000000001637 - 1.574\times10^{-13}~i$ & $0.3434099852474118 - 6.6763296857\times10^{-6}~i$  \\
		$E_2$&$2.688483613229089 + 0.341756680360593~i$ & $1.0000000000000234 + 0  \times 10^{-17}~i$  \\
		$E_3$&$4.451648618080736 + 1.283188377158743~i$ & $1.586043124901770 + 0.003705584739338~i$  \\
		$E_4$&$6.432370193567029 + 2.450993751714253~i$ & $2.062600732900113 + 0.083831268728741~i$  \\
		\hline
		&$\theta_3=0.8681208139536 + 5.74677\times10^{-8}~i$& $\theta_4=1.2053409964720 - 2.008\times10^{-10}~i$\\
		\hline
		$E_1$&$0.2073287370633492 - 1.17610749\times10^{-8}~i $ & $0.1485189632836222 + 2.95560\times10^{-11}~i$ \\
		$E_2$& $0.6124169659248588 - 3.40110555\times10^{-8}~i$ & $0.4408849681561786 + 8.66967\times10^{-11}~i$ \\
		$E_3$& $1.0000000000000076 + 1  \times 10^{-16}~i$ & $0.7251525764412156 + 1.398999\times10^{-10}~i$ \\
		$E_4$&$1.3646723729078898 + 0.0000165285082619~i$ & $0.9999999999999799 + 8.5\times10^{-15}~i$ \\
		\hline
	\end{tabular}
		\caption{Voros Spectrum of the $\theta_n$ for $n\leq4$ solving the EQC  \refeq{eq:EQCcubicOsc} for the potential $q^2/2-g x^3$ with $g=1$ and $g=1/10$ respectively, together with the -- normalized to one -- control spectrum of energies obtained for these values of $\hbar_n=\exp(-\theta_n)$ using the complex dilatation method described in the appendix \ref{app:CDil} (using dilatation angles slightly above or below the real positive axis with the appropriate $\pm$ prescription for the EQC). The $\epsilon$-functions involved are obtained numerically using the iterative integration method with $16000$ and $8000$ Gaussian distributed points, with cutoffs $\in [-75,22]$ and $\in [-75,18]$ respectively. Since the energy (as a WKB curve parameter) was chosen too be $E=1/200$ and $E=1$, the diagonal elements of this table should be $200 E=1$ and $E=1$, according to \refeq{eq:checkE}. For $g=1$ and $E=1/200$, the non-perturbatively small imaginary part is inaccessible with our precision for the third and forth levels of the Voros spectrum (the two first levels are well within our precision though). This is why we are providing a second example with $g=1/10$ and $E=1$, for which the non-perturbative effects are stronger.}
		\label{tab:SpectrumCubicMinC}
	\end{center}
\end{table}

\subsubsection{The PT cubic potential}
The next example is the maximal chamber extension of the potential \refeq{eq:cubicOscPot}. Let us focus on the interesting PT cubic potentials. For concreteness, let us chose
\begin{equation}
\label{eq:PTcubicPot}
V(q)=i q^3- i q~,~~~E=1
\end{equation}
Of course, \refeq{eq:PTcubicPot} can always be linked back to \refeq{eq:cubicOscPot} by the appropriate rescalings and shifts in the energy and $q$, and by promoting the coupling constant $g$ to a complex number.\\

This potential is an interesting one to study since it is the special case of ${V(q)=i q^3- i \lambda q}$ (with $\lambda=1$), which has a been a rich playground for investigating spontaneous PT-symmetry breaking caused by non-perturbative effect arising in that case for $\lambda$ positive and sufficiently large, producing intricate patterns of coalescing levels at the so-called Bender-Wu branch-points (see \cite{PhysRev.184.1231,doi:10.1063/1.532860,Bender:2007nj,Delabaere_2000,Emery:2019znd}).\\
The TBA system for this cubic in the maximal chamber is consisting of 3 TBA equations, encoded in the intersection matrix
\begin{equation}
\langle (a),(b)\rangle = \begin{pmatrix}
0 & -1 & -1\\
1 & 0 & 1\\
1 & -1 & 0
\end{pmatrix}
\end{equation} 
through \refeq{eq:ultimateTBA} -- with kernel \refeq{eq:kernelSinh} -- as can be read from  Fig. \ref{fig:TBAmaxC} (in that case, $(a),(b) \in S_3$ with $S_3 = \{(1,2),(2,3),(1,3)\}$). Explicitly:
\begin{align*}
\tilde{\epsilon}_{1,2}(\theta) =& \abs{\Pi^{(0)}_{1,2}} \exp(\theta)-\frac{1}{2\pi i} \int_{\mathbb{R}}  \frac{\tilde{L}_{2,3}(\bar{\theta})~d\bar{\theta}}{\sinh(\theta-\bar{\theta}+ i \varphi_{(1,2),(2,3)})}-\frac{1}{2\pi i} \int_{\mathbb{R}}  \frac{\tilde{L}_{1,3}(\bar{\theta})~d\bar{\theta}}{\sinh(\theta-\bar{\theta}+ i \varphi_{(1,2),(1,3)})}\\
\tilde{\epsilon}_{2,3}(\theta) =& \abs{\Pi^{(0)}_{2,3}} \exp(\theta)+\frac{1}{2\pi i} \int_{\mathbb{R}}  \frac{\tilde{L}_{1,2}(\bar{\theta})~d\bar{\theta}}{\sinh(\theta-\bar{\theta}+ i \varphi_{(2,3),(1,2)})}+\frac{1}{2\pi i} \int_{\mathbb{R}}  \frac{\tilde{L}_{1,3}(\bar{\theta})~d\bar{\theta}}{\sinh(\theta-\bar{\theta}+ i \varphi_{(2,3),(1,3)})}\\
\tilde{\epsilon}_{1,3}(\theta) =& \abs{\Pi^{(0)}_{1,3}} \exp(\theta)+\frac{1}{2\pi i} \int_{\mathbb{R}}  \frac{\tilde{L}_{1,2}(\bar{\theta})~d\bar{\theta}}{\sinh(\theta-\bar{\theta}+ i \varphi_{(1,3),(1,2)})}-\frac{1}{2\pi i} \int_{\mathbb{R}}  \frac{\tilde{L}_{2,3}(\bar{\theta})~d\bar{\theta}}{\sinh(\theta-\bar{\theta}+ i \varphi_{(1,3),(2,3)})}
\end{align*}
This TBA system can be read from the $d=3$ maximal TBA graph in figure \ref{fig:TBAmaxC}. 

Once the $\epsilon$-functions are extracted from the TBA system above, one can compute the quantum corrections to the periods using \refeq{eq:WKBTBA}. Our results can be found in the table \ref{tab:QperiodsCubic} and is compared with the WKB periods computed with standard quantum mechanical techniques described in the appendix \ref{app:qperiods}.\\

Ultimately, we want to find the Voros spectrum for this Hamiltonian. In order to compute this spectrum, we need to find the $\theta_n =\exp(-\hbar_n)$ that are solving the exact quantization condition found using the techniques in the appendix \ref{app:EQC}:
\begin{equation}
\label{eq:EQCcubic}
	2 \cos\left(\frac{1}{\hbar}\Pi_{\text{p}}(\theta)\right) + \exp\left(-\frac{1}{\hbar}\Pi_{\text{np}}(\theta)\right) = 0
\end{equation}
where $\frac{1}{\hbar}\Pi_{\text{p}}$ and $\frac{1}{\hbar}\Pi_{\text{np}}$ are respectively the real and imaginary part of the all order resummed period $\frac{1}{\hbar}\mathcal{B}\left(\Pi_{1,2}\right)$, obtained using \refeq{eq:regEpsilon}. In that case, it is Borel summable and we can compute it without using the median resummation. Explicitly, 
\begin{align*}
\frac{1}{\hbar}\mathcal{B}\left(\Pi_{1,2}\right)(\theta) &= -i \epsilon_{1,2}\left(\theta + \frac{i \pi}{2}\right) =-i \tilde{\epsilon}_{1,2}\left(\theta + \frac{i \pi}{2} + i \varphi_{1,2}\right)\\
& = \abs{\Pi_{1,2}^{(0)}}\exp(\theta-i \varphi_{1,2}) -\frac{i}{2\pi} \int_{\mathbb{R}} \frac{L_{2,3}(\bar{\theta})~d\bar{\theta}}{\cosh(\theta-\bar{\theta}+i\varphi_{1,2})}-\frac{i}{2\pi} \int_{\mathbb{R}} \frac{L_{1,3}(\bar{\theta})~d\bar{\theta}}{\cosh(\theta-\bar{\theta})}
\end{align*}
where we used $\varphi_{1,2} = -\varphi_{2,3}$ and $\varphi_{1,3}=0$. This function (thus the functions $\Pi_{\text{p}}$ and $\Pi_{\text{np}}$) can be computed provided the pseudo-energies $\epsilon_{2,3}(\theta)$ and $\epsilon_{1,3}(\theta)$. We compute them numerically solving the TBA system above using the methods in appendix \ref{app:num}.\\
The Voros spectrum of the $\theta_n$ resulting from the solutions of \refeq{eq:EQCcubic} can be found in the table \ref{tab:SpectrumCubic}, together with the control energies obtained using appendix \ref{app:CDil}, which is providing a numerical check of the theory developed in the previous sections.

\begin{table}[tb]
	\begin{center}
		\begin{tabular}{|c|l|l|}\hline
			\multicolumn{1}{|c}{$\gamma$} & \multicolumn{1}{|c}{$\Pi_\gamma^{(1)}$ (TBA)}  &  \multicolumn{1}{|c|}{$\Pi_\gamma^{(2)}$ (TBA)} \\ \hline
			$(1,2)$ & $-0.0254496388280+0.0858852570741 ~i$ & $0.01178213058270-0.00323811512906 ~i$ \\
			$(2,3)$ & $-0.0254496388280-0.0858852570741 ~i$ & $0.01178213058270+0.00323811512906 ~i$ \\
			$(1,3)$ & $-0.0508992776559$ & $0.0235642611654$ \\
			\hline
			\multicolumn{1}{|c}{$\gamma$} & \multicolumn{1}{|c}{$\Pi_\gamma^{(3)}$ (TBA)}  &  \multicolumn{1}{|c|}{$\Pi_\gamma^{(4)}$ (TBA)} \\ \hline
			$(1,2)$ & $-0.00697694900501-0.00887243467638 ~i$ & $-0.0111043012723+0.0254040465020 ~i$ \\
			$(2,3)$ & $-0.00697694900501+0.00887243467638 ~i$ & $-0.0111043012723-0.0254040465020 ~i$ \\
			$(1,3)$ & $-0.0139538980100$ & $-0.0222086025445$ \\
			\hline
			\hline	
			\multicolumn{1}{|c}{$\gamma$} & \multicolumn{1}{|c}{$\Pi_\gamma^{(1)}$ (Diff. Op.)}  &  \multicolumn{1}{|c|}{$\Pi_\gamma^{(2)}$ (Diff. Op.)} \\ \hline
			$(1,2)$ & $-0.0254496388279+0.0858852570738 ~i$ & $0.01178213058258-0.00323811512903 ~i$ \\
			$(2,3)$ & $-0.0254496388279-0.0858852570738 ~i$ & $0.01178213058258+0.00323811512903 ~i$ \\
			$(1,3)$ & $-0.0508992776557$ & $0.0235642611652$ \\
			\hline
			$\gamma$ & \multicolumn{1}{|c}{$\Pi_\gamma^{(3)}$ (Diff. Op.)}  &  \multicolumn{1}{|c|}{$\Pi_\gamma^{(4)}$ (Diff. Op.)} \\ \hline
			$(1,2)$ & $-0.00697694900495-0.00887243467629 ~i$ & $-0.0111043012761+0.0254040465099 ~i$\\
			$(2,3)$ & $-0.00697694900495+0.00887243467629 ~i$ & $-0.0111043012761-0.0254040465099 ~i$  \\
			$(1,3)$ & $-0.0139538980099$ & $-0.0222086025523$ \\
			\hline		
		\end{tabular}
		
		\caption{WKB periods for the PT-cubic potential \refeq{eq:PTcubicPot}, computed using the $\epsilon$-functions in turn obtained by solving the appropriate TBA system numerically. We used the iterative integration method with $7000$ Gaussian distributed points and cutoff $\in [-75,17]$ described in appendix \ref{app:num} (TBA). We compare our results with the exact values obtained using the differential operator method described in appendix \ref{app:qperiods} (Diff. Op.).}
		\label{tab:QperiodsCubic}
	\end{center}
\end{table}

\begin{table}[tb]
	\begin{center}
		\begin{tabular}{|l|c|c|c|}\hline
			&$\theta_1=-0.579094293628$ & $\theta_2=0.402954599488$ & $\theta_3=0.919905483211$\\
			\hline
		$E_1$&$0.9999999999997$ & $0.24129916905739$ & $0.27497315649894-0.09697464892851 i$ \\
		$E_2$&$4.4029261713438$ & $0.9999999999933$ & $0.27497315649894+0.09697464892851 i$ \\
		$E_3$&$8.7160417651226$ & $2.2443924820007$ & $0.9999999999992$ \\
		$E_4$&$13.479530478517$ & $3.6352172068964$ & $1.7090585088940$ \\
		$E_5$&$18.572772223641$ & $5.1401027548585$ & $2.4860951506208$ \\
		\hline		
		\end{tabular}
	\begin{tabular}{|l|c|c|}\hline
		&$\theta_4=1.25497406313$ & $\theta_5=1.50614755426$ \\
		\hline
		$E_1$&$0.1780573134710+0.1988974382848 i$ & $0.1396122721567+0.2382926559611 i$ \\
		$E_2$&$0.1780573134710-0.1988974382848 i$ & $0.1396122721567-0.2382926559611 i$ \\
		$E_3$&$0.5551089608920$ & $0.3615952688865$ \\
		$E_4$&$0.999999999999$ & $0.6406923442807$ \\
		$E_5$&$1.500747102596$ & $0.999999999997$ \\
		\hline		
	\end{tabular}
\caption{Voros Spectrum of the $\theta_n$ solving the EQC  \refeq{eq:EQCcubic}, together with the control spectrum of energies obtained for these values of $\hbar_n=\exp(-\theta_n)$ using the complex dilatation method described in the appendix \ref{app:CDil}. The $\epsilon$-functions involved are obtained numerically using the iterative integration method with $7000$ Gaussian distributed points and cutoff $\in [-75,17]$. Since the energy (as a WKB curve parameter) was chosen too be $E=1$, the diagonal elements of this table should be $E=1$ according to \refeq{eq:checkE}. Increasing $\theta$, one can observe the aforementioned coalescing of levels and spontaneous breaking of the PT symmetry.}
\label{tab:SpectrumCubic}
\end{center}
\end{table}

\subsection{The quartic potential}

Our next example is the double well corresponding to the potential
\begin{equation}
\label{eq:quarticPot}
V(q)= \frac{1}{2}q^4-\frac{1}{4}q^2
\end{equation}
which was already analyzed in \cite{Ito:2018eon}. When $-1/32<E<0$, we are in the minimal chamber regime since we have 4 distinct and real turning points (energy below the wells). When the energy increase beyond $E>0$, we are left with only two real turning points, corresponding to the maximal chamber regime (energy above the wells). We chose $E=-1/64$ and $E=1$ in the examples below.

\subsubsection{Symmetric quartic in the minimal chamber}
The TBA system for a symmetric quartic potential in the minimal chamber is simply
\begin{align}
\label{eq:TBAmincQuartic}
\begin{split}
\epsilon_{1,2}(\theta)=& \abs{\Pi_{1,2}}\exp(\theta) - K\star L_{2,3}(\theta)\\
\epsilon_{2,3}(\theta)=& \abs{\Pi_{2,3}}\exp(\theta) - 2K\star L_{1,2}(\theta)
\end{split}
\end{align}
where the convolution with the kernel $K$ is defined in \refeq{eq:CoshKernel} and where we used the symmetry \refeq{eq:symRedundancies}. The corresponding (bi)chromatic TBA graph is
\begin{center}
	\begin{tikzpicture}
	\coordinate[label = above:$1$] (1) at (0,0);
	\coordinate[label = above left:$2$] (2) at (2,0);
	\coordinate[label = above:$3$] (3) at (2,1);
	\coordinate[label = above left:$4$] (4) at (4,1);
	
	\draw[line width=0.3mm,black] (1) -- (2);
	\draw[line width=0.3mm,black!10!red] (2) -- (3);
	\draw[line width=0.3mm,black] (3) -- (4);
	
	\node at (1)[circle,fill,inner sep=1.5pt]{};
	\node at (2)[circle,fill,inner sep=1.5pt]{};
	\node at (3)[circle,fill,inner sep=1.5pt]{};
	\node at (4)[circle,fill,inner sep=1.5pt]{};
	\end{tikzpicture}
\end{center}

After solving this TBA system, one can compute the WKB periods using \refeq{eq:WKBTBA}. Our results can be found in the table \ref{tab:QperiodsQuarticMinC} together with standard quantum mechanical results (described in the appendix \ref{app:qperiods}) for comparison.\\
The relevant EQC for this system is obtained using the techniques in the appendix \ref{app:EQC} and we find
\begin{equation}
\label{eq:EQCquarticMinClat}
e^{-\frac{i}{\hbar}\mathcal{B}\left(\Pi_{2,3}\right)(\theta)}+\left(1+e^{\pm\frac{i}{\hbar}\mathcal{B}_{\pm}\left(\Pi_{1,2}\right)(\theta)} \right)^2 = 0
\end{equation}
The pseudo-energies obtained from the TBA system above are encoding the median resummation $\mathcal{B}_{\text{med}}$ defined in \refeq{eq:meidanBorelResum}. Therefore, we still have to extract this quantity from \refeq{eq:EQCquarticMinClat}. Solving the EQC for $\mathcal{B}_{\pm}\left(\Pi_{1,2}\right)$, one finds
\begin{equation}
\mathcal{V}_{1,2}^{\pm} = \left(-1- i\sigma_\pm \sqrt{\mathcal{V}_{2,3}}\right)^{\pm 1}
\end{equation}
where $\mathcal{V}_{1,2}^{\pm} = \exp\left(-\frac{i}{\hbar} \mathcal{B}_{\pm}\left(\Pi_{1,2}\right)\right)$, $\mathcal{V}_{2,3} = \exp\left(-\frac{i}{\hbar} \mathcal{B}\left(\Pi_{2,3}\right)\right)$ and the $\sigma_\pm \in \{1,-1\}$ are encoding the choice of square root. In fact, one can show that $\sigma_\pm = \pm p$ where $p$ is the parity. Using \refeq{eq:meidanBorelResum}, it follows that
\begin{equation}\label{eq:UmedQUarticMinC}
\mathcal{V}_{\text{med}}^2 = \exp\left(
2 \frac{i}{\hbar} \mathcal{B}_{\text{med}}(\Pi_{1,2})
\right)=\mathcal{V}_{1,2}^{+} \mathcal{V}_{1,2}^{-} = \frac{1+i\sigma_+ \sqrt{\mathcal{V}_{2,3}}}{1+i\sigma_{-} \sqrt{\mathcal{V}_{2,3}}}
\end{equation}
thus,
\begin{equation}
\frac{1}{\hbar}\mathcal{B}_{\text{med}}(\Pi_{1,2}) = \pi k -\frac{i}{2}\log\left(\frac{1+i\sigma_+ \sqrt{\mathcal{V}_{2,3}}}{1+i\sigma_{-} \sqrt{\mathcal{V}_{2,3}}}\right) = \pi k + p \arctan\left(\sqrt{\mathcal{V}_{2,3}}\right)
\end{equation}
with $k\in \mathbb{Z}$ and $p\in\{-1,1\}$ the parity, such that
\begin{equation}
\label{eq:ZinnEQCk}
\cos\left(\frac{1}{\hbar}\mathcal{B}_{\text{med}}(\Pi_{1,2})\right) =\frac{(-1)^k}{\sqrt{1+\mathcal{V}_{2,3}}}
\end{equation}
The all order Bohr-Sommerfeld quantization approximation dictates the perturbative behavior and fix the parity of $k$. Indeed, when we set $\mathcal{V}_{2,3}\to 0$ (effectively neglecting exponentially suppressed terms), one finds
\begin{equation}
\lim_{\mathcal{V}_{2,3} \to 0} \cos\left(\frac{1}{\hbar} \mathcal{B}_{\text{med}}(\Pi_{1,2})\right) = (-1)^k = \cos\left(\pi(2n + 1)\right) = -1
\end{equation}
(because $n \in \mathbb{N}$) and $k$ must be an odd number. Then, \refeq{eq:ZinnEQCk} yields the Zinn-Justin's EQC:
\begin{equation}
\label{eq:ZinnEQC}
	\cos\left(\frac{1}{\hbar}\mathcal{B}_{\text{med}}(\Pi_{1,2})\right) + \frac{1}{\sqrt{1+\exp\left(-\frac{i}{\hbar} \mathcal{B}\left(\Pi_{2,3}\right)\right)}} = 0
\end{equation}
where the all order resummed periods involved in the EQC above are obtained using \refeq{eq:regEpsilon}:
\begin{align}
\begin{split}
\label{eq:BorelToEspsilonQuartic}
\frac{1}{\hbar} \mathcal{B}_{\text{med}}\left(\Pi_{1,2}\right)(\theta)=&-\frac{i}{2}\lim_{\delta \to 0}\left(
\epsilon_{1,2}\left(\theta+\frac{i \pi}{2} + i \delta\right)+\epsilon_{1,2}\left(\theta+\frac{i \pi}{2} - i \delta\right)
\right) \\
\frac{i}{\hbar} \mathcal{B}\left(\Pi_{2,3}\right)(\theta)=&~\epsilon_{2,3}\left(\theta\right)
\end{split}
\end{align}
Using the TBA system \refeq{eq:TBAmincQuartic}, we can rewrite the median resummation as
\begin{equation}
\label{eq:BmedQuarticMinC}
\frac{1}{\hbar} \mathcal{B}_{\text{med}}\left(\Pi_{1,2}\right)(\theta)= \abs{\Pi_{1,2}} \exp(\theta) + \frac{1}{2\pi}\mathcal{P}\left(\int_{\mathbb{R}} \frac{L_{2,3}(\bar{\theta})d\bar{\theta}}{\sinh(\theta-\bar{\theta})}\right)
\end{equation}
where $\mathcal{P}\left(\int_{\mathbb{R}}\right) = \lim_{\delta \to 0} \left(\int_{-\infty}^{-\delta}+\int_{\delta}^{\infty}\right)$ is the Cauchy principal value.

In conclusion, the rusummed WKB periods can be computed provided the $\epsilon$-functions, which can be done numerically as explained in appendix \ref{app:num}. The Voros spectrum of the quantum mechanical system of interest can then be extracted form the all-orders WKB periods by solving \refeq{eq:ZinnEQC}. This is what has been done numerically in table \ref{tab:SpectrumQuarticMinC}, and we found a precise numerical check -- by using a standard technique in QM described in appendix \ref{app:CDil} together with \eqref{eq:checkE} -- of the methods described in the present paper.\\

\begin{table}[tb]
	\begin{center}
		\begin{tabular}{|c|c|c|c|}\hline
			$\gamma$&$\Pi_\gamma^{(1)}$ (TBA FD)&$\Pi_\gamma^{(2)}$ (TBA FD)&$\Pi_\gamma^{(3)}$ (TBA FD)  \\ \hline
			$(1,2)$&$2.730837607957584$ & $246.8391875779934$ & $131821.5581804492$ \\
			$(2,3)$&$-6.213720604370592\,i$ & $1187.382247975386\,i$ & $-1.323232764659763\times 10^6\,i$ \\ \hline
		\end{tabular}
		\begin{tabular}{|c|c|c|c|}\hline
			$\gamma$&$\Pi_\gamma^{(1)}$ (TBA II)&$\Pi_\gamma^{(2)}$ (TBA II)&$\Pi_\gamma^{(3)}$ (TBA II)  \\ \hline
			$(1,2)$&$2.730837608311631$ & $246.8391875790927$ & $131821.5581790699$ \\
			$(2,3)$&$-6.213720604930958\,i$ & $1187.382247979388\,i$ & $-1.323232764649346\times 10^6\,i$ \\ \hline
		\end{tabular}
		\begin{tabular}{|c|c|c|c|}\hline
			$\gamma$&$\Pi_\gamma^{(1)}$ (Diff. Op.)&$\Pi_\gamma^{(2)}$ (Diff. Op.)&$\Pi_\gamma^{(3)}$ (Diff. Op.)  \\ \hline
			$(1,2)$&$2.730837608310337$ & $246.8391875787170$ & $131821.5581806956$ \\
			$(2,3)$&$-6.213720604927630\,i$ & $1187.382247977231\,i$ & $-1.323232764661000\times 10^6\,i$ \\ \hline
		\end{tabular}
	\caption{WKB periods for the symmetric quartic potential \refeq{eq:quarticPot} with $E=-1/64$ computed using the $\epsilon$-functions obtained by solving the appropriate TBA system numerically using the Fourier discretization method with $2^{14}$ points and cutoff $\in [-21,21]$ (TBA FD), the iterative integration method with $14000$ Gaussian distributed points and cutoff $\in [-75,22]$ (TBA II) -- both described in appendix \ref{app:num} -- compared with exact values obtained using the differential operator method described in appendix \ref{app:qperiods} (Diff. Op.).}
	\label{tab:QperiodsQuarticMinC}
	\end{center}
\end{table}

\begin{table}[tb]
	\begin{center}
		\begin{tabular}{|l|c|c|}\hline
			&$\theta_1=3.3247808343910237030$ & $\theta_2=3.4138911118682158018$ \\
			\hline
			$E_1$&$0.9999999999998636$ & $0.936971570538917$  \\
			$E_2$&$1.0977171436197144$ & $0.999999999999831$  \\
			$E_3$&$2.4436026336522710$ & $2.272417879374312$  \\
			$E_4$&$3.3059489687350466$ & $2.964197286216196$  \\
			\hline
			 &$\theta_3=4.5039585292309019509$& $\theta_4=4.5042465001441240264$\\
			\hline
			$E_1$&$0.34580527442056$ & $0.34570823839933$ \\
			$E_2$& $0.34580730798057$ & $0.34571026291026$ \\
			$E_3$& $0.99999999999999$ & $0.99973326612985$ \\
			$E_4$&$1.00026781752251$ & $1.00000000000000$ \\
			\hline
		\end{tabular}
		\caption{Voros Spectrum of the $\theta_n$ solving the EQC  \refeq{eq:ZinnEQC}, together with the -- normalized to one -- control spectrum of energies obtained for these values of $\hbar_n=\exp(-\theta_n)$ using the complex dilatation method described in the appendix \ref{app:CDil} on the shifted potential $q^4/2-q^2/4+1/32$. The $\epsilon$-functions involved are obtained numerically using the iterative integration method with $12000$ Gaussian distributed points and cutoff $\in [-75,22]$. Since the energy (as a WKB curve parameter) was chosen too be $E=-1/64$, the shifted potential has an energy $E=1/64$ and the diagonal elements of this table should be $64 E=1$ according to \refeq{eq:checkE}. We can observe that the purely non-perturbative splitting of levels is correctly reproduced.}
		\label{tab:SpectrumQuarticMinC}
	\end{center}
\end{table}

\subsubsection{Symmetric quartic in the maximal chamber}
We already figured out the explicit TBA system associated to a symmetric quartic potential in the maximal chamber with \refeq{eq:SymQuarticMaxChamber} (where the tildes have been omitted). Alternatively, one could look at the following TBA graph with four colors,

\begin{center}
	\begin{tikzpicture}
	\coordinate[label = left:$1$] (1) at (-2,0);
	\coordinate[label = below:$2$] (2) at (0,-2);
	\coordinate[label = above:$3$] (3) at (0,2);
	\coordinate[label = right:$4$] (4) at (2,0);
	
	\draw[line width=0.3mm,black] (1) -- (2);
	\draw[line width=0.3mm,black!10!red] (2) -- (3);
	\draw[line width=0.3mm,black] (3) -- (4);
	\draw[line width=0.3mm,blue] (1) -- (3);
	\draw[line width=0.3mm,blue] (2) -- (4);
	\draw[line width=0.3mm,black!30!green] (1) -- (4);
	
	\node at (1)[circle,fill,inner sep=1.5pt]{};
	\node at (2)[circle,fill,inner sep=1.5pt]{};
	\node at (3)[circle,fill,inner sep=1.5pt]{};
	\node at (4)[circle,fill,inner sep=1.5pt]{};
	\end{tikzpicture}
\end{center}
where the matching colors are parallel lines, as it is typically the case because of the symmetry \eqref{eq:symRedundancies}.\\

The quantum corrections to the periods we found using this TBA system together with \refeq{eq:WKBTBA} are found in table \ref{tab:QperiodsQuarticMaxC}, and is compared with values obtained using standard quantum mechanical computations presented in appendix \ref{app:qperiods}.\\

As usual, the starting point is the EQC for the lateral resummed WKB periods. Using the theory developed in the appendix \ref{app:EQC}, one can find
\begin{equation}
\label{eq:EQCquarticMaxClat}
e^{\frac{i}{2\hbar}\mathcal{B}_{\pm}\left(\Pi_{1,4}\right)(\theta)}+e^{-\frac{i}{2\hbar}\mathcal{B}_{\pm}\left(\Pi_{1,4}\right)(\theta)}+e^{\pm \frac{i}{2\hbar}\mathcal{B}_{\pm}\left(\Pi_{1,4}\right)(\theta)+ \frac{i}{\hbar}\mathcal{B}\left(\Pi_{2,3}\right)(\theta)}+2 e^{\frac{i}{2\hbar}\mathcal{B}\left(\Pi_{2,3}\right)(\theta)} = 0
\end{equation}
where the $\mathcal{B}_{\pm}$ are the lateral Borel resummations defined in \refeq{eq:LateralBorelResum} (the sign $\pm$ is encoding the choice of lateral resummation). We denote $\mathcal{V}_{1,4}^\pm= e^{\frac{i}{2\hbar}\mathcal{B}_{\pm}\left(\Pi_{1,4}\right)}$ and $\mathcal{V}_{2,3}= e^{\frac{i}{2\hbar}\mathcal{B}\left(\Pi_{2,3}\right)}$ such that the EQC takes the compact form 
\begin{equation}\label{eq:EQCquarticMaxCU}
\mathcal{V}_{1,4}^\pm+(\mathcal{V}_{1,4}^\pm)^{-1}+\mathcal{V}_{2,3}^2(\mathcal{V}_{1,4}^\pm)^{\pm 1}+2\mathcal{V}_{2,3} = 0
\end{equation}
and resolves to
\begin{equation}
\mathcal{V}_{1,4}^\pm= \left(-\sigma_\pm i - \mathcal{V}_{2,3}\right)^{\mp 1}
\end{equation}
Where $\sigma_\pm \in \{1,-1\}$ encodes the choice of the square root. This allows us to extract $\mathcal{V}_{\text{med}}^2$:
\begin{equation}\label{eq:UmedQUarticMaxC}
\mathcal{V}_{\text{med}}^2 = \exp\left(
\frac{i}{\hbar} \mathcal{B}_{\text{med}}(\Pi_{1,4})
\right)=(\mathcal{V}_{1,4}^{+} \mathcal{V}_{1,4}^{-})^2 = \left(\frac{\sigma_- i + \mathcal{V}_{2,3}}{\sigma_+ i + \mathcal{V}_{2,3}}\right)^2
\end{equation}
thus, taking into account the perturbative behaviour at large $\mathcal{V}_{2,3}$,
\begin{equation}
\frac{1}{\hbar}\mathcal{B}_{\text{med}}(\Pi_{1,4}) = \pi k -i\log\left(\frac{\sigma_- i + \mathcal{V}_{2,3}}{\sigma_+ i + \mathcal{V}_{2,3}}\right) = \pi (2n+1) + 2 \sigma \arctan\left(\mathcal{V}_{2,3}^{-1}\right)
\end{equation}
where $\sigma \in \{1,-1\}$ is related to the parity of the state and $n \in \mathbb{N}$. Applying $f(x) = \cos(x/2)$ on each side yields
\begin{equation}
\cos\left(\frac{1}{2\hbar}\mathcal{B}_{\text{med}}(\Pi_{1,4})\right) =\frac{-\sigma (-1)^n}{\sqrt{1+\mathcal{V}_{2,3}^2}}
\end{equation}
Finally, since $\sigma (-1)^n = p^2$ where $p$ is the parity, one can write the following EQC:
\begin{equation}
\label{eq:ZinnMaxC}
\cos\left(\frac{1}{2\hbar}\mathcal{B}_{\text{med}}(\Pi_{1,4})\right)+\frac{1}{\sqrt{1+e^{\frac{i}{\hbar}\mathcal{B}\left(\Pi_{2,3}\right)}}} = 0
\end{equation}
where the all order resummed periods involved in the EQC above are obtained using \refeq{eq:regEpsilon} on the TBA system \refeq{eq:SymQuarticMaxChamber} and are given as
\begin{align}\label{eq:epsQuarticMaxC}
\begin{split}
\frac{1}{\hbar} \mathcal{B}_{\text{med}}\left(\Pi_{1,4}\right)(\theta)=&
\abs{\Pi_{1,4}} \exp(\theta) - \frac{i}{\pi} \int_{\mathbb{R}} \frac{L_{1,2}(\bar{\theta}) d\bar{\theta}}{\cosh\left(\theta-\bar{\theta}+i \varphi_{1,2}\right)}\\
&+ \frac{i}{\pi} \int_{\mathbb{R}} \frac{L_{1,3}(\bar{\theta}) d\bar{\theta}}{\cosh\left(\theta-\bar{\theta}-i \varphi_{1,2}\right)}-\frac{1}{\pi}\mathcal{P}\left(\int_{\mathbb{R}} \frac{L_{2,3}(\bar{\theta})d\bar{\theta}}{\sinh(\theta-\bar{\theta})}\right)
  \\
\frac{i}{\hbar} \mathcal{B}\left(\Pi_{2,3}\right)(\theta)=&~\epsilon_{2,3}\left(\theta\right)
\end{split}
\end{align}
where we used $\varphi_{1,2}= -\varphi_{1,3}$, $\varphi_{1,4}=0$ (which implies that the shifted functions along the cycle $(1,4)$ are equal to the unshifted ones), the identity
\begin{align}
K_{(a),(b)}^\text{reg}(\theta) &= -\frac{i}{2} \lim_{\delta \to 0}\left(K_{(a),(b)}\left(\theta+i\left(\varphi_{(a),(b)}+\frac{\pi}{2}+\delta\right)\right)+K_{(a),(b)}\left(\theta+i\left(\varphi_{(a),(b)}+\frac{\pi}{2}-\delta\right)\right)\right)\nonumber\\
&=\frac{i}{2\pi}\frac{1}{\cosh\left(\theta+i \varphi_{(a),(b)}\right)}
\end{align}
for the two first kernels of \refeq{eq:SymQuarticMaxChamber} and \refeq{eq:BmedQuarticMinC} for the last term of \refeq{eq:SymQuarticMaxChamber}.\\

The Voros spectrum obtained solving \refeq{eq:ZinnMaxC} with the $\epsilon$-functions \refeq{eq:epsQuarticMaxC} can be found in table \ref{tab:SpectrumQuarticMaxC} and provides a high precision check of the theory developed in the previous sections.

\begin{table}[tb]
	\begin{center}
			\begin{tabular}{|c|c|c|}\hline
			$\gamma$&$\Pi_\gamma^{(1)}$ (TBA)&$\Pi_\gamma^{(2)}$ (TBA) \\ \hline
			$(1,2)$&$-0.0819566304122-0.0927891470157\,i$ & $0.00922113508254-0.00383952256644\,i$ \\ \hline
			$(2,3)$&$0.185578294031\,i$ & $0.00767904513289\,i$ \\ \hline
			$(1,3)$&$-0.0819566304122+0.0927891470157\,i$ & $0.00922113508254+0.00383952256644\,i$ \\ \hline
			$(1,4)$&$-0.163913260824$ & $0.0184422701651$ \\ \hline
			$\gamma$&$\Pi_\gamma^{(3)}$ (TBA)&$\Pi_\gamma^{(4)}$ (TBA) \\ \hline
			$(1,2)$&$0.00011483490925+0.00927329580341\,i$ & $-0.0189708755177-0.0008042208294\,i$ \\ \hline
			$(2,3)$&	$-0.0185465916068\,i$ & $0.00160844165886\,i$ \\ \hline
			$(1,3)$&$0.00011483490925-0.00927329580341\,i$ & $-0.0189708755177+0.0008042208294\,i$ \\ \hline
			$(1,4)$&	$0.000229669818505$ & $-0.0379417510353$ \\ \hline
			\end{tabular}
		\begin{tabular}{|c|c|c|}\hline
			$\gamma$&$\Pi_\gamma^{(1)}$ (TBA Diff. Op.)&$\Pi_\gamma^{(2)}$ (TBA Diff. Op.) \\ \hline
			$(1,2)$&$-0.0819566304085 - 0.0927891470118\,i$ & $0.00922113508228 - 0.00383952256633\,i$ \\ \hline
			$(2,3)$&$0.185578294024\,i$ & $0.00767904513265\,i$ \\ \hline
			$(1,3)$&$-0.0819566304085 + 0.0927891470118\,i$ & $0.00922113508228 + 0.00383952256633 \,i$ \\ \hline
			$(1,4)$&$-0.163913260817$ & $0.0184422701646$ \\ \hline
			$\gamma$&$\Pi_\gamma^{(3)}$ (TBA Diff. Op.)&$\Pi_\gamma^{(4)}$ (TBA Diff. Op.) \\ \hline
			$(1,2)$&$0.00011483490920 + 0.00927329580314\,i$ & $-0.0189708755226 - 0.0008042208295\,i$ \\ \hline
			$(2,3)$&	$-0.0185465916063\,i$ & $0.00160844165909\,i$ \\ \hline
			$(1,3)$&$0.00011483490920 - 0.00927329580314\,i$ & $-0.0189708755226 + 0.0008042208295 I\,i$ \\ \hline
			$(1,4)$&	$0.000229669818406$ & $-0.0379417510451$ \\ \hline
		\end{tabular}
		\caption{WKB periods for the symmetric quartic potential \refeq{eq:quarticPot} with $E=1$ computed using the $\epsilon$-functions obtained by solving the appropriate TBA system numerically using the the iterative integration method with $6000$ Gaussian distributed points and cutoff $\in [-75,18]$ (TBA) described in appendix \ref{app:num} and compared with the exact digits obtained using the differential operator method described in appendix \ref{app:qperiods} (Diff. Op.).}
		\label{tab:QperiodsQuarticMaxC}
	\end{center}
\end{table}

\begin{table}[tb]
	\begin{center}
		\begin{tabular}{|l|c|c|}\hline
			&$\theta_1=-0.57225864053120$ & $\theta_2=0.35356561525223$ \\
			\hline
			$E_1$&$0.9999999999938$ & $0.25462266336339$  \\
			$E_2$&$3.7371023450083$ & $0.9999999999995$  \\
			$E_3$&$7.5336347645547$ & $2.0748925796367$  \\
			$E_4$&$11.911068138176$ & $3.3201614161345$  \\
			\hline
			&$\theta_3=0.86207899679463$& $\theta_4=1.1962025160788$\\
			\hline
			$E_1$&$0.11204126359296$ & $0.061626521076274$ \\
			$E_2$& $0.46717617140001$ & $0.27594762108696$ \\
			$E_3$& $0.9999999999997$ & $0.61040199466660$ \\
			$E_4$&$1.6193800753561$ & $0.9999999999999$ \\
			\hline
		\end{tabular}
		\caption{Voros Spectrum of the $\theta_n$ solving the EQC  \refeq{eq:ZinnMaxC}, together with the control spectrum of energies obtained for these values of $\hbar_n=\exp(-\theta_n)$ using the complex dilatation method described in the appendix \ref{app:CDil}. The $\epsilon$-functions involved are obtained numerically using the iterative integration method with $6000$ Gaussian distributed points and cutoff $\in [-75,18]$. Since the energy (as a WKB curve parameter) was chosen too be $E=1$, the diagonal elements of this table should be $E=1$ according to \refeq{eq:checkE}.}
		\label{tab:SpectrumQuarticMaxC}
	\end{center}
\end{table}

\subsection{The quintic potential}
For the quintic example, we purposefully choose a generic potential with no simplifying symmetries, \emph{i.e.}
\begin{equation}
\label{eq:quinticPot}
V(q)=i q^5+\frac{1}{2}q^2-\frac{1}{10}q~,~~~E=1
\end{equation}
in order to check if our TBA machinery is accommodating for this difficult example. In this case we are in the maximal chamber and there are 10 TBA equations, encoded in the TBA graph or equivalently the intersection matrix of figure \ref{fig:TBAmaxC}. The results for the first four quantum correction to the periods can be found in the table \ref{tab:QperiodsQuinticTBA} and compared to the exact digits in table \ref{tab:QperiodsQuinticDiffOp}. We did not compute the EQC and Voros spectrum for this highly unphysical example. 

\begin{table}[tb]
	\begin{center}
		\begin{tabular}{|c|l|l|}\hline
			$\gamma$ & \multicolumn{1}{|c}{$\Pi_\gamma^{(1)}$}  &  \multicolumn{1}{|c|}{$\Pi_\gamma^{(2)}$} \\ \hline
			$(1,2)$ & $-0.063331399371+0.158332153661 ~i$ & $0.00263596614100+0.00300064152677 ~i$ \\
			$(2,3)$ & $-0.114195904636-0.247868478439 ~i$ & $-0.0378946345086-0.0190247965603 ~i$ \\
			$(3,4)$ & $-0.108813639853+0.253065980340 ~i$ & $-0.0283897195873+0.0210123204610 ~i$ \\
			$(4,5)$ & $-0.062412272993-0.161508879916 ~i$ & $0.00883111672216+0.00173626896725 ~i$ \\
			$(1,3)$ & $-0.177527304007-0.089536324778 ~i$ & $-0.0352586683676-0.0160241550335 ~i$ \\
			$(2,4)$ & $-0.223009544489+0.005197501901 ~i$ & $-0.0662843540959+0.0019875239008 ~i$ \\
			$(3,5)$ & $-0.171225912845+0.091557100425 ~i$ & $-0.0195586028651+0.0227485894283 ~i$ \\
			$(1,4)$ & $-0.286340943860+0.163529655562 ~i$ & $-0.0636483879549+0.0049881654275 ~i$ \\
			$(2,5)$ & $-0.285421817482-0.156311378015 ~i$ & $-0.0574532373738+0.0037237928680 ~i$ \\
			$(1,5)$ & $-0.348753216853+0.002020775646 ~i$ & $-0.0548172712328+0.0067244343948 ~i$ \\
			\hline
				$\gamma$ & \multicolumn{1}{|c}{$\Pi_\gamma^{(3)}$}  &  \multicolumn{1}{|c|}{$\Pi_\gamma^{(4)}$} \\ \hline
			$(1,2)$ & $0.0215449471635+0.0061879789899 ~i$ & $0.0001022445395-0.0779436678457 ~i$ \\
			$(2,3)$ & $-0.1188590024120-0.0485822988532 ~i$ & $-0.624697423577-0.326466722045 ~i$ \\
			$(3,4)$ & $-0.0323241469577+0.0404502847208 ~i$ & $-0.089863922798+0.137565588891 ~i$ \\
			$(4,5)$ & $0.00984641379283+0.00085651279493 ~i$ & $-0.0199211857295+0.0491073059654 ~i$ \\
			$(1,3)$ & $-0.0973140552485-0.0423943198633 ~i$ & $-0.624595179038-0.404410389891 ~i$ \\
			$(2,4)$ & $-0.151183149370-0.008132014132 ~i$ & $-0.714561346376-0.188901133154 ~i$ \\
			$(3,5)$ & $-0.0224777331649+0.0413067975157 ~i$ & $-0.109785108528+0.186672894857 ~i$ \\
			$(1,4)$ & $-0.1296382022063-0.0019440351425 ~i$ & $-0.714459101836-0.266844800999 ~i$ \\
			$(2,5)$ & $-0.141336735577-0.007275501337 ~i$ & $-0.734482532105-0.139793827188 ~i$ \\
			$(1,5)$ & $-0.1197917884134-0.0010875223476 ~i$ & $-0.734380287566-0.217737495034 ~i$ \\
			\hline			
		\end{tabular}
		\vspace{3mm}
		\caption{WKB periods for the quintic potential \refeq{eq:quinticPot} computed using the $\epsilon$-functions obtained by solving the appropriate TBA system numerically using the iterative integration method with $3000$ Gaussian distributed points and cutoff $\in [-75,15]$.}
		\label{tab:QperiodsQuinticTBA}
	\end{center}
\end{table}

\begin{table}[tb]
	\begin{center}
		\begin{tabular}{|c|l|l|}\hline
			$\gamma$ & \multicolumn{1}{|c}{$\Pi_\gamma^{(1)}$}  &  \multicolumn{1}{|c|}{$\Pi_\gamma^{(2)}$} \\ \hline
			$(1,2)$ & $-0.063331399367+0.158332153648 ~i$ & $0.00263596614060+0.00300064152624 ~i$ \\
			$(2,3)$ &$-0.114195904626-0.247868478417 ~i$ & $-0.0378946345010-0.0190247965564 ~i$ \\
			$(3,4)$ &$-0.108813639847+0.253065980320 ~i$ & $-0.0283897195813+0.0210123204569 ~i$ \\
			$(4,5)$ &$-0.062412272985-0.161508879901 ~i$ & $0.00883111672023+0.00173626896691 ~i$ \\
			$(1,3)$ &$-0.177527303993-0.089536324769 ~i$ & $-0.0352586683604-0.0160241550301 ~i$ \\
			$(2,4)$ &$-0.223009544473+0.005197501903 ~i$ & $-0.0662843540822+0.0019875239005 ~i$ \\
			$(3,5)$ &$-0.171225912833+0.091557100418 ~i$ & $-0.0195586028611+0.0227485894238 ~i$ \\
			$(1,4)$ &$-0.286340943840+0.163529655551 ~i$ & $-0.0636483879416+0.0049881654267 ~i$ \\
			$(2,5)$ &$-0.285421817458-0.156311377999 ~i$ & $-0.0574532373620+0.0037237928674 ~i$ \\
			$(1,5)$ &$-0.348753216825+0.002020775650 ~i$ & $-0.0548172712214+0.0067244343936 ~i$ \\
			\hline
			$\gamma$ & \multicolumn{1}{|c}{$\Pi_\gamma^{(3)}$}  &  \multicolumn{1}{|c|}{$\Pi_\gamma^{(4)}$} \\ \hline
			$(1,2)$ & $0.0215449471560+0.0061879789883 ~i$ & $0.0001022445458-0.0779436677730 ~i$ \\
			$(2,3)$ & $-0.1188590023681-0.0485822988357 ~i$ & $-0.624697423623-0.326466722149 ~i$ \\
			$(3,4)$ & $-0.0323241469479+0.0404502847087 ~i$ & $-0.089863922924+0.137565589266 ~i$ \\
			$(4,5)$ & $0.00984641378954+0.00085651279437 ~i$ & $-0.0199211857393+0.0491073059514 ~i$ \\
			$(1,3)$ & $-0.0973140552121-0.0423943198474 ~i$ & $-0.624595179077-0.404410389922 ~i$ \\
			$(2,4)$ & $-0.151183149316-0.008132014127 ~i$ & $-0.714561346547-0.188901132883 ~i$ \\
			$(3,5)$ & $-0.0224777331583+0.0413067975031 ~i$ & $-0.109785108663+0.186672895218 ~i$ \\
			$(1,4)$ & $-0.1296382021600-0.0019440351387 ~i$ & $-0.714459102001-0.266844800656 ~i$ \\
			$(2,5)$ & $-0.141336735526-0.007275501333 ~i$ & $-0.734482532286-0.139793826931 ~i$ \\
			$(1,5)$ & $-0.1197917883704-0.0010875223443 ~i$ & $-0.734380287740-0.217737494704 ~i$ \\
			\hline
		\end{tabular}
		\vspace{3mm}
		\caption{WKB periods for the quintic potential \refeq{eq:quinticPot} computed using the differential operator method described in \ref{app:qperiods}. All the presented digits are exact.}
		\label{tab:QperiodsQuinticDiffOp}
	\end{center}
\end{table}

\subsection{The sextic potential}
For the next example, we will be looking at the triple well potential
\begin{equation}
\label{eq:sexticPot}
V(q) = q^2-2 q^4+ q^6
\end{equation}
For $0<E<4/27$, we are in the minimal chamber regime since we have 6 real turning points (energy below the wells). When $E>4/27$, we have two real turning points (the other four are complex and responsible for complex instantons) and we are in the maximal chamber (energy above the wells). We chose $E=1/16$ and $E=1$ in the examples below.
\subsubsection{Symmetric sextixc in the minimal chamber}
The (tri)chromatic TBA graph encoding the TBA system for the symmetric sextic potential is
\begin{center}
\begin{tikzpicture}
\coordinate[label = above:$1$] (1) at (0,0);
\coordinate[label = above left:$2$] (2) at (2,0);
\coordinate[label = above:$3$] (3) at (2,1);
\coordinate[label = above left:$4$] (4) at (6,1);
\coordinate[label = above:$5$] (5) at (6,2);
\coordinate[label = above:$6$] (6) at (8,2);

\draw[line width=0.3mm,black] (1) -- (2);
\draw[line width=0.3mm,blue] (2) -- (3);
\draw[line width=0.3mm,red] (3) -- (4);
\draw[line width=0.3mm,blue] (4) -- (5);
\draw[line width=0.3mm,black] (5) -- (6);

\node at (1)[circle,fill,inner sep=1.5pt]{};
\node at (2)[circle,fill,inner sep=1.5pt]{};
\node at (3)[circle,fill,inner sep=1.5pt]{};
\node at (4)[circle,fill,inner sep=1.5pt]{};
\node at (5)[circle,fill,inner sep=1.5pt]{};
\node at (6)[circle,fill,inner sep=1.5pt]{};
\end{tikzpicture}
\end{center}
furthermore, because of our choice of potential, we can add the additional symmetry ${2\epsilon_{1,2}=\epsilon_{3,4}}$ (red = 2 black in the TBA graph) such that we only have two independent $\epsilon$-functions to compute.\\

As in the previous examples, one can use \refeq{eq:WKBTBA} and compute the WKB periods. We found they are matching with the exact results , found for the sextic potential in \ref{app:qperiods}. We compared the quantum corrections to the periods to the order $O(\hbar^{12})$ and the results are matching within approximately 9 digits when using the iterative integration method with $12000$ Gaussian distributed points and cutoff $\in [-75,19]$. In order to not overcharge the present paper and because we already shown similar results, we do not provide the table of the WKB periods, the ultimate goal being the Voros spectrum.

As in the quartic case, the periods corresponding to cycles along the classically allowed regions \emph{i.e.} $(1,2)$ and $(3,4)$, are not Borel summable and must therefore be resummed using the same procedure as before. Using the results of the appendix \ref{app:EQC}, the EQC in terms of the lateral resummed periods is
\begin{equation}
\left(1+\mathcal{V}_{\text{p}}^{\pm 1}\right)^{2} \left(1+\mathcal{V}_{\text{p}}^{\pm 2}\right)+2\left(1+\mathcal{V}_{\text{p}}^{\pm 1}\right)\mathcal{V}_{\text{np}}+\mathcal{V}_{\text{np}}^2 =0
\end{equation}
where $\mathcal{V}_{\text{p}}^{\pm} = \mathcal{V}_{1,2}^{\pm} = \mathcal{V}_{5,6}^{\pm} $, $(\mathcal{V}_{\text{p}}^{\pm})^2 = \mathcal{V}_{3,4}$, $\mathcal{V}_{\text{np}} = \mathcal{V}_{2,3}= \mathcal{V}_{4,5}$ and $\mathcal{V}_{a,b}^{(\pm)} = e^{\frac{i}{\hbar} \mathcal{B}^{(\pm)}(\Pi_{a,b})}$. Proceeding as usual, \emph{i.e.} solving for $\mathcal{V}_{\text{p}}^{\pm}$ then rewriting an EQC for the median-resummed period, one finds the ``Zinn-Justin''-type EQC:
\begin{align}
\begin{split}
\label{eq:EQCsexticMinC}
\left(\cos\left(\frac{1}{\hbar}\mathcal{B}_{\text{med}}(\Pi_{1,2})\right) + \frac{1}{\sqrt{1+\left(e^{\frac{i}{\hbar} \mathcal{B}\left(\Pi_{2,3}\right)}\left(\sqrt{1+2e^{-\frac{i}{\hbar} \mathcal{B}\left(\Pi_{2,3}\right)}}-1\right)-1\right)^2}}\right) & \\
\left(\cos\left(\frac{1}{\hbar}\mathcal{B}_{\text{med}}(\Pi_{1,2})\right) - \frac{e^{-\frac{i}{\hbar} \mathcal{B}\left(\Pi_{2,3}\right)}}{\sqrt{2\left(1+e^{-\frac{i}{\hbar} \mathcal{B}\left(\Pi_{2,3}\right)}\right)\left(1+e^{-\frac{i}{\hbar} \mathcal{B}\left(\Pi_{2,3}\right)}+\sqrt{1+2 e^{-\frac{i}{\hbar} \mathcal{B}\left(\Pi_{2,3}\right)}}\right)}}\right)&= 0
\end{split}
\end{align}
Where the zeroes of the two factors are selecting the levels localized in the outer or inner wells. As in our previous examples, we can relate the Borel resummed periods to our $\epsilon$-functions using the Delabaere-Pham formula~\refeq{eq:regEpsilon}, which leads to relations of the type~\refeq{eq:BorelToEspsilonQuartic}. \\

Solving the EQC \refeq{eq:EQCsexticMinC} numerically yields the Voros spectrum presented in table \ref{tab:SpectrumSexticMinC}, together with the energies computed with the associated value of $\hbar$ using a standard technique in QM.
\begin{table}[tb]
	\begin{center}
		\begin{tabular}{|l|c|c|}\hline
			&$\theta_1=2.2550882179631$ & $\theta_2=3.0440517343238$ \\
			\hline
			$E_1$&$0.99999999998162$ & $0.50911268535197$  \\
			$E_2$&$1.8538029210438$ & $0.99999999999583$  \\
			$E_3$&$2.0241786738382$ & $1.0001945622010$  \\
			$E_4$&$3.0907877329610$ & $1.4538807749127$  \\
			\hline
			&$\theta_3=3.0442659923975$& $\theta_4=3.4612023941484$\\
			\hline
			$E_1$&$0.50901079803411$ & $0.34265110005140$ \\
			$E_2$& $0.99980601394390$ & $0.67835953623940$ \\
			$E_3$& $0.99999999999504$ & $0.67835966437968$ \\
			$E_4$&$1.4536124515081$ & $0.99999999999840$ \\
			\hline
		\end{tabular}
		\caption{Voros Spectrum of the $\theta_n$ solving the EQC  \refeq{eq:EQCsexticMinC}, together with the -- normalized to one -- control spectrum of energies obtained for these values of $\hbar_n=\exp(-\theta_n)$ using the method described in the appendix \ref{app:CDil} on the potential \refeq{eq:sexticPot}. The $\epsilon$-functions involved are obtained numerically using the iterative integration method with $12000$ Gaussian distributed points and cutoff $\in [-75,19]$. Since the energy (as a WKB curve parameter) was chosen too be $E=1/16$, the diagonal elements of this table should be $16 E=1$ according to \refeq{eq:checkE}. One can observe the non-perturbative spliting of the levels 2 and 3 mod 4, corresponding to the energies of states localized in the outer wells, as one can observe from the degeneracies of the EQC \refeq{eq:EQCsexticMinC} when neglecting the non-perturbative contribution, or from the analysis in \cite{Basar:2017hpr,Dunne:2020gtk}.}
		\label{tab:SpectrumSexticMinC}
	\end{center}
\end{table}

\subsubsection{Symmetric sextic in the maximal chamber}
The TBA system for the symmetric sextic in the maximal chamber can be read from the figure \ref{fig:TBAmaxCsym}.\\

As usual, one can use \refeq{eq:WKBTBA} and compute the WKB periods. For the same reasons as in our precedent examples, we will not provide such big tables in the present paper. The quantum corrections obtained using the TBA procedure are reproducing the exact WKB periods of appendix \ref{app:qperiods} to approximately 10 digits of precision when using the iterative integration method with $4000$ Gaussian distributed points and cutoff $\in [-75,16]$ and up to order $O(\hbar^{12})$.\\

As in the maximal quartic case, the period $\Pi_{1,6}$ is not Borel summable. We must therefore follow the same procedure we went through before. Using the results of the appendix \ref{app:EQC}, one can find the EQC for the sextic potential in the maximal chamber as a function of the lateral resummed periods. For compactness, we express the EQC in term of the Voros multiplier-like quantities $\mathcal{V}_{\text{p}}^\pm$ and $\mathcal{V}_{\text{np}}$:
\begin{equation}
	\left(\mathcal{V}_{\text{p}}^{\pm 1}+\mathcal{V}_{\text{np}}\right)^{2}\left(\mathcal{V}_{\text{p}}^{\pm 2}+\mathcal{V}_{\text{np}}^2\right)+2\left(\mathcal{V}_{\text{p}}^{\pm 1}+\mathcal{V}_{\text{np}}\right)\mathcal{V}_{\text{np}}+1 = 0
\end{equation}
where $\mathcal{V}_{\text{p}}^{\pm} = (\mathcal{V}^{\pm}_{1,2})^{\frac{1}{4}} = (\mathcal{V}^{\pm}_{5,6})^{\frac{1}{4}} = (\mathcal{V}^{\pm}_{3,4})^{\frac{1}{4}}$, $\mathcal{V}_{\text{np}} = \mathcal{V}^{\frac{1}{2}}_{2,3} = \mathcal{V}^{\frac{1}{2}}_{4,5}$ and $\mathcal{V}_{a,b}^{(\pm)} = e^{\frac{i}{\hbar} \mathcal{B}^{(\pm)}(\Pi_{a,b})}$. Proceeding as usual, i.e. solving for $\mathcal{V}_{\text{p}}^{\pm}$ then rewriting an EQC for the median-resummed period, one finds the ``Zinn-Justin''-type EQC for the sextic potential in the maximal chamber:
\begin{equation}\label{eq:EQCsexticMaxC}
\cos\left(\frac{1}{\hbar}\mathcal{B}_{\text{med}}(\Pi_{1,6})\right) + \frac{1-2e^{-\frac{i}{\hbar} \mathcal{B}\left(\Pi_{2,3}\right)}-e^{-\frac{2i}{\hbar} \mathcal{B}\left(\Pi_{2,3}\right)}}{\left(1+e^{-\frac{i}{\hbar} \mathcal{B}\left(\Pi_{2,3}\right)}\right)^2}= 0
\end{equation}
One can relate the resummed periods in the EQC \eqref{eq:EQCsexticMaxC} and our $\epsilon$-functions similarly to the quartic potential in the maximal chamber case, except it is involving a more complicated TBA system. Explicitly, we get
\begin{align}
\begin{split}
\frac{1}{\hbar} \mathcal{B}_{\text{med}}\left(\Pi_{1,4}\right)(\theta)=&
\abs{\Pi_{1,6}} \exp(\theta) \\
&- \frac{i}{\pi} \int_{\mathbb{R}} \frac{L_{1,2}(\bar{\theta})+L_{3,4}(\bar{\theta})}{\cosh\left(\theta-\bar{\theta}-i \varphi_{1,2}\right)}d\bar{\theta} + \frac{i}{\pi} \int_{\mathbb{R}} \frac{L_{1,3}(\bar{\theta})+L_{2,5}(\bar{\theta}) }{\cosh\left(\theta-\bar{\theta}+i \varphi_{1,2}\right)}d\bar{\theta}\\\\
&-\frac{i}{\pi} \int_{\mathbb{R}} \frac{L_{1,4}(\bar{\theta}) d\bar{\theta}}{\cosh\left(\theta-\bar{\theta}-i \varphi_{1,4}\right)}+\frac{i}{\pi} \int_{\mathbb{R}} \frac{L_{1,5}(\bar{\theta}) d\bar{\theta}}{\cosh\left(\theta-\bar{\theta}+i \varphi_{1,4}\right)}\\
&-\frac{2}{\pi}\mathcal{P}\left(\int_{\mathbb{R}} \frac{L_{2,3}(\bar{\theta})d\bar{\theta}}{\sinh(\theta-\bar{\theta})}\right)
\\
\frac{i}{\hbar} \mathcal{B}\left(\Pi_{2,3}\right)(\theta)=&~\epsilon_{2,3}\left(\theta\right)
\end{split}
\end{align}
where we used $\varphi_{1,2}=\varphi_{3,4}= -\varphi_{1,3}= -\varphi_{2,5}$, $\varphi_{1,4}=-\varphi_{1,5}$, $\varphi_{2,4}=\varphi_{1,6}=0$. Solving the EQC \refeq{eq:EQCsexticMaxC} numerically yields the Voros spectrum presented in table \ref{tab:SpectrumSexticMaxC}.\\

\begin{table}[tb]
	\begin{center}
		\begin{tabular}{|l|c|c|}\hline
			&$\theta_1=-0.44704899967438$ & $\theta_2=0.28554431921498$ \\
			\hline
			$E_1$&$0.99999999996147$ & $0.30434860059971$  \\
			$E_2$&$3.7248850413244$ & $0.99999999998805$  \\
			$E_3$&$8.0432344259844$ & $2.0812165495837$  \\
			$E_4$&$13.782680825803$ & $3.6165604373697$  \\
			\hline
			&$\theta_3=0.83651755572084$& $\theta_4=1.1554539882629$\\
			\hline
			$E_1$&$0.17975871852441$ & $0.13025425612694$ \\
			$E_2$& $0.51313989900212$ & $0.32321423269094$ \\
			$E_3$& $0.99999999999801$ & $0.57486198218456$ \\
			$E_4$&$1.7447971706055$ & $0.99999999999894$ \\
			\hline
		\end{tabular}
		\caption{Voros Spectrum of the $\theta_n$ solving the EQC  \refeq{eq:EQCsexticMaxC}, together with the control spectrum of energies obtained for these values of $\hbar_n=\exp(-\theta_n)$ using the complex dilatation method described in the appendix \ref{app:CDil} on the potential \refeq{eq:sexticPot}. The $\epsilon$-functions involved are obtained numerically using the iterative integration method with $4000$ Gaussian distributed points and cutoff $\in [-75,16]$. Since the energy (as a WKB curve parameter) was chosen too be $E=1$, the diagonal elements of this table should be $E=1$ according to \refeq{eq:checkE}.}
		\label{tab:SpectrumSexticMaxC}
	\end{center}
\end{table}

%

\section{Conclusion and outlook}
\label{chap:conclusion}

After reviewing briefly the theory underlying the present work in section \ref{chap:WKB&TBA}, we presented in section \ref{chap:wallCrossingDiagram} a very useful diagrammatic method that is allowing us to analytically continue a TBA system into a more complicated one (wall-crossing), inspired by Toledo's thesis \cite{ToledoJonathan2016} in the context of minimal surfaces in $AdS$. In the next couple of sections, we used this diagramatic method in order to obtain some general results, especially for maximal chamber configurations for which we cannot wall-cross anymore. Indeed, in this section, we uncovered the general structure of TBA systems in the maximal chamber and explicitly computed the associated intersection matrix (encoding the TBA equations) for arbitrary degree polynomial potentials. In section \ref{chap:simplifyTBA} and \ref{chap:purePot}, we simplified the general TBA system using the symmetries of the potential. In particular, in \ref{chap:purePot}, we are putting an emphasis on pure polynomial potential of the form $V(q) = a q^d$, preparing for the section \ref{chap:DTasSpecialCase}, where we are giving some very strong arguments indicating that Dorey and Tateo TBA equations, appearing in the seminal paper \cite{Dorey:1998pt}, are indeed a special case of the TBA equations in \cite{Ito:2018eon}. In section \ref{chap:res}, we applied the theory developed previously to interesting Quantum Mechanical problems (from the cubic oscillator to the sextic triple well). We were able to verify numerically it was matching with alternative technique to a great accuracy for every examples, demonstrating that the TBA equations are not only a powerful and exact analytic tool, but are also providing useful computational techniques. We derived the exact quantification conditions in term of the (lateral then, when imposed by the lack of Borel summability, median) Borel resummed periods, and together with the TBA results, we were able to solve very subtle spectral problems.

Of course, the techniques we presented in this paper can be used in the context of Quantum Mechanics in order to solve a large class of spectral problems. But, as a side effect, they can also be useful in holography. Indeed, one can reverse the conformal limit in figure \ref{fig:TBAdiagram} by substituting ${\exp(\theta)\mapsto 2\cosh(\theta)}$ in the mass/period term, hence obtaining TBA equations of the form
\begin{equation}
\label{eq:ultimateTBAAMSV}
\tilde{\epsilon}_{(a)}(\theta) = 2 \abs{Z_{(a)}}\cosh(\theta) +\sum_{(b) \in S_d} K_{(a),(b)}\star \tilde{L}_{(b)}(\theta)
\end{equation}
which is the TBA system governing the area of minimal surfaces in $AdS_3$ delimited by a polygonal closed contour (with $2d+4$ sides) as described in \cite{Alday:2009yn,Alday:2010vh,Toledo:2014koa,ToledoJonathan2016}.\\

Concerning the unresolved issues, there is still at least a small one in this paper. We found the exact TBA system for pure polynomial potentials of arbitrary degree, and it should in principle be equivalent to the Dorey and Tateo TBA system in \cite{Dorey:1998pt}. We made a lot of progress in order to make that equivalence clear by boiling it down to a pure mathematical identity. We provided strong evidence that the conjecture of equivalence was holding, by systematically testing it up to large degrees; but there is still a lack of a definitive proof that the identity $\mathcal{K}_{[k],[l];d}(\theta) =\mathcal{K}^{\text{DT}}_{[k],[l];d}(\theta)$ holds in order to close this matter definitively.

An obvious amelioration we could bring to the method described in this paper would be to extend its validity outside polynomial potentials. The original ODE/IM correspondence in \cite{Dorey:1998pt,Dorey:2007zx} was already accommodating for an angular momentum term $\propto 1/q^2$ for example. Accommodating this particular centrifugal term in the context of the quantum mechanical generalized ODE/IM correspondence was done recently in \cite{Ito:2019llq}. In \cite{Hollands:2019wbr}, they are using the integral equation in \cite{Gaiotto:2014bza} (that can be interpreted as a generalization of the ODE/IM correspondence) in order to accomodate for the Mathieu (and modified Mathieu) equation (i.e. with $V(q)\propto\cos(q)$ and $V(q)\propto\cosh(q)$ respectively) in the context of abelianization. A related analysis was carried out in \cite{Grassi:2019coc}, where they are explicitly deriving TBA equations for the modified Mathieu operator, notably in the resurgent framework presented in this paper. Each time we are extending the generalized ODE/IM correspondence to even more complicated potentials, we are getting closer to the exact resolution of the general Schrödinger equation \eqref{eq:Schro}.\\

\section*{Acknowledgments}
We would like to thank Jie Gu for enlightening discussions and especially Marcos Mariño for many helpful comments and careful reading of the draft. Many of the computations reported in this paper were performed at University of Geneva on the Baobab cluster. This work is supported by the Fonds National Suisse, subsidies 200020-175539.

\appendix

\section{Solving the TBA equations numerically}
\label{app:num}

In this section, we will discuss the two main methods we used in order to solve the TBA integral equations numerically and extract the $\epsilon$-functions. The basic idea in both of the methods is to start from the asymptotic exponential behavior at $\theta\to \infty$,
\begin{equation}
\epsilon_{a}^{(0)}(\theta) = \abs{\Pi_{a}}\exp(\theta)
\end{equation}
and integrate the functions iteratively according to the recurrence relation
\begin{equation}
\label{eq:rec1TBA}
\epsilon_{a}^{(n)}(\theta) = \abs{\Pi_{a}}\exp(\theta) +\sum_{b} K_{a,b}\star L_{b}^{(n-1)}(\theta)
\end{equation}
where $L_{a}^{(n)}(\theta) = \log(1+\exp(-\epsilon_{a}^{(n)}(\theta)))$. For $d\leq 4$, \emph{i.e.} potentials defining a WKB curve \refeq{eq:WKBcurve} the Riemann surface of which has genus $0$ or $1$, the $\epsilon^{(n)}$-functions are converging to the exact $\epsilon$-functions in the $n\to\infty$ limit. For genus $>1$ however, \refeq{eq:rec1TBA} is converging to 2 different values according to the parity of $n$. Thus, one can use the modified recurrence relation 
\begin{equation}
\label{eq:rec2TBA}
\epsilon_{a}^{(n)}(\theta) = \frac{1}{2}\left(\epsilon_{a}^{(n-1)}(\theta)+\abs{\Pi_{a}}\exp(\theta) +\sum_{b} K_{a,b}\star L_{b}^{(n-1)}(\theta)\right)
\end{equation}
in order to mix even and odd parity and make the $\epsilon^{(n)}$-functions converge to the appropriate $\epsilon$-functions in the $n\to\infty$ limit. Note that \refeq{eq:rec2TBA} is converging more slowly than \refeq{eq:rec1TBA}.

\subsection{Fourier discretization}

A convenient way to compute convolutions numerically is to use the discrete Fourier transform as in \cite{Ito:2018eon}. For this purpose, we discretize our functions and kernels into finite dimensional vectors in an interval (defined by a cutoff) then Fourier transform the resulting vectors, knowing that convolutions are transforming into products under Fourier transformation according to the convolution theorem, \emph{i.e.}
\begin{equation}
\mathcal{F}(f\star g) = \mathcal{F}(f)\cdot \mathcal{F}(g) 
\end{equation}
We get back the discretized version of the convolution function by inverse Fourier transform. In the limit where the number of points and cutoff go to infinity, we recover the exact functions.

However, this method has some shortcomings. Indeed, the discrete convolutions are forced to be periodic functions, which is introducing errors around the cutoff. Also, in our implementation of the algorithm, we need a power of 2 number of uniformly distributed sampling points to get the correct phase. There surely are some fancy ways to circumvent these problems. For example, it is possible to fix the behavior at $\theta\to -\infty$ by adding to the TBA equation of the form
\begin{equation}
	\epsilon(\theta) = f_0(\theta)-\frac{1}{2\pi} \int_{\mathbb{R}} \frac{L_0(\bar{\theta}) d\bar{\theta}}{\cosh(\bar{\theta}-\theta)}
\end{equation}
the (by construction) null term
\begin{equation}
\label{eq:zeroterm}
0=f_1(\theta)-\frac{1}{2\pi} \int_{\mathbb{R}} \frac{L_1(\bar{\theta}) d\bar{\theta}}{\cosh(\bar{\theta}-\theta)}
\end{equation}
Solving \refeq{eq:zeroterm} yields the condition $ L_1(\theta) = (f_1(\theta+i \pi/2)+f_1(\theta-i \pi/2))/2$. Furthermore, we want to impose the asymptotic behavior $f_0(\theta)+f_1(\theta) = \epsilon_\star$ at large negative $\theta$, where $\epsilon^\star$ is the know value of the the $\epsilon$-function at $\theta \to -\infty$\footnote{It is easy to show that the $Y$-functions are obeying a functional equation at large negative $\theta$ which solves to $Y_a^\star = \frac{\sin(\pi a/(d+2))\sin(\pi (a+2)/(d+2))}{\sin[2](\pi/(d+2))}$, and $\epsilon_a^\star = -\log(Y_a^\star)$, where $d$ is the degree of the polynomial. See \cite{Alday:2010vh,Ito:2018eon}.}, and $f_0(\theta)+f_1(\theta)=\abs{\Pi}\exp(\theta)$ at large positive $\theta$. Since $f_0(\theta)=\abs{\Pi}\exp(\theta)$ already, we must have $f_1(\theta\to\infty) \to~0$ and $f_1(\theta\to-\infty)~\to~\epsilon^\star$. In other words, we can chose any ``step''-like function we please for $f_1$ as long as it is going to 0 at positive infinity and $\epsilon^\star$ at negative infinity. For example, if we chose the sigmoid function $f_1(\theta)=\epsilon^\star/(1+c \exp(\theta))$, it implies that the appropriate kernel is $L_1(\theta)=2\epsilon^\star/(1+c^2 \exp(2\theta))$.

\subsection{Numeric integration and interpolation}
Another way to solve the TBA equations numerically and efficiently is to sample the convolution $(K\star L) (\theta)$ at some discrete set of points by integrating numerically, then interpolate the resulting values in order to build the $\epsilon$-function. This can be easily implemented in \verb+Mathematica+ using the \verb+NIntegrate+ and \verb+Interpolation+ build-in functions. Just like the previous example, we can introduce a cutoff and a number of points, and the numerical  $\epsilon$-function are converging to the exact ones in the large cutoff and number of points limit. See for \cite{Alday:2010vh,Gromov:2008gj} for examples. In this paper, we used similar code, but with two little twists. First, because of the exponential behavior of the $\epsilon$-function at large $\theta$: it is very time consuming to extend the cutoff in the positive theta direction. In order to optimize the algorithm, it is then interesting to have two different cutoff, in the negative and positive direction. Extending the cutoff in the negative direction does not cost too much additional time. Finally, we can distribute our sample points anyway we want with this method. Since the extremities (near the cutoff) are contributing less than the center (near $\theta=0$), we decided to distribute our points according to a Gaussian distribution. In the main text, we will refer to this method as the \emph{Iterative Integration} or II method.

\subsection{$\epsilon$-functions in the complex plane}
Using the methods mentioned above, one gets a function interpolated on the real line. For some computations, it is useful to have access to complex arguments, analytically continuing these data in the full complex plane. A brute force solution would be to interpolate the $\epsilon$-functions on a grid, but this is very costly. Fortunately, it is unecessary and quite easy to analytically continue these data in the complex $\theta$ plane by simply using the associated TBA equation and integrate at shifted kernels. This is in fact what we are doing in the main text when we are evaluating $\epsilon(\theta+i\pi/2)$ when we want to relate the $\epsilon$-functions and the resummed WKB periods. Explicitly,
\begin{equation}
\epsilon_{(a)}(\theta_R+i \theta_I) = \abs{\Pi_{(a)}}\exp(\theta_R+i \theta_I) +\sum_{(b)} \int_\mathbb{R} K_{(a),(b)}(\theta_R+i\theta_I) L_{(b)}(\theta \in \mathbb{R}) d\theta
\end{equation}
such that we only need the knowledge of the function $L_{(b)} = \log(1+\exp(-\epsilon_{(b)}))$ on the real line, at the cost of an integration. For some values of $\theta_I$, one can encounter singularities. It is possible to avoid them using the Cauchy principal value.

\begin{figure}
	\center
	\includegraphics[width=0.88\textwidth]{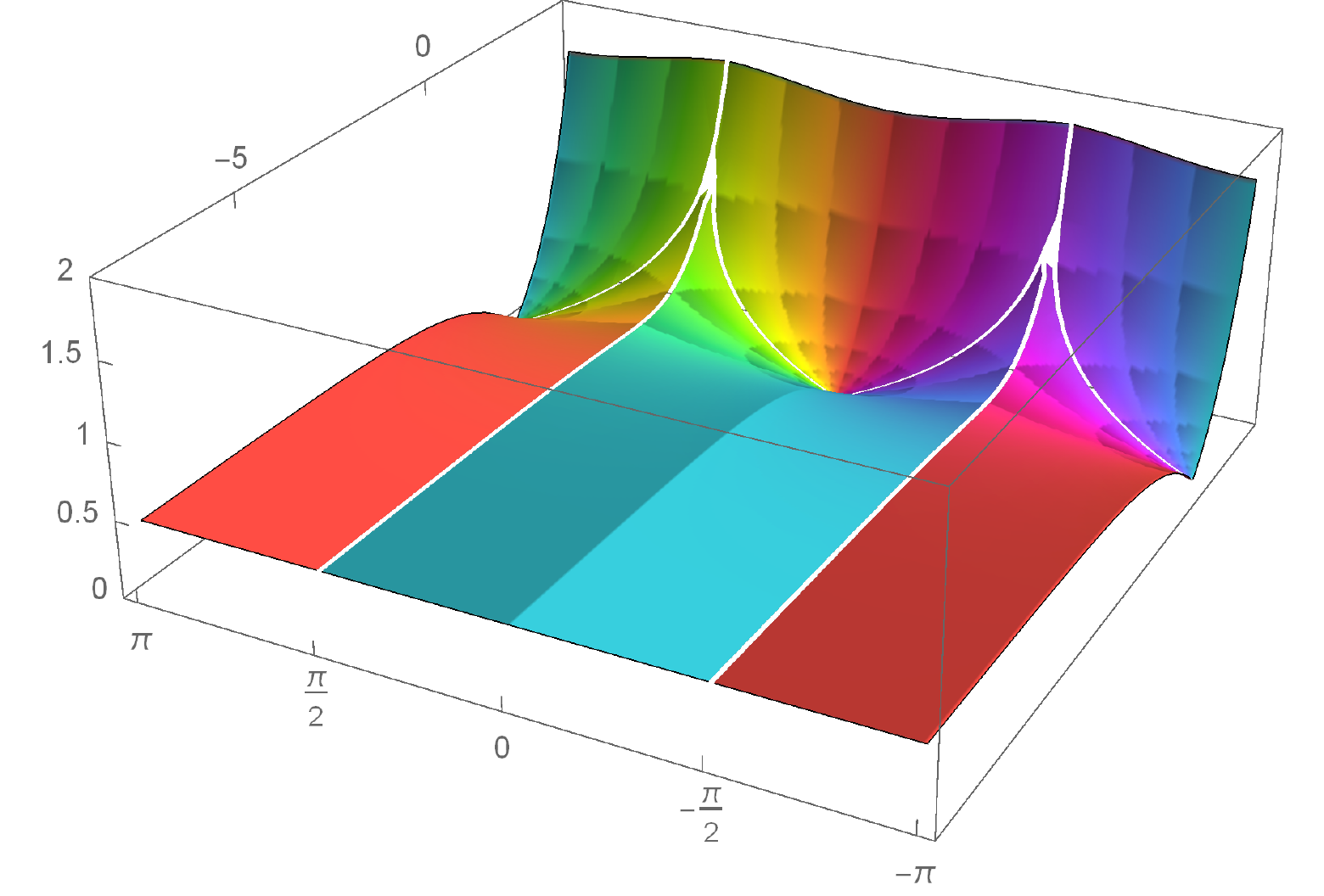}
	\includegraphics[width=0.11\textwidth]{Figures/legendCplot}
	\caption{Typical form of the $\epsilon$-functions in the complex plane.}
	\label{fig:epscf}
\end{figure}

The typical form of these analytically continued $\epsilon$-functions are drawn in figure \ref{fig:epscf}.

\section{Deriving the Exact Quantization Conditions}
\label{app:EQC}
Using connection formulae -- arising from the necessity of having recessive or dominant WKB functions in different regions of the complex $q$-plane, together with the need to connect these regions in a coherent way -- it is possible to derive powerful and exact quantization conditions. One viable road for deriving EQC is to use Voros' matrix approach, known as the Knoll-Schaeffer connection method, later refined by Delabaere, Dillinger and Pham \cite{voros1981spectre,AIF_1993__43_1_163_0,doi:10.1063/1.532206,AIHPA_1999__71_1_1_0}, looking at the Stokes lines (i.e. the codimension one loci on the complex q-plane defined by $\Re(p)=0$) in order to find the EQC of interest. However, in this paper, we used an equivalent method: the Voros-Silverstone connection formula described in \cite{MarcosAQM}, which we are going to summarize (with much less detail) in the following section.

\subsection{The Voros-Silverstone connection formula}

Near a turning point $q_\text{tp}$ (defined by $p(q_\text{tp})=0$ or equivalently $V(q_\text{tp})=E$), the potential appearing in the Schrödinger equation can be approximated as
\begin{equation}
V(q) - E = -k(q-q_\text{tp})
\end{equation}
and $q_\text{tp}$ is a first order turning point if $k\neq 0$, which will be assuming to be the case in the following. Also, we will assume that the classicaly allowed region (where $V(q)<E$) is to the right of the turning point (i.e. $q>q_\text{tp}$) when the classically forbidden region is to the right of the turning point, such that $k$ is positive. Locally, the Schrödinger equation is corresponding to a linear potential problem. This problem has an exact solution: the wavefunction is a linear combination of the Airy functions
\begin{equation}
\psi(q) = \tilde{a} \operatorname{Ai}\left(-\hbar^{-\frac{2}{3}} \phi(q)\right)+\tilde{b} \operatorname{Bi}\left(-\hbar^{-\frac{2}{3}} \phi(q)\right)
\end{equation}
where $\phi(q)=p(q)^2 = \sqrt[3]{2 m k}(q-q_\text{tp})$. Of course, since $\phi(q)$ is linear, we can also write
\begin{equation}
\label{eq:unifWKBlin}
\psi(q) =\frac{1}{\sqrt{\phi'(q)}}\left( a \operatorname{Ai}\left(-\hbar^{-\frac{2}{3}} \phi(q)\right)+b \operatorname{Bi}\left(-\hbar^{-\frac{2}{3}} \phi(q)\right)\right)
\end{equation}
in order to prepare the wavefunction for the following. Also, doing so, we are taking this problem in the more general context of the uniform WKB method, where we use the ansatz 
\begin{equation}
\label{eq:WKBansatz3}
\psi(q) = \frac{1}{\sqrt{\phi'(q)}}f\left(\phi(q)\right)
\end{equation}
with $f$ and $\phi$ are constrained by a second order differential equation and a non-linear differential equation respectively instead of the standard and less flexible WKB ansatz \refeq{eq:WKBansatz2}. See \cite{MarcosAQM} for the detail. In the classically allowed region, $E>V(q)$ such that $\phi(q)>0$ and the argument of the Airy functions in \refeq{eq:unifWKBlin} is negative. The asymptotic expansion of \refeq{eq:unifWKBlin} is in that case, up to an overall constant,
\begin{equation}
\psi(q) \sim \frac{1}{\sqrt{\phi'(q)}\phi(q)^{\frac{1}{4}}} \left(
(b-i a)e^{\frac{i \pi}{4}} \beta\left(\frac{2 i}{3 \hbar} \phi^{\frac{3}{2}}(q)\right)+(b+i a)e^{-\frac{i \pi}{4}} \beta\left(-\frac{2 i}{3 \hbar} \phi^{\frac{3}{2}}(q)\right)
\right)
\end{equation}
where
$$\beta(x) = e^x \sum_{k \geq 0} c_k x^{-k} \quad\text{and}\quad c_k = \frac{1}{2\pi}  \frac{\Gamma\left(k+\frac{1}{6}\right)\Gamma\left(k+\frac{5}{6}\right)}{2^k k!} $$
as shown in \cite{MarcosAQM}. We can proceed mutatis mutandis in the classically forbidden region, where $E<V(q)$ and the asymptotic expansion reads
\begin{equation}
\psi(q) \sim \frac{1}{\sqrt{\phi'(q)}(-\phi(q))^{\frac{1}{4}}} \left(
2 b \beta\left(\frac{2}{3 \hbar}(- \phi^{\frac{3}{2}}(q))\right)+(a\pm i b) \beta\left(-\frac{2}{3 \hbar} (-\phi^{\frac{3}{2}}(q))\right)
\right)
\end{equation}
where the choice of $\pm$ is corresponding to the choice of lateral Borel resummation as explained in \cite{MarcosAQM}. Noticing that 
\begin{equation*}
\frac{1}{\sqrt{\phi'(q)}\phi(q)^{\frac{1}{4}}} \,\beta\left(\pm\frac{2 i}{3 \hbar} \phi^{\frac{3}{2}}(q)\right) = \frac{1}{\sqrt{P(q)}}\exp\left(\pm \frac{i}{\hbar} \int^{q}_{q_{\text{tp}}} P(\bar{q}) d\bar{q}\right) = \frac{1}{\sqrt{P(q)}}\exp\left(\pm \frac{i}{\hbar} \Pi_{q_{\text{tp}},q}\right)
\end{equation*}
with $P(q)$ defined in the WKB context presented in the section \ref{chap:WKB&ResurgentQM},  defined as a formal power series in \refeq{eq:formalSeriesP}, one can restate the two previous asymptotic expansions in a WKB friendly manner. In the classically allowed region,
\begin{equation}
\label{eq:part1VS}
\psi(q) \sim \frac{1}{\sqrt{P(q)}} \left(
(b-i a)e^{\frac{i}{\hbar}\Pi_{q_{\text{tp}},q}+\frac{i \pi}{4}} +(b+i a) e^{-\frac{i}{\hbar}\Pi_{q_{\text{tp}},q}-\frac{i \pi}{4}}
\right)
\end{equation}
In the classically forbidden region, one has to be careful with the choice of signs and branch cut implementation. The resulting asymptotic expansion is
\begin{equation}
\label{eq:part2VS}
\psi(q) \sim \frac{1}{\sqrt{\tilde{P}(q)}} \left(
2b\,e^{\frac{1}{\hbar}\tilde{\Pi}_{q,q_{\text{tp}}}} +(a\pm i b) e^{-\frac{1}{\hbar}\tilde{\Pi}_{q,q_{\text{tp}}}}
\right)
\end{equation}
where $\tilde{P}(q)$ is obtained from $P(q)$ by a multiplication of $\pm i$ (it is of the form $\tilde{P}(q) = \tilde{p}(q) + O(\hbar^2)$ with $\tilde{p}(q)= \sqrt{2m(V(q)-E)}$ when the standard momentum is $p(q)= \sqrt{2m(E-V(q))}$ as in the main text) and $\tilde{\Pi}_{a,b} = \int_{a}^{b}\tilde{P}(q) dq$. Putting \refeq{eq:part1VS} and \refeq{eq:part2VS} together, we get the \emph{Voros-Silverstone connection formula}
\begin{equation}
\label{eq:VSforbToAllab}
\begin{gathered}
\frac{1}{\sqrt{P(q)}} \left(
(b-i a)e^{\frac{i}{\hbar}\Pi_{q_{\text{tp}},q}+\frac{i \pi}{4}} +(b+i a) e^{-\frac{i}{\hbar}\Pi_{q_{\text{tp}},q}-\frac{i \pi}{4}}
\right)\\
\xleftarrow{q>q_{\text{tp}}} \psi(q) \xrightarrow{q<q_{\text{tp}}}\\
\frac{1}{\sqrt{\tilde{P}(q)}} \left(
2b\,e^{\frac{1}{\hbar}\tilde{\Pi}_{q,q_{\text{tp}}}} +(a\pm i b) e^{-\frac{1}{\hbar}\tilde{\Pi}_{q,q_{\text{tp}}}}
\right)
\end{gathered}
\end{equation}
that is allowing us to connect the wavefunction before and after the crossing of a turning point in the case where we are going from the classically forbidden region to the classically allowed region when increasing $q$, as assumed earlier. We can repeat this analysis verbatim in the case we are going from the classically allowed region to the classically forbidden region, and one finds a similar formula:
\begin{equation}\label{eq:VSallToForbab}
\begin{gathered}
\frac{1}{\sqrt{P(q)}} \left(
(b-i a)e^{\frac{i}{\hbar}\Pi_{q,q_{\text{tp}}}+\frac{i \pi}{4}} +(b+i a) e^{-\frac{i}{\hbar}\Pi_{q,q_{\text{tp}}}-\frac{i \pi}{4}}
\right)\\
\xleftarrow{q<q_{\text{tp}}} \psi(q) \xrightarrow{q>q_{\text{tp}}}\\
\frac{1}{\sqrt{\tilde{P}(q)}} \left(
2b\,e^{\frac{1}{\hbar}\tilde{\Pi}_{q_{\text{tp}},q}} +(a\pm i b) e^{-\frac{1}{\hbar}\tilde{\Pi}_{q_{\text{tp}},q}}
\right)
\end{gathered}
\end{equation}

\subsection{Worked out example: the quartic potential}

Let us illustrate the general method for deriving the EQC used in the present paper by explicitly finding the EQC for the quartic case in the minimal chamber (double well with energy bellow the wells). Then, we will extend this result to the maximal chamber configuration (double well with energy above the wells and, also, pure quartic) by deforming the path of the cycles, transforming the periods in order to accommodate the EQC obtained using the Voros-Silverstone formula. The method presented below for the quartic example can of course be generalized without too much effort to more complicated potentials.

\subsubsection{Minimal chamber}
In the minimal chamber, we can order the turning points that are all along the real line by $q_1<q_2<q_3<q_4$. This is defining five regions, $I = ]-\infty,q_1[$, $II=]q_1,q_2[$, $III=]q_2,q_3[$, $IV=]q_3,q_4[$ and $V=]q_4,\infty [$. The wavefunction must decay at infinity. It fix the behavior of $\psi(q)$ in the region $I$ and $V$:
\begin{align}
\psi_I(q)=& \frac{A}{\sqrt{\tilde{P}(q)}} e^{-\frac{1}{\hbar}\tilde{\Pi}_{q,q_1}} \label{eq:psiI}\\
\psi_V(q)=&\frac{B}{\sqrt{\tilde{P}(q)}} e^{-\frac{1}{\hbar}\tilde{\Pi}_{q_4,q}} \label{eq:psiV}
\end{align}
where $A$ and $B$ are normalization constants. \refeq{eq:psiI} and \refeq{eq:psiV} are simply the wavefunction in the classically forbidden region described in \refeq{eq:VSforbToAllab} and \refeq{eq:VSallToForbab} respectively, $q_{\text{tp}} = q_1$ and $q_4$ respectively, $a=A$ and $B$ respectively, and $b=0$  in both cases. Now, our goal is clear: we want to connect \refeq{eq:psiI} and \refeq{eq:psiV} together by the repeated use of \refeq{eq:VSforbToAllab} and \refeq{eq:VSallToForbab}, alternating between classically allowed and forbidden region until we have two expression for the same wavefunction. We have multiple possible equivalent way to do that. We could straightforwardly alternate between \refeq{eq:VSforbToAllab} and \refeq{eq:VSallToForbab} and get \eqref{eq:psiI} from the region $I$ to the region $V$ for example (which also yields the EQC \refeq{eq:appendixEQCminC}, as it should be). In this notes, we chose to go from $I$ to $III$ and from $V$ to $III$ for \refeq{eq:psiI} and \refeq{eq:psiV} respectively, meeting in the middle. Using \refeq{eq:VSforbToAllab}, we are going from $I$ to $II$ and we find
\begin{align}
\psi_{II}(q)&= \frac{i A}{\sqrt{P(q)}} \left(
 e^{-\frac{i}{\hbar}\Pi_{q_1,q}-\frac{i \pi}{4}}-e^{\frac{i}{\hbar}\Pi_{q_1,q}+\frac{i \pi}{4}}
\right)\\
&= \frac{i A}{\sqrt{P(q)}} \left(
e^{\frac{i}{\hbar}\Pi_{q,q_2}-\frac{i}{\hbar}\Pi_{q_1,q_2}-\frac{i \pi}{4}}-e^{-\frac{i}{\hbar}\Pi_{q,q_2}+\frac{i}{\hbar}\Pi_{q_1,q_2}+\frac{i \pi}{4}}
\right)
\end{align}
In the last step, using $\Pi_{a,c} = \Pi_{a,b}+\Pi_{b,c}$, we made the exponentiated period $\Pi_{1,2}=\Pi_{q_1,q_2}$ appear explicitly, which we could rewrite as a Voros multiplier \eqref{eq:VorosMultiplier}. It is a general fact that these EQC are in general functions of these exponentiated periods, as stated in \eqref{eq:genericEQC}. Now, we are in the classically allowed region $II$ and we want to cross $q_2$ ending up up in the classicaally forbidedn region $III$. The appropriate connection formula is \refeq{eq:VSforbToAllab}; setting $a= A \sin\left(\frac{1}{\hbar}\Pi_{1,2}\right)$ and $b= A \cos\left(\frac{1}{\hbar}\Pi_{1,2}\right)$ yields
\begin{align}
\psi_{III}(q)&= \frac{A}{\sqrt{\tilde{P}(q)}} \left(
2 \cos\left(\frac{1}{\hbar}\Pi_{1,2}\right) \,e^{\frac{1}{\hbar}\tilde{\Pi}_{q_{2},q}} \pm i e^{\mp \frac{i}{\hbar} \Pi_{1,2}} e^{-\frac{1}{\hbar}\tilde{\Pi}_{q_{2},q}}
\right)\\
&= \frac{A}{\sqrt{\tilde{P}(q)}} \left(
2 \cos\left(\frac{1}{\hbar}\Pi_{1,2}\right) e^{\frac{1}{\hbar}\tilde{\Pi}_{2,3}} e^{-\frac{1}{\hbar}\tilde{\Pi}_{q,q_{3}}} \pm i e^{\mp \frac{i}{\hbar} \Pi_{1,2}-\frac{1}{\hbar}\tilde{\Pi}_{2,3}} e^{\frac{1}{\hbar}\tilde{\Pi}_{q,q_{3}}}
\right) \label{eq:psiIII}
\end{align}
Repeating mutatis mutandis the steps detailed above for \eqref{eq:psiV}, going this time from $V$ to $III$, we find
\begin{equation}
\tilde{\psi}_{III}(q)= \frac{B}{\sqrt{\tilde{P}(q)}} \left(
2 \cos\left(\frac{1}{\hbar}\Pi_{3,4}\right) e^{\frac{1}{\hbar}\tilde{\Pi}_{q,q_{3}}} \pm i e^{\mp \frac{i}{\hbar} \Pi_{3,4}} e^{-\frac{1}{\hbar}\tilde{\Pi}_{q,q_{3}}}
\right)\label{eq:psiIII2}
\end{equation}
Identifying the coefficients of the dominant and recessive part in \eqref{eq:psiIII} and \eqref{eq:psiIII2}, i.e. the coefficients of the exponentially growing and decaying exponential, one gets the following system of two equations
\begin{align}
2 A \cos\left(\frac{1}{\hbar}\Pi_{1,2}\right) e^{\frac{1}{\hbar}\tilde{\Pi}_{2,3}}&= \pm i B e^{\mp \frac{i}{\hbar} \Pi_{3,4}} \\
\pm i A e^{\mp \frac{i}{\hbar} \Pi_{1,2}-\frac{1}{\hbar}\tilde{\Pi}_{2,3}} &= 2 B \cos\left(\frac{1}{\hbar}\Pi_{3,4}\right)
\end{align}
By dividing them together and simplifying, we get the EQC for this class of problems
\begin{equation}
\label{eq:appendixEQCminC}
e^{-\frac{i}{\hbar}\Pi_{2,3}}+\left(1+e^{\pm\frac{i}{\hbar}\Pi_{1,2}} \right)\left(1+e^{\pm\frac{i}{\hbar}\Pi_{3,4}} \right) = 0
\end{equation}
which is \eqref{eq:EQCquarticMinClat} indeed in the symmetric case, i.e. when $\Pi_{1,2}=\Pi_{3,4}$. 

\subsubsection{Maximal chamber}
We can easily deform our minimal chamber problem into a maximal chamber one by analytical continuation. By increasing the energy above the wells such that the turning points are no longer close to the real line or by sending the coupling constant in front of the quadratic term to zero, we are deforming the problem into the maximal chamber. Doing so, the periods in the symmetric case are transforming as
\begin{equation}
\Pi_{2,3} \mapsto - \Pi_{2,3}~, \quad \Pi_{1,2}^{\pm} \mapsto \frac{1}{2} \Pi_{1,4}^{\pm} \mp \frac{i}{2}\Pi_{2,3}
\end{equation}
which yields \eqref{eq:EQCquarticMaxClat} when applied to \eqref{eq:EQCquarticMinClat}.

\section{Computation of the quantum periods and differential operators}
\label{app:qperiods}
In section \ref{chap:res}, we compare the quantum corrections to the WKB periods obtained from the TBA side to standard QM results. The goal of this appendix is to present the standard QM computation we we used: the differential operators approach. We must note there are alternative methods one could have used in order to obtain the all-order WKB periods and check with the TBA results. Namely, the Holomorphic Anomaly equation \cite{Codesido:2016dld,santiThesis,Fischbach:2018yiu}, originating from topological strings.

\subsection{Classical periods}

The basic ingredients for finding the quantum corrections to the periods $\Pi^{(n)}_{(a)}$ are the classical periods $\Pi^{(0)}_{(a)}$. Even if one does not knows how to compute the exact classical periods in closed form for a given polynomial, one can still use the differential operators technique shown below numerically. However, it is always nice to have closed form results, so let us explicitly write the classical periods for the quartic and sextic potentials. In those cases, one can use Ramanujan's theory of elliptic functions in order to express our classical periods as the hypergeometric function $_2 F_1$ through the identity
\begin{equation}
_2 F_1 \left( \frac{1}{p}, 1-\frac{1}{p},1;z \right) = \frac{2}{\pi} \int_{0}^{\arcsin(\sqrt{z})} \frac{\cos\left(\left(\frac{2}{p}-1\right) \theta\right)}{\sqrt{z-\sin^2\left(\theta\right)}} d\theta
\end{equation}
as shown in \cite{Basar:2017hpr}. We also are providing the exact classical periods in closed form for pure potential in section \ref{chap:ExactClassPerPurePot}.

\subsubsection{Quartic} 
For the quartic of the form $V(q)=g q^4-q^2$ and energy $E$, we found
\begin{equation}
\Pi^{(0)}_{\text{p}}(g,E) = \frac{\pi  (4 E g+1)}{4 g} \, _2F_1\left(\frac{1}{4},\frac{3}{4};2;4 E g+1\right)
\end{equation}
\begin{equation}
\Pi^{(0)}_{\text{np}}(g,E) = i \sqrt{2} \pi  E \, _2F_1\left(\frac{1}{4},\frac{3}{4};2;-4 E g\right)
\end{equation}

\subsubsection{Sextic} 
For the sextic of the form $V(q)=q^2 \left(g_2^2 q^2-g_1^2\right)^2$ and energy $E$, we found
\begin{equation}
\Pi^{(0)}_{\text{p}}(g_1,g_2,E) = \frac{\pi  E  }{\sqrt{2} g_1^2} \, _2F_1\left(\frac{1}{3},\frac{2}{3};2;\frac{27 g_2^2 E }{4 g_1^6}\right)
\end{equation}

\begin{equation}
\Pi^{(0)}_{\text{np}}(g_1,g_2,E) =\frac{i \pi  g_1^4 }{9 \sqrt{6} g_2^2}\left(4-\frac{27 g_2^2 E }{g_1^6}\right)\, _2F_1\left(\frac{1}{3},\frac{2}{3};2;1-\frac{27 g_2^2 E }{4g_1^6}\right)
\end{equation}

\subsection{Quantum corrections to the periods}
The WKB method allows us to express the quantum corrections as derivative of the classical periods. For any polynomial potential of degree $d$, we can define
\begin{equation}
\tilde{p}^2(q,u)=V(q)-E = \sum_{k=0}^{d} u_k q^k
\end{equation}
where $u$ is the vector of the $u_k$. It is easy to show that 
\begin{equation}
\partial_{u_k} \tilde{p}(q,u) = \frac{q^k}{2 \tilde{p}(q,u)}
\end{equation}
where we will be omitting the dependency on the moduli from now on. As a result, one can write the $p^{(n)}$ obtained using the WKB expansion in \refeq{eq:formalSeriesP} (where they are denoted as $p_n$) as an ansatz plus a total derivative :
\begin{equation}
p^{(n)}(q) = \sum_{k=0}^{d} b_k^{(n)}\frac{q^k}{2 \tilde{p}(q)} + \partial_q \left(\frac{\mathcal{P}^{(n)}(q)}{\tilde{p}^{6n-3}(q)}\right)
\end{equation}
with $\mathcal{P}^{(n)}(q)= \sum_{k=0}^{d(3n-1)+1} c_k q^k$ a degree $(3n-1)d+1$ polynomial. By solving for the $b_k^{(n)}$ and $c_k$, one finds the differential operators that one needs in order to express the quantum corrections to the periods through
\begin{equation}
\Pi^{(n)} = \sum_{k=0}^{d} b_k^{(n)}\partial_{u_k} \Pi^{(0)}
\end{equation}
Notice that the system we solve for the $b_k^{(n)}$ and $c_k$ does not contain enough equation in order to fix every $b_k^{(n)}$. For example, in the following, we are expressing all the coefficients as functions of the free parameter $b_0^{(n)}$. Of course, we are free kill this symmetry by setting the remaining coefficient to the most convenient value for our purpose (often setting it to a value that is nullifying itself or another coefficient). In fact, it is always possible to put the Schrödinger equation in the simpler form \eqref{eq:SchroODEIM}, by shifting and rescaling, taking care of the spurious degree of freedom that is $u_{d-1}$ in this context, which in turn is effectively setting $b_0$ to some set value. One could argue it would have been cleaner to work with $u_{d-1}=0$ such that the $b_k^{(n)}$ coefficient are functions of the moduli only and that one can go back to any desired potential by a new scaling and shift; we preferred working with a general $u_{d-1}$ though, since the previous analysis is not harder to work out doing so and the resulting parameter allows us to put the $b_k^{(n)}$ coefficient we want to zero very easily.\\

As we said previously, it is not necessary to work out the classical periods in closed form in order to compute the quantum corrections: $\Pi^{(0)}$ is a function of the moduli (including the energy) $u_k$ which can be computed numerically as the integral of the momentum \eqref{eq:WKBperiodsSeries}, such that the derivative of the classical period can in turn be computed numerically using the definition of the derivative 
$$ \partial_{u_k} \Pi^{(0)} = \lim_{h \to 0} \frac{\Pi^{(0)}(\ldots,u_k+h,\ldots) -\Pi^{(0)}(\ldots,u_k,\ldots)}{h}  $$
then choosing a finite but small $h$, thus obtaining thousands of stable digits in seconds for the quantum corrections to the periods of any potentials, once the coefficients $b_k^{(n)}$ are computed.

\subsubsection{Quartic} 
Let's give a concrete example for the quartic $V(q)=g q^4-q^2$ with energy $E$. The first $b_k^{(n)}$ coefficients are
\begin{equation*}
b_2^{(1)} = \frac{8 b_0^{(1)}-\frac{g (4 E g+3)}{4 E g+1}}{4 E}~,~~~b_4^{(1)} = \frac{g \left(\frac{g}{4 E g+1}-3 b_0^{(1)}\right)}{E}
\end{equation*}
\begin{equation*}
b_2^{(2)} = \frac{2 b_0^{(2)}}{E}+\frac{g^2 (40 E g (6 E g+89)-21)}{1920 E^2 (4 E g+1)^3}~,~~~b_4^{(2)} = \frac{g^2 (15 E g (8 E g (22 E g-27)-9)-14)}{960 E^3 (4 E g+1)^3}-\frac{3 b_0^{(2)} g}{E}
\end{equation*}
\begin{align*}
b_2^{(3)} &= \frac{2 b_0^{(3)}}{E}+\frac{g^2 (7 E g (8 E g (16 E g (E g (9394 E g+11345)-7305)-3433)-3841)-1488)}{215040
	E^4 (4 E g+1)^5}\\
b_4^{(3)} &= -\frac{3 b_0^{(3)} g}{E}-\frac{g^2 (E g (7 E g (4 E g (4 E g (70836 E
	g-14465)+10235)+10413)+9479)+496)}{53760 E^5 (4 E g+1)^5}
\end{align*}
and so on. The unspecified (odd in that example) coefficients are zero, with the exception of the $b_0^{(n)}$ that are free parameters. Now, let's write explicitly and in closed form the first correction to the perturbative quantum period $\Pi^{(1)}_{\text{p}}$. Since we made the choice to put the coupling constant on the degree four term, it will be easier to chose $b_0^{(n)}$ by imposing $b_2^{(n)}=0$, such that the derivative with respect to $E$ and $g$ (modulo signs) will get us the correct corrections to the periods.
\begin{equation*}
\Pi^{(1)}_{\text{p}}(g,E) = \frac{\pi  \left((8 E g-2) \, _2F_1\left(\frac{1}{4},\frac{3}{4};2;4 E g+1\right)+3 E g (4 E g+1) \,
	_2F_1\left(\frac{5}{4},\frac{7}{4};3;4 E g+1\right)\right)}{128 E}
\end{equation*}
One can use the  $b_k^{(n)}$ coefficients computed above in order to compute the next corrections, or the non perturbative period in closed form. One can repeat this computation mutatis mutandis for the sextic potential and obtain a closed form result for an arbitrary number of quantum corrections. We computed them up to the order $O(\hbar^{12})$, but the expressions are becoming too involved too quickly to be contained in a readable manner in the present pages.
\section{Numerical method in standard Quantum Mechanics}
\label{app:CDil}

There is a plethora of numerical techniques in standard Quantum Mechanics we can use in order to solve the spectral problem described by \refeq{eq:Schro}. In the present work, we focus on bounded and resonant states in the context of polynomial potentials, for which expressing the Hamiltonian in the harmonic oscillator basis is easy and for which the truncation to a finite size Hamiltonian and diagonalization yield a good numerical approximation to the spectra and eigenvectors. This technique will be described in detail during the next section. The second section of this appendix extend this numerical method to resonant or Gamov states using the complex dilatation method.

\subsection{Bounded states and sprectra obtained by diagonalizing the Hamiltonian in the harmonic oscillator basis}
Anyone who has studied Quantum Mechanics knows that the eigenfunctions of the Schrödinger equation \refeq{eq:Schro}, when the potential is of the form $V(q) = \frac{1}{2} m \omega^2 q^2$, amount to the Hermite functions
\begin{equation}
\psi_n(x) = \frac{1}{\sqrt{2^n\,n!}} \left(\frac{m\omega}{\pi \hbar}\right)^{1/4}  e^{
	- \frac{m\omega x^2}{2 \hbar}}  H_n\left(\sqrt{\frac{m\omega}{\hbar}} x \right), \qquad n = 0,1,2,\ldots
\end{equation}
where
\begin{equation}
H_n(z)=(-1)^n~ e^{z^2}\frac{d^n}{dz^n}\left(e^{-z^2}\right)
\end{equation}
are the Hermite polynomials. In this basis, one can easily write the matrix elements of the Hamiltonian $H=H^\text{free}+H^\text{int} = - \frac{\hbar^2}{2m} \partial_q^2+V(\hat{q})$ with arbitrary polynomial potential $V(q) = \sum_{k=0}^{d} a_k q^k$. The free part is
\begin{equation}
H^\text{free}_{ij}=\frac{\hbar m \omega }{4} \begin{cases} 
2 i + 1 & i=j \\
-\sqrt{(i+1)(i+2)} & \abs{i-j}=2
\end{cases}
\end{equation}
and interaction part is
\begin{equation}
H^\text{int}_{ij}=\sum_{k=0}^{d} a_k \langle \psi_i | q^k | \psi_j \rangle
\end{equation}
where $\langle \psi_i | q^k | \psi_j \rangle = 0$ if $\abs{i-j}>k$ or if $i-j+k$ is odd,
\begin{equation}
\langle \psi_i | q^k | \psi_j \rangle = \frac{\sqrt{i! j!} \left(\frac{m \omega }{\hbar }\right)^{-k/2}}{\sqrt{2^{i+j}}} \sum_{l=0}^{L_{ijk}} \frac{2^{\frac{1}{2} (j-i-k+2 l)+i-l} k!  (i-j+k-2 l-1)\text{!!}}{l! (i-l)! (j-i+l)! (i-j+k-2 l)!}
\end{equation}
otherwise, $L_{ijk} = \min
\left(i,\left[\frac{1}{2} (i-j+k)\right]\right)$ and $[x]$ the integer part of $x$.  In order to get the exact spectrum, one must in principle diagonalize the infinite dimensional operator with elements $H_{ij}$, $i,j \in \mathbb{N}$. However, if only a numerical approximation is needed, it is sufficient to compute the Hamiltonian truncated to a finite dimension $N$, \emph{i.e.} the $N^2$ elements $H_{ij}^{(N)}$ with $i,j \in {0,\ldots,N}$. Since we are recovering the appropriate eigenvalue problem in the large $N$ limit, \emph{i.e.} $\lim_{N \to \infty} H^{(N)} = H$, the numerical approximation for an eigenvalue $E_{n\ll N}$ is getting closer and closer to the exact value as $N$ increase. Furthermore, since the matrix elements are mostly zeroes for $N \gg d$\footnote{To be precise, the truncated Hamiltonian is a symmetric band matrix with a bandwith of $d+1$ such that the number of a priori nonzero elements are increasing linearly with $N$ (as $(3+2d)N - (1+d)(2+d)$ in fact), which loses to $N^2$ eventually such that the matrix density goes to zero in the large $N$ limit.}, it is possible to use sparse matrix technique in order to solve the eigenvalue problem with algorithms that are a lot faster than their dense counterpart. As a result, we were able to compute hundreds of eigenvalues $E_n(\hbar)$ with a thousand of stable digits within minutes, given a fixed polynomial potential with reasonable maximum degree $d$.
\subsection{Resonant states and the complex dilatation method}
\label{app:CDilres}
We can accommodate the method described in the previous section for potentials that are no longer bounded but have \emph{resonances}. In Quantum Mechanics, we are typically sorting states into two distinguished classes: the scattering states and the bounded states. However, in many situations, we encounter scattering states that happen to be long lived; for example, a particle localized around a local minima of an unbounded potential: classically, the particle is confined in the well, but ultimately tunneling effects allow the particle to escape after a sufficiently long time. The spectrum resulting of such a problem is resonant \emph{i.e.} its levels has a nonzero imaginary part and are described by wave functions called \emph{Gamov states}. The physical interpretation being that, provided an energy in the form $E = E_R \pm i \Gamma/2$ with $\Gamma$ small enough, the Gamov state can be considered as a ``quasi-bound state'', exponentially decaying because of the factor $\exp\left(-\Gamma t/2\hbar\right)$ arising when we are considering its time evolution. In other words, the exponential decay law of a resonant state~--~arising because of the imaginary part of the energy associated with it~--~is identifiable with the exponential decay law of a metastable state with lifetime $\tau = 1/\Gamma$. One can interpret the resonances as the poles of the resolvent $G = (H-E)^{-1}$, that can be written as a function of the wave vector $k$ through the Jost functions, arising when $k$ is analytically continued in the lower complex plane $\Im k <0$. For the full story, see \cite{MarcosAQM}. \\

In order to extend the numerical method described above to resonant states, let's first describe the action of the group of dilatation on $\mathbb{R}^n$ with element $U_\theta$ on a wavefunction $\psi$:
\begin{equation}
	(U_\theta \psi)(q) = e^{\theta n/2} \psi(e^{\theta}q)
\end{equation}
Since we are interested in the one dimensional case, we will chose $n=1$ in the flowing. $U_\theta$ is a unitary transformation, acting on the Heisenberg operators $\hat{q}$ et $\hat{p}$ as 
\begin{equation}
U_\theta \hat{q} U_\theta^{-1} =  e^{\theta} \hat{q} \qquad\text{and}\qquad U_\theta \hat{p} U_\theta^{-1} =  e^{-\theta} \hat{p}
\end{equation}
which implies that its action on the Hamiltonian, if of the standard form $H=\frac{\hat{p}^2}{2m}+V(\hat{q})$, is
\begin{equation}
H(\theta) = U_\theta H U_\theta^{-1} =  e^{-2\theta} \frac{\hat{p}^2}{2m}+V( e^{\theta}\hat{q})
\end{equation}
This can be easily realized in the previous method since it is equivalent to the very simple transformation $H(\hbar,m \omega) \mapsto H(\hbar, e^{-2\theta} m \omega)$, such that our previous formulae still hold. Now, let's promote $\theta$ to a complex number. As a result, $U_\theta \ket{\psi}$ may not be in $L^2(\mathbb{R})$ anymore. At the condition that $\cos\left(2 \Im \theta\right)>0$ however, the dilated wavefunction of the form
\begin{equation}
(U_\theta \psi)(q) = e^{\theta n} \mathcal{P}(e^{\theta}q) \exp\left(- \alpha e^{2 \theta} q^2\right)
\end{equation}
with $\mathcal{P}$ a polynomial stays in $L^2(\mathbb{R})$. Physically, we can interpret this rotation of the Hamiltonian in term of the analytical continuation of the wave vector $k = \abs{k} e^{-\Im \theta} $. As a result, the complex dilated resolvent $G(\theta) = (H(\theta)-E)^{-1}$ provide an analytical continuation of the resolvent in the complex band $0>k>\Im -\theta$, whose poles are the resonances. It is equivalent to diagonalize the complex dilated Hamiltonian $H(\theta)$, which can be done numerically using the previously described method. For example, in table \ref{tab:SpectrumCubic}, we are using the method described in the previous section with the Hamiltonian rotated by the transformation $ \omega \mapsto e^{\frac{i \pi}{16}}\omega $.
\medskip

\bibliographystyle{unsrtnat}
\bibliography{biblio}

\end{document}